\documentclass[structabstract]{aa}  
\usepackage{graphicx}
\usepackage{txfonts}
\usepackage{natbib}
\usepackage{longtable}
\begin{document}
\authorrunning{V.~Mainieri et al.}
\titlerunning{Black hole accretion and host galaxies of obscured quasars in XMM-COSMOS}
   \title{Black hole accretion and host galaxies of obscured quasars in XMM-COSMOS.}

%   \subtitle{I. Overviewing the $\kappa$-mechanism}

   \author{V.~Mainieri\inst{1},
           A.~Bongiorno\inst{2},   
           A.~Merloni\inst{2,3},
           M.~Aller\inst{4},
           M.~Carollo\inst{4},
           K.~Iwasawa\inst{5},
           A.M.~Koekemoer\inst{6},
           M.~Mignoli\inst{7},
           J.D.~Silverman\inst{8},
           M.~Bolzonella\inst{7},
           M.~Brusa\inst{2},
           A.~Comastri\inst{7},
           R.~Gilli\inst{7},
           C.~Halliday\inst{9},
           O.~Ilbert\inst{10},
           E.~Lusso\inst{7,11},
           M.~Salvato\inst{12},
           C.~Vignali\inst{11},
           G.~Zamorani\inst{7},
           T.~Contini\inst{13},
	   J.-P.~Kneib\inst{10},
	   O.~Le F\`{e}vre\inst{10},
	   S.~Lilly\inst{4},
	   A.~Renzini\inst{14},
	   M.~Scodeggio\inst{15},
           I.~Balestra\inst{2},
	   S.~Bardelli\inst{7},
	   K.~Caputi\inst{16},
	   G.~Coppa\inst{7},
	   O.~Cucciati\inst{15},
	   S.~de la Torre\inst{16},
	   L.~de Ravel\inst{10},
	   P.~Franzetti\inst{15},
	   B.~Garilli\inst{15},
	   A.~Iovino\inst{17},
	   P.~Kampczyk\inst{4},
	   C.~Knobel\inst{4},
	   K.~Kova\v{c}\inst{18},
	   F.~Lamareille\inst{13},
	   J.-F.~Le Borgne\inst{13},
	   V.~Le Brun\inst{10},
	   C.~Maier\inst{4},
           P.~Nair\inst{7}
	   R.~Pello\inst{13},
	   Y.~Peng\inst{4},
	   E.~Perez Montero\inst{13},
           L.~Pozzetti\inst{7},
	   E.~Ricciardelli\inst{19},
	   M.~Tanaka\inst{8},
	   L.~Tasca\inst{10},
	   L.~Tresse\inst{10},
	   D.~Vergani\inst{7},
	   E.~Zucca\inst{7},
           H. Aussel\inst{26}
           P.~Capak\inst{20},
           N.~Cappelluti\inst{7},
	   M.~Elvis\inst{21},
           F.~Fiore\inst{23},
           G.~Hasinger\inst{12},
	   C.~Impey\inst{22},
           E. Le Floc'h\inst{27},
	   N.~Scoville\inst{20},
           Y.~Taniguchi\inst{24},
	   J.~Trump\inst{25}
 %         \fnmsep\thanks{Just to show the usage
 %         of the elements in the author field}
          }

   \institute{ESO, Karl-Schwarschild-Strasse 2, D--85748 Garching 
     bei M\"unchen, Germany
     \and
     Max-Planck-Institute f\"ur Extraterrestrische Physik,
     Postfach 1312, 85741, Garching bei M\"unchen, Germany
     \and
     Excellence Cluster Universe, TUM, Boltzmannstr. 2,
     D-85748 Garching  bei M\"unchen, Germany
     \and
     Institute for Astronomy, ETH Zurich, 8093, Zurich, Switzerland
     \and
     ICREA and Institut de Ci\`encies del Cosmos (ICC), Universitat de Barcelona (IEEC-UB), Mart\'i i Franqu\`es, 1, 08028 Barcelona, Spain 
     \and
     Space Telescope Science Institute, Baltimore, Maryland 21218, USA
     \and
     INAF$-$Osservatorio Astronomico di Bologna, Via
     Ranzani 1, I--40127 Bologna, Italy
     \and
     Institute for the Physics and Mathematics of the Universe (IPMU), University of Tokyo, Kashiwanoha 5-1-5, Kashiwa-shi, Chiba 277-8568, Japan
     \and
     INAF$-$Osservatorio Astrofisico di Arcetri, Largo Enrico Fermi 5, 50125 Firenze, Italy
     \and
     Laboratoire d'Astrophysique de Marseille, Marseille, France
     \and
     Dipartimento di Astronomia, Universita` di Bologna, via Ranzani 1, 40127 Bologna, Italy
     \and
     Max-Planck-Institute f\"ur Plasma Physis, Boltzmann Strasse 2, 85748 Garching bei M\"unchen, Germany
     \and
     Laboratoire d'Astrophysique de Toulouse-Tarbes, Universit\'{e} de Toulouse, CNRS, 14 avenue Edouard Belin, F-31400 Toulouse, France
     \and
     Dipartimento di Astronomia, Universita di Padova, Padova, Italy
     \and
     INAF$-$IASF Milano, Milan, Italy
     \and
     SUPA, Institute for Astronomy, The University of Edinburgh, Royal Observatory, Edinburgh EH9 3HJ, UK 
     \and
     INAF$-$Osservatorio Astronomico di Brera, Milan, Italy
     \and
     Max-Planck-Institute f\"ur Astrophysik, 85748 Garching bei M\"unchen, Germany
     \and
     Dipartimento di Astronomia, Universita di Padova, Padova, Italy
     \and
     California Institute of Technology, MC 105-24, 1200
     East California Boulevard, Pasadena, CA 91125 UA
     \and
     Harvard-Smithsonian Center for Astrophysics, 60
     Garden St., Cambridge, MA 02138 USA
     \and
     Steward Observatory, University of Arizona, 933 North Cherry Avenue, Tucson, AZ 85721
  \and
     INAF$-$Osservatorio Astronomico di Roma, via Frascati 33, 00040 Monteporzio-Catone, Italy 
     \and
     Research Center for Space and Cosmic Evolution,
     Ehime University, Bunkyo-cho 2-5, Matsuyama 790-8577, Japan
     \and
     UCO/Lick Observatory, University of California, Santa Cruz, CA 95064
     \and
     AIM Unit\'e Mixte de Recherche CEA  CNRS Universit\'e Paris VII UMR n158
     \and
     CEA-Saclay, Service d'Astrophysique, Orme des Merisiers, Bat.709, 91191
     Gif-sur-Yvette, France
             }

   \date{Received ...; accepted ...}

% \abstract{}{}{}{}{} 
% 5 {} token are mandatory
 
  \abstract
   {}
   {We explore the connection between black hole growth at the center of obscured quasars selected from the XMM-COSMOS survey and the physical properties of their host galaxies. We study a bolometric regime ($<$L$_{bol} > 8 \times 10^{45}$ erg s$^{-1}$) where several theoretical models invoke major galaxy mergers as the main fueling channel for black hole accretion.}
   {To derive robust estimates of the host galaxy properties, we use an SED fitting technique to distinguish the AGN and host galaxy emission. We evaluate the effect on galaxy properties estimates of being unable to remove the nuclear emission from the SED. The superb multi-wavelength coverage of the COSMOS field allows us to obtain reliable estimates of the total stellar masses and star formation rates of the hosts. We supplement this information with a morphological analysis of the ACS/HST images, optical spectroscopy, and an X-ray spectral analysis.}
   {We confirm that obscured quasars mainly reside in massive galaxies (M$_\star>10^{10}$ M$_\odot$) and that the fraction of galaxies hosting such powerful quasars monotonically increases with the stellar mass. We stress the limitation of the use of rest-frame color-magnitude diagrams as a diagnostic tool for studying galaxy evolution and inferring the influence that AGN activity can have on such a process. We instead use the correlation between star-formation rate and stellar mass found for star-forming galaxies to discuss the physical properties of the hosts. We find that at z$\sim 1$, $\approx 62\%$ of Type-2 QSOs hosts are actively forming stars and that their rates are comparable to those measured for normal star-forming galaxies. The fraction of star-forming hosts increases with redshift: $\approx 71 \%$ at z$\sim 2$, and $100\%$ at z$\sim 3$. We also find that the evolution from z$\sim 1$ to z$\sim 3$ of the specific star-formation rate of the Type-2 QSO hosts is in excellent agreement with that measured for star-forming galaxies. From the morphological analysis, we conclude that most of the objects are bulge-dominated galaxies, and that only a few of them exhibit signs of recent mergers or disks. Finally, bulge-dominated galaxies tend to host Type-2 QSOs with low Eddington ratios ($\lambda<0.1$), while disk-dominated or merging galaxies have at their centers BHs accreting at high Eddington ratios ($\lambda > 0.1$).}
   {}


   \maketitle

%
%________________________________________________________________

\begin{figure}
% \vspace*{-2.0 cm}
\begin{center}
 \includegraphics[width=3.4in]{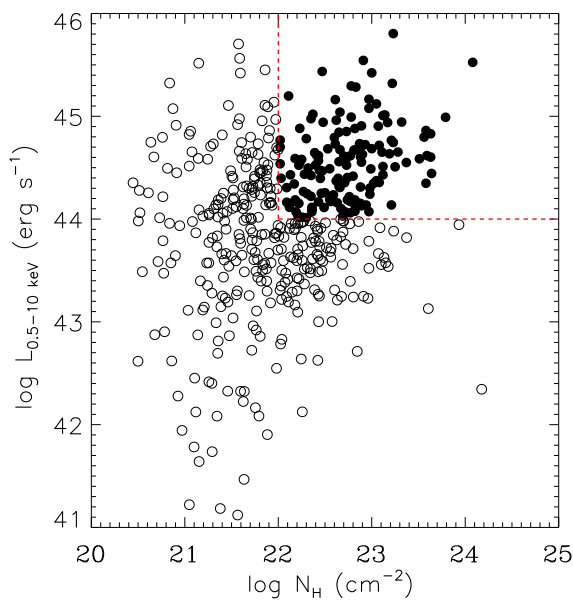}
% \vspace*{-1.0 cm}
 \caption{X-ray luminosity in the [0.5-10 keV] band corrected for
   obscuration plotted against the N$_{\rm H}$ column density for the
   X-ray sources in the XMM-COSMOS survey with a redshift. The 142
   Type-2 QSOs studied in this paper are indicated by filled
   circles. The dashed lines delimit the area where L$_{[0.5-10
       keV]}>10^{44}$ erg s$^{-1}$ and N$_{\rm H}>10^{22}$ cm$^{-2}$.}
   \label{lx_nh}
\end{center}
\end{figure}

\section{Introduction}

There is strong observational evidence that the formation and growth
of supermassive black holes (SMBHs) and their host galaxies are
related processes: e.g. the tight correlation between the SMBH mass
and the luminosity (\citealt{kormendy95}; \citealt{mclure02};
\citealt{marconi03}), mass (\citealt{magorrian98}), velocity
dispersion (\citealt{ferrarese00}; \citealt{gebhardt00};
\citealt{merritt01}; \citealt{tremaine02}), or the structure
(\citealt{graham01}) of the bulge\footnote{See \citet{peng07} and
  \citet{jahnke10} for alternative ideas on the origin of these
  scaling relations.}. Additional evidence of the connection between
the two phenomena is provided by the similarity between the global
star-formation rate (SFR) and active galactic nuclei (AGN) activity:
both peak at z$\approx 2$ and decline rapidly at lower
(e.g. \citealt{dickinson03}; \citealt{merloni04}) and higher
(e.g. {\citealt{wilkins08}; \citealt{brusa09}) redshifts. From the
  theoretical side, a number of models have been proposed to link the
  formation and evolution of SMBHs to the structure formation over
  cosmic time (e.g \citealt{kauffmann00}; \citealt{somerville01};
  \citealt{granato04}; \citealt{monaco05}; \citealt{granato06};
  \citealt{croton06}; \citealt{lapi06}; \citealt{hopkins06};
  \citealt{cen11}). Some of these semi-analytical models and
  hydrodynamical simulations (e.g. \citealt{springel05}) invoke major
  mergers of gas-rich galaxies as the mechanism enabling the fueling
  of the central SMBHs and the building of the galaxy's
  bulge. Alternative fueling mechanisms have been discussed in the
  literature, including minor-mergers (e.g. \citealt{johansson09}),
  bars (e.g. \citealt{jogee04}), disk instabilities
  (e.g. \citealt{genzel08}), and recycled gas from dying stars
  (e.g. \citealt{ciotti10}). \citet{hopkins09} argued that such
  alternative processes may be enough to fuel Seyfert-like AGNs, but
  that extreme mechanisms such as major mergers are necessary to
  provide the gas supply needed for bright quasars. Further more, some
  observational studies indicate that there is a morphological
  transition in AGN hosts around $L_{\rm bol}\approx
  10^{12}$L$_{\odot}$ between a mixture of disk and elliptical
  galaxies to exclusively bulge-dominated galaxies (see Fig. 1 of
  \citealt{hopkins09} and references therein). In this framework, the
  interest in studying the host galaxy properties of bright quasars is
  clear. These objects also dominate the space density of AGN at
  z$\approx 2$ (e.g. \citealt{ueda03}; \citealt{hasinger05}). An
  obvious complication in the study of the host galaxy properties of
  quasars is usually the emission of the central AGN which outshines
  the galaxy light making it extremely difficult to derive constraints
  on the colors, stellar populations, and morphologies of the
  host. The situation clearly improves if the powerful quasar is
  obscured by gas and dust on parsec scales: we refer to these objects
  as Type-2 QSOs. Although the fraction of obscured AGN is found to
  decrease with luminosity in several studies (X-ray:
  \citealt{ueda03}, \citealt{steffen03}, \citealt{hasinger04},
  \citealt{hasinger08}, \citealt{brusa10}; optical:
  \citealt{simpson05}; MIR: \citealt{maiolino07};
  \citealt{treister08}), a non-negligible population of obscured QSOs
  is still required by X-ray background synthesis models
  (e.g. \citealt{gilli07}). Radio-loud Type-2 QSOs have been observed
  for many years thanks to radio surveys (see \citealt{mccarthy93} for
  a comprehensive review), while radio-quiet Type-2 QSOs were detected
  more recently in {\it Chandra} and XMM-{\it Newton} X-ray surveys
  (\citealt{dawson01,norman02,mainieri02,stern02,dellaceca03,fiore03,perola04,tozzi06,barger05,mateos05,krumpe08,vignali10}),
  optical surveys (SDSS, \citealt{zakamska03}, \citealt{reyes08}), and
  mid-IR observations (\citealt{martinez06}; \citealt{polletta08};
  \citealt{lanzuisi09}). The aim of this paper is to extend to quasar
  luminosities the study of the host galaxy properties of AGN, using a
  large sample of X-ray selected Type-2 QSOs. In Section 2, we
  describe the selection criteria, while in Sections 3 and 4 we
  present the average spectral properties of the sample in the X-ray
  and optical band, respectively. Section 5 describes the method used
  to separate the host galaxy from the AGN emission. In Section 6, we
  illustrate the derived properties for the host galaxies, and in
  Section 7 discuss their morphological appearance. Finally, in
  Section 8 we summarize the results of this work. We quote in this
  paper magnitudes in the AB system, we use a Chabrier Initial Mass
  Function (\citealt{chabrier03}), and we assume a cosmology with H$_0
  = 70$ km s$^{-1}$ Mpc$^{-1}$, $\Omega _{\rm M} = 0.3$, and $\Omega
  _\Lambda = 0.7$.

\begin{figure}
% \vspace*{-2.0 cm}
\begin{center}
 \includegraphics[width=3.4in]{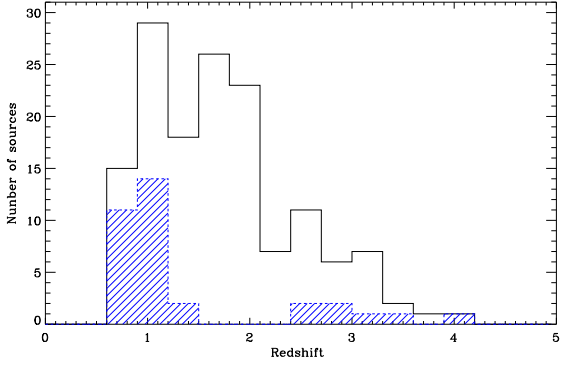} 
% \vspace*{-1.0 cm}
 \caption{Redshift distribution of the Type-2 QSO sample. The hatched
   histogram is for spectroscopic redshifts, while the empty one
   includes both spectroscopic and photometric redshifts. }
   \label{z_histo}
\end{center}
\end{figure}

\section{Sample selection}
\label{sample}

We started from the sample of $\sim$1800 point-like X-ray sources
detected in the XMM-COSMOS survey (\citealt{cappelluti09}) for which
optical identifications, multiwavelength photometry, and a compilation
of available redshifts from ongoing spectroscopic campaigns in the
COSMOS field (Lilly et al. 2009; Trump et al. 2009) were presented in
\citet{brusa10}. For the remaining sources very accurate photometric
redshifts are available (\citealt{salvato09}). For each X-ray source,
we performed a detailed X-ray spectral analysis to quantify their
X-ray luminosity and obscuration. The procedure adopted to extract
source and background spectra is equivalent to that described by
\cite{mainieri07}. We used
XSPEC\footnote{http://heasarc.gsfc.nasa.gov/docs/xanadu/xspec/}
(v11.3.2) to perform our spectral fitting analysis. We first fitted
the EPIC-pn data with two basic input models: a single {\it powerlaw}
({\it PL}) and a {\it powerlaw} modified by intrinsic absorption at
the redshift of the source ({\it APL}). The PL model consists of two
XSPEC components {\it wabs*zpowerlw}, while the APL model consists of
the combination of three different components {\it
  wabs*zwabs*zpowerlw}. In both models, the {\it wabs} component
accounts for the photoelectric absorption due to the Galactic column
density and is fixed to the average value in the COSMOS region
(N$_{\rm H}^{\rm Gal}\sim 2.7 \times 10^{20}$ cm$^{-2}$,
\citealt{dickey90})\footnote{This is an average value for the Galactic
  N$_{\rm H}$ in the COSMOS area, where N$_{\rm H}^{\rm Gal}$ is in
  the range [2.5-2.9] $\times 10^{20}$ cm$^{-2}$.  This range in
  Galactic column density does not affect the results of our spectral
  analysis.}. The {\it zwabs} component describes the photoelectric
absorption using Wisconsin cross-sections \citep{morrison83} and the
only two parameters are the equivalent hydrogen column density and the
source redshift. {\it zpowerlw} is a single power-law parametrized by
the photon index, the redshift, and a normalization factor.\footnote{
  We refer the reader to
  http://heasarc.gsfc.nasa.gov/docs/xanadu/xspec/ for further details
  on the spectral models.} Additional components are included to
describe the presence of soft-excess or the Fe K$\alpha$ emission
line. The model fits yield the power-law photon index $\Gamma$, and
the X-ray luminosity in both the [0.5-2] and [2-10] keV rest-frame
bands, and the {\it APL} model also provides the intrinsic column
density ${\rm N_H}$.

We used the following selection criteria to select Type-2 QSOs: a)
de-absorbed L$_{\rm X} [0.5-10~keV] > 10^{44}$ erg s$^{-1}$; b)
N$_{\rm H} > 10^{22}$ cm$^{-2}$ (see also
\citealt{mainieri02,szokoly04,tozzi06,krumpe08}). We excluded four
sources (XID$=82, 215, 2618, 5208$) that would formally match these
criteria but for which \citet{brusa10} proved that the XMM-{\it
  Newton} emission is the blend of two Chandra sources. The final
sample of X-ray selected Type-2 QSOs consists of 142
objects\footnote{Source XID-2028 discussed in \citet{brusa10} is not
  included in this paper because a re-analysis of the X-ray spectrum
  returned a column density value of log (N$_{\rm H}$)$=21.7\pm0.1$
  cm$^{-2}$.}.

\begin{figure}
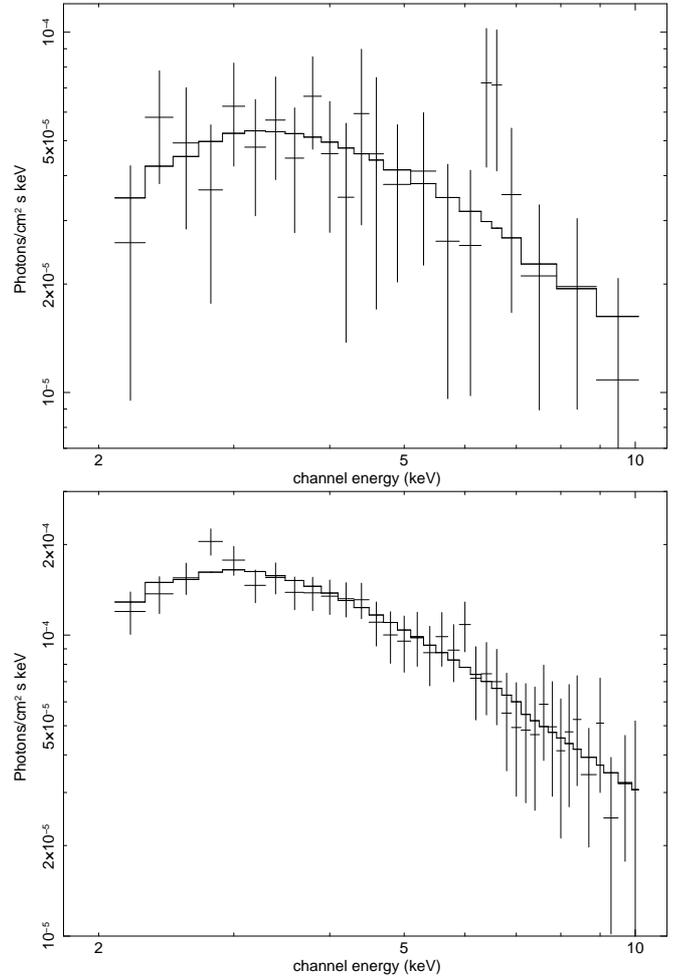

% \vspace*{-2.0 cm}
\begin{center}
 \includegraphics[angle=-90,width=3.4in]{fig3a.ps}
 \includegraphics[angle=-90,width=3.4in]{fig3b.ps} 
% \vspace*{-1.0 cm}
 \caption{Stacked X-ray spectrum of the 33 Type-2 QSOs with a
   spectroscopic redshift (left) and 79 Type-2 QSOs with only a
   photometric redshift (right). Adjacent bins were combined until
   they had a significant detection at least as large as
   1.5$\sigma$. The solid line is the best fit to the continuum with a
   {\it wabs*zwabs*powerlaw} XSPEC model. In the left panel, clear residuals
   are present around the rest-frame 6.4 keV energy.}
   \label{xstack}
\end{center}
\end{figure}

In Fig. \ref{lx_nh}, these candidates are plotted in the X-ray
luminosity versus column density plane. The redshift distribution of
the Type-2 QSO sample is shown in Fig. \ref{z_histo}: 33
($\approx23\%$) of them have a spectroscopic redshift. For this
sub-sample of sources, we verified that no broad lines (FWHM$>2000$ km
s$^{-1}$) are present in the optical spectrum. For the remaining
$77\%$ of the sample, we used the very accurate photometric redshifts
(\citealt{salvato09}).  The X-ray selected sample of Type-2 QSOs
presented in this work is complementary to the SDSS optically selected
one presented by \cite{zakamska03} and \citet{reyes08} in terms of
redshift range covered: the SDSS sample based on [OIII] luminosities
cannot be extended to redshifts higher than z$\approx 0.8$, while our
sample is almost entirely at z$>0.8$.

From the absorption-corrected luminosities in the rest-frame [2-10
  keV] energy band, we estimated the bolometric luminosities for the
Type-2 QSOs using the luminosity-dependent bolometric correction
factor ($k_{2-10 keV}$) of \citet{marconi04}. For the X-ray
luminosities sampled by our Type-2 QSOs, $k_{2-10 keV}$ is in the
range [25,120], and the bolometric luminosities cover the range
L$_{\rm bol}=[1.2 \times 10^{45}, 4.5 \times 10^{47}]$ erg s$^{-1}$
with a mean value of $<$L$_{\rm bol}> \approx 8 \times 10^{45}$ erg
s$^{-1}$. A limitation of this approach is the intrinsic spread in AGN
SEDs, which could lead to significant differences in the bolometric
correction from that predicted based on a mean energy distribution for
AGN (see \citealt{vasudevan07}; \citealt{lusso10}). From these
bolometric luminosities, and assuming a standard radiative efficiency
of $\epsilon_{\rm rad}=0.1$, we derive a median accretion rate onto
the black holes, $\dot M = L_{\rm bol}/\epsilon_{\rm rad}c^2$, of
approximately $\approx 1 $M$_\odot/$year, similar to what observed for
optically selected Type-1 QSOs
(e.g. \citealt{kollmeier06,hopkins09b}).

\section{Stacked X-ray spectrum}

In order to study the average X-ray properties of our Type-2 QSO
sample, we obtain a stacked X-ray spectrum using the EPIC-pn data. To
avoid the possibility of the blurring and broadening of the Fe
K$\alpha$ line caused by the uncertainties in the photometric
redshifts, we first considered only the 33 sources with a
spectroscopic redshift. The background-subtracted data of individual
objects in the energy range corresponding to the rest-frame 2-10 keV
were divided into 40 intervals so that each interval corresponds to a
rest-frame 200 eV. These spectra were corrected for the detector
response and summed together. The details of the procedure will be
presented in a future paper (Iwasawa et al. in preparation). The
smallest net count recorded in the rest-frame 2-10 keV for the
spectroscopic sample is 28 photons. The resulting stacked spectrum
(2682 net counts) is shown in the left panel of Fig. \ref{xstack}. It
was further rebinned for display purposes: adjacent bins were combined
until they had a significant detection at least as large as
1.5$\sigma$. From the stacked spectrum, we measured a column density
N$_{\rm H}=5.6^{8.3}_{3.3}\times 10^{22}$ cm$^{-2}$ (quoted errors are
$1\sigma$), which is consistent with the average value of $<$N$_{\rm
  H}>=5.4\times 10^{22}$ cm$^{-2}$ obtained from the fits to the
single X-ray spectra. As is clear from the left panel of
Fig. \ref{xstack}, an absorbed power-law model leaves large residuals
around the expected energy of the Fe K$\alpha$ line. If we add a
gaussian model centered on 6.4 keV in the XSPEC fitting procedure, the
quality of the fit improves significantly with $99\%$ confidence
according to an F-test. For the Fe K$\alpha$ equivalent width, we
measured a value of EW$=104$ eV. Taking into account the known
anti-correlation between the Fe K$\alpha$ and the X-ray luminosity
(\citealt{iwasawa93}), this value is consistent with typical EW values
for Compton-thin AGNs (\citealt{bianchi07};
\citealt{chaudhary10}).\\ We repeated this procedure for the
photometric redshift sample selecting objects that have more than 30
counts in the rest-frame 2-10 keV band. This selection leaves 79
objects out of 112 and the corresponding stacked X-ray spectrum (6,005
net counts) is shown in the right panel of Fig. \ref{xstack}. Owing to
the larger number of objects, the final spectrum has more counts and
consequently the uncertainties in the column density values are
smaller, N$_{\rm H}=4.7^{5.5}_{3.9}\times 10^{22}$ cm$^{-2}$. At the
same time, the significance of Fe K$\alpha$ is significantly lower,
and we ascribed this to the uncertainties, although small, in the
photometric redshifts that tend to blur out the signal. In fact, an
uncertainty in the photometric redshift value as low as
$\Delta_z=0.02\times$(1+z) would move the centroid of the 6.4 keV line
outside the rest-frame energy bin of 200 eV used in Fig. \ref{xstack}.

%\onecolumn
\begin{figure}
% \vspace*{-2.0 cm}
\begin{center}
 \includegraphics[width=3.4in]{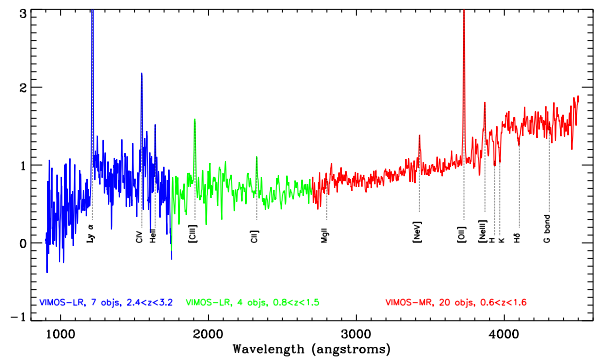} 
% \vspace*{-1.0 cm}
 \caption{Stacked optical spectrum of 31 Type-2 QSOs from the zCOSMOS
   survey. The LR and MR spectra have been grouped in redshift bins,
   as specified in the legend inside the figure, in order to have a
   common wavelength coverage between them. The more evident
   emission/absorption lines are labeled in the figure.}
   \label{optstack}
\end{center}
\end{figure}

\section{Optical spectroscopy}

Of the 33 Type-2 QSOs with a spectroscopic redshift, 31 have an
optical spectrum from the zCOSMOS survey, while the remaining two were
observed with the IMACS spectrograph at Magellan. We used the 31
zCOSMOS spectra to construct an average optical spectrum for the
Type-2 QSO sample. We grouped the spectra into redshift bins to ensure
that each bin had a common spectral range. The VIMOS low-resolution
(LR) spectra obtained with the LR-blue grism (3600-6800 \AA) were
divided into two redshift bins: seven objects in $2.4<{\rm z}<3.2$ and
four objects in $0.8<$z$<1.5$. An additional twenty medium-resolution
(MR) spectra (5550-9650 \AA) covering the redshift range $0.6<{\rm
  z}<1.6$ were stacked together. The stacked optical spectrum covers
the rest-frame wavelength range from 900 to 4500 \AA~ (see
Fig. \ref{optstack}).\\ Both [NeV]3426 and [OII]3727 are clearly
detected, and the ratio of the two lines is [OII]/[NeV]$\sim 4.1$. A
similar value for the lines ratio was measured for a [OIII]5007
selected sample of SDSS Type-2 QSOs by \citet{zakamska03}. In
addition, \citet{gilli10} found a ratio [OII]/[NeV]$>4$ for most of
the obscured quasars selected from the SDSS using the ratio of the
2-10 keV flux to the [NeV]3426 emission line flux.\\ We considered two
features in the stacked optical spectrum of Fig. \ref{optstack}: the
4000 \AA~ break, D$_n$(4000), and the H$\delta$ absorption
line. Because of the wavelengths of these two features, the following
discussion is limited to the 20 Type-2 QSOs in the redshift range
$0.6<{\rm z}<1.6$ with a VIMOS-MR spectrum, hence the mean optical
spectra may not be representative of the overall population of Type-2
QSOs. The D$_n$(4000) is a rough stellar age indicator, with small
values of the break for young stellar populations and large values of
the index for old stellar populations. Following the definition of
\citet{balogh99}, we derived the D$_n$(4000) as the ratio of the
fluxes measured in the [3850-3950] \AA~ to [4000-4100] \AA~ bands. In
the first band, we subtracted from the stacked spectrum the emission
associated with the [NeIII] line. The second age indicator is based on
the H$\delta$ absorption line, i.e.  the H$\delta_{\rm A}$ index
defined by \citet{worthey97}: the continuum is estimated from the
average flux in the two bands [4041.60,4079.75] \AA~ and
[4128.50,4161.00] \AA, and the index is expressed in terms of an
equivalent width integrating within the band [4083.50,4122.25] \AA. A
strong H$\delta$ absorption line is a signature of a large fraction of
young stars, hence a recent burst of star formation. We measured
D$_n$(4000)$=1.19\pm0.02$ and H$\delta_{\rm A}=4.7\pm0.4$. We can
compare these values with Fig. 7 of \citet{winter10}. These authors
used the two spectral features as age indicators in the study of local
AGN selected in the 14-195 keV band with SWIFT. Our values fall in a
region where the spectral models have a significant ($>30\%$)
contribution from a young stellar populations.

\begin{figure}
% \vspace*{-2.0 cm}
\begin{center}
 \includegraphics[width=3.4in]{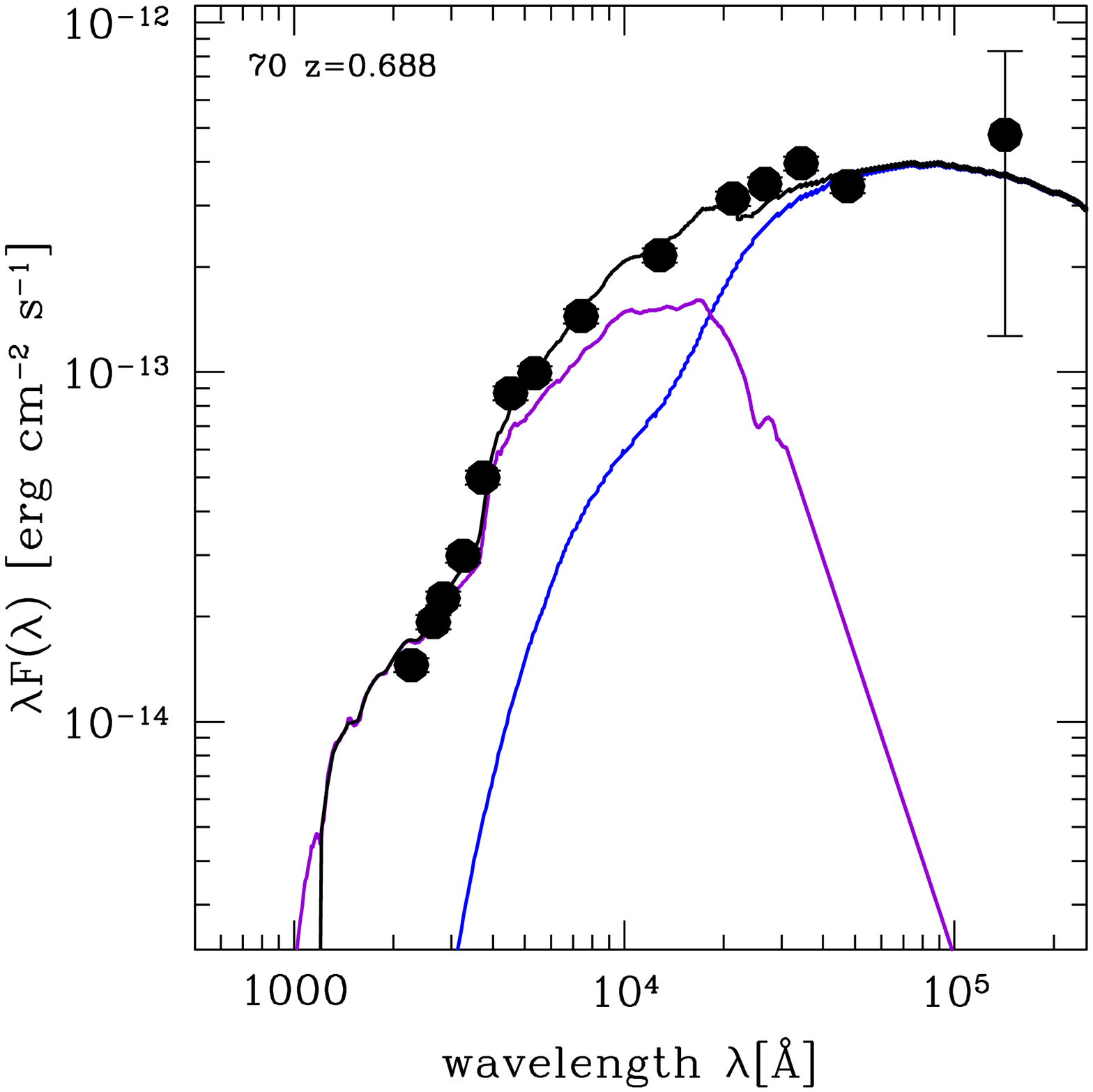} 
 \includegraphics[width=3.4in]{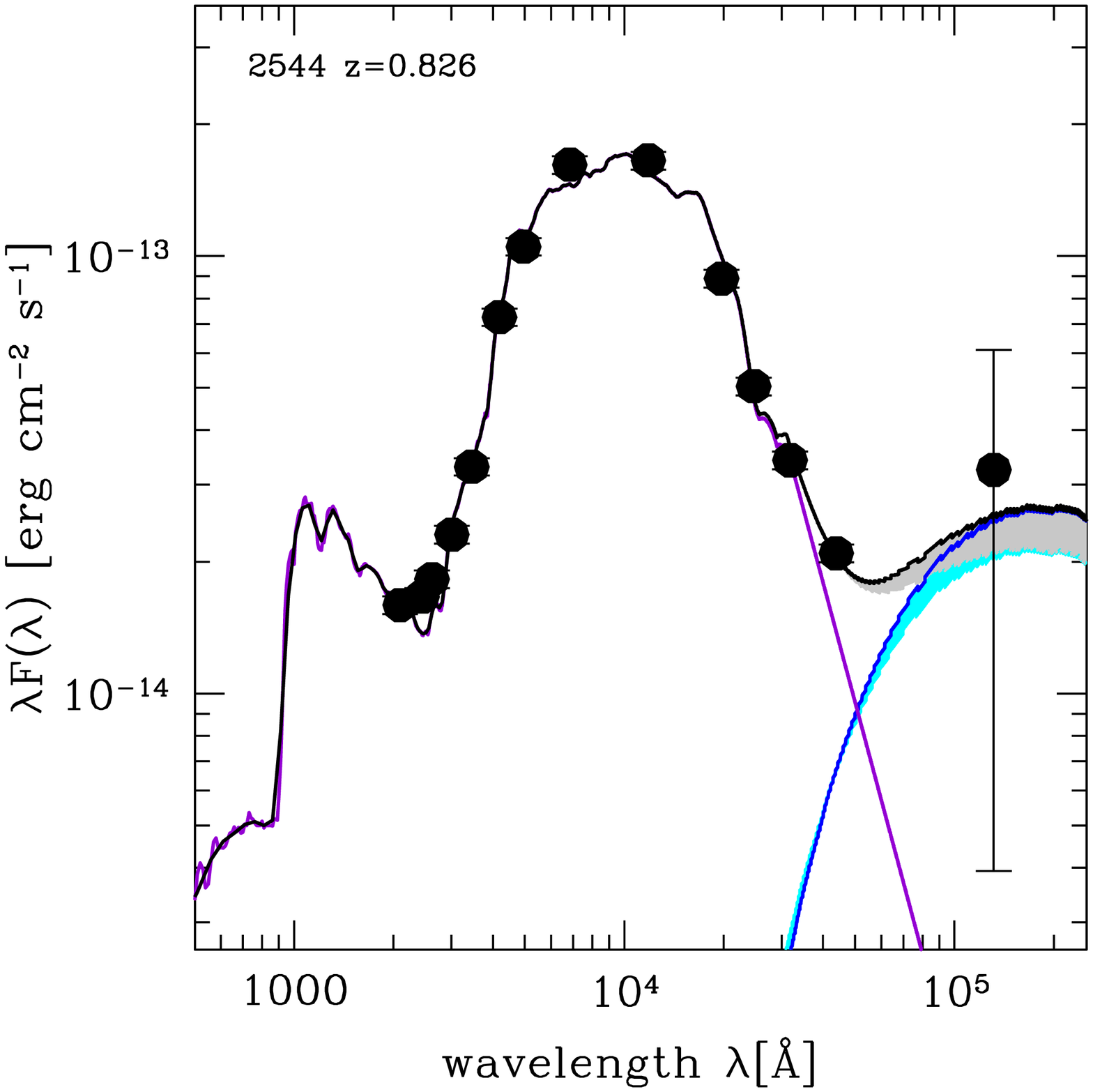} 
% \vspace*{-1.0 cm}
 \caption{Two examples of SED decomposition. Black circles are the
   observed photometry in the rest-frame. Purple and blue lines
   correspond respectively to the galaxy and AGN template found as the
   best fit solution, while the black line shows their sum.}
   \label{sed}
\end{center}
\end{figure}

\section{Distinguishing the AGN and host galaxy emission with SED fitting}
\label{sec:SED}

The main goal of our study is to use the rich multiwavelength coverage
of the COSMOS field to constrain the properties of the galaxies
hosting Type-2 QSOs. In particular, we wish to estimate their stellar
mass and star-formation rate (SFR). The technique we used is a
detailed model fitting of the total SED of the Type-2 QSOs for 14
different photometric bands that encompass optical to MIR wavelengths:
six SUBARU bands (B, V, g, r, i, z); U, J, and K bands from CFHT; four
Spitzer/IRAC channels, and 24$\mu m$ from Spitzer/MIPS (see
\citealt{brusa10} and references therein). This allows us to sample a
wide wavelength interval ranging from $\sim 3800 \AA$ (U$_{\rm CFHT}$)
to 24 $\mu m$. As already presented in \citet{merloni10}, we fitted
the observed SED with a grid of models compiled by combining AGN and
host galaxy templates. For the AGN component, we adopted the
\citet{richards06} mean QSO SED derived from the study of 259 IR
selected quasars with both Sloan Digital Sky Survey and Spitzer
photometry.  We allow for extinction of the nuclear AGN light by
applying a SMC-like dust-reddening law (\citealt{prevot84}) of the
form: $A_{\lambda}/E(B-V) = 1.39 \ \lambda_{\mu m}^{-1.2}$. The
$E(B-V)$ values for the AGN component are allowed to vary in the range
$1\le E(B-V) \le 9$. Assuming a Galactic gas-to-dust ratio, our
selection criteria N$_{\rm H}> 10^{22}$ cm$^{-2}$ would correspond to
$E(B-V)>1.7$, but there is evidence (e.g. \citealt{maiolino01}) that
the E(B-V)/N$_{\rm H}$ ratio in the circumnuclear region of an AGN may
be lower than the Galactic value, at least for part of the AGN
population. Our choice to set the lower boundary to $E(B-V)=1$ is a
compromise between allowing lower than Galactic values for the
dust-to-gas ratio and avoid degeneracies in the SED fitting procedure
caused by the inclusion of an unobscured AGN
template. Spectropolarimetry observations of Type-2 QSOs from the SDSS
\citep{zakamska05} and of narrow line radio galaxies
\citep{tadhunter02} show that in some sources there could be a
non-negligible constribution from scattered quasar light. Assuming
that the scattering process is independent of wavelength
\citep{vernet01}, the SED of an unobscured quasars should be a good
representation of this additional component. To quantify the effect of
AGN scattered light on the estimates of the host galaxies properties
presented in the next section, we repeated the SED fitting with $0\le
E(B-V) \le 9$ for the AGN component. The values for the host stellar
mass are essentially unchanged: for $97\%$ of the objects the
difference in stellar mass is less than $5\%$. Stellar population
synthesis (SPS) models, as the one of \citet{bruzual03} used in this
paper, combine individual simple stellar population spectra to match
the observed broad band photometry of a galaxy. Old stellar
populations are the main contributor to the total galaxy stellar mass,
and their SED dominates at wavelength above $\approx 600$ nm. This
explains why adding the SED of an unobscured AGN, that would
contribute significantly mostly in the UV, does not affect the stellar
mass estimates based on SPS models. As for the star-formation rate, we
find that scattered AGN light would change significantly the values of
SFR for $\approx 15\%$ of the objects. Therefore, in spite of the
possible multi-‐component nature of the optical continuum, the SED
fits unambiguously require the presence of a young stellar population
for the majority, $\approx 85\%$, of the objects.  We also verified
that the spread on SFR values, generated from the inclusion or not of
an unobscured AGN component in the SED fitting procudure, does not
affect the overall properties of the host galaxies population (see
Fig. 8, 9, 10, 11). As for the upper boundary in the reddening for the
AGN component, $E(B-V)<9$, it corresponds to the average X-ray column
density of our Type-2 QSO sample of N$_{\rm H}\sim5 \times 10^{22}$
cm$^{-2}$ for a Galactic dust-to-gas ratio. We verified that allowing
higher values of E(B-V) would not affect the values for the physical
quantities derived from the SED fitting, since an AGN template with
$E(B-V)=9$ has practically no significant impact below $\approx 2
\mu$m.\\ While \citet{merloni10}, dealing with unobscured QSOs,
assumed that all the 24 $\mu m$ flux is due to the AGN, we tried to
estimate the AGN contribution at those wavelengths using the results
of \citet{gandhi09}. These authors found a strong correlation between
the core luminosity at 12.3 $\mu m$ and the X-ray luminosity in the
[2-10 keV] band for a sample of local Seyfert galaxies. In the
mid-infrared, they used high-resolution observations with the
VISIR/VLT instrument, and owing to the superb angular resolution of
those images were able to minimize the contamination by unresolved
extended emission to the core fluxes. We used their Eq. (2) for the
best-fit correlation obtained by considering only the 22 well-resolved
sources

\begin{eqnarray}
log(\frac{L_{12.3 \mu m}}{10^{43}})=(0.19 \pm 0.05)+(1.11 \pm
0.07)log(\frac{L_{2-10 keV}}{10^{43}}).\nonumber
\end{eqnarray}

Using the formula above, we estimated the AGN flux at 12.3 $\mu m$
from the L[2-10 keV] luminosities corrected for obscuration obtained
from the X-ray spectral analysis. We associated an error in this
mid-infrared flux using the dispersion observed by \citet{gandhi09} in
the ratio of mid-infrared to X-ray luminosity, $\sigma (log
L_{MIR}/L_X)=0.18$ (see their Table 2 and Fig. 2). Finally, we
included in the SED fitting procedure the constraint that the AGN
component matches this predicted flux in the mid-infrared.\\ For the
host galaxy, we created a library of synthetic spectra using the
well-known stellar population synthesis models of
\citet{bruzual03}. We compiled ten exponentially declining
star-formation histories (SFH) $SFR\propto e^{-t_{\rm age}/\tau}$ with
e-folding times ($\tau$) ranging from 0.1 to 30 Gyr, plus a model with
constant star formation. For each of these SFH, we calculated the
synthetic spectrum at different ages, $t_{\rm age}$, ranging from 50
Myr to 5 Gyr, subject only to the constraint that the age should be
younger than the age of the Universe at the redshift of the
source. Finally, we included dust extinction, modeled using the
Calzetti's law (\citealt{calzetti00}), with values in the range $0\le
E(B-V) \le 0.5$.  Following \citet{fontana04} and \citet{pozzetti07},
we imposed the prior $E(B-V) < 0.15$ if $t_{\rm age}/\tau > 4$, a
significant extinction being allowed only for galaxies with a high
SFR. We adopted a \citet{chabrier03} initial mass function
(IMF)\footnote{Different choices of IMF lead to systematic shifts in
  the estimated stellar masses for any given SED, with the maximum
  shift for a Salpeter IMF (\citealt{salpeter55}) given by
  $M_{\star,Salpeter}\approx M_{\star,Chabrier}\times 1.7$ (see
  e.g. Pozzetti et al. 2007; Ilbert et al. 2010).}.\\ Finally, an
additional component that could contribute to the UV continuum of the
SED is nebular continuum ( see \citealt{dickson95}). For a sample of
ten narrow line radio galaxies, possibly radio-loud Type-2 QSOs,
\citet{tadhunter02} calculated the fractional contribution of this
componenent to the UV flux in the range $3-30\%$. In this work, we
will not consider the possible contribution of the nebular
continuum. We note that at least for stellar mass estimates the effect
of this component should be negligible: the intensity of the nebular
continuum drops quickly with wavelength and becomes extremelly faint
at $\lambda> 500$nm, where we expect to be more sensitive to the
galaxy stellar mass.

\begin{figure}
% \vspace*{-2.0 cm}
\begin{center}
 \includegraphics[width=3.4in]{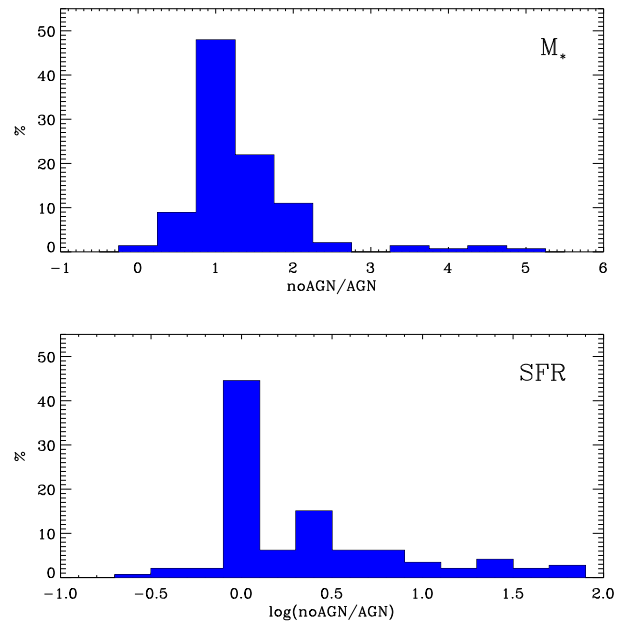} 
% \vspace*{-1.0 cm}
 \caption{Comparison between the estimates of M$_\star$ (top panel)
   and SFR (bottom panel) obtained with the SED fitting technique
   using only galaxies templates ({\it noAGN}) or a combination of
   galaxies and AGN templates ({\it AGN}).}
   \label{mstar_qso2}
\end{center}
\end{figure}

For each one of the Type-2 QSOs, we derived the best fit SED from the
minimization of the $\chi^2$ comparing the observed and template
fluxes at the redshift of the source. The combination of templates
allowed us to disentangle the emission of the host galaxy from the
contribution of the central black hole. We show in Fig. \ref{sed} two
examples of this composite fit. From the best fit SED, we used the
galaxy component to estimate both the stellar mass and SFR of the
host. For each one of these physical quantities, we estimated the
confidence region corresponding to $68\%$ of confidence level for one
interesting parameter, imposing that $\chi^2-\chi^2_{BEST-FIT}\le 1$
(\citealt{avni76}, and Table 1 therein). The best-fit values and
errors obtained are reported in Table 1. We flagged in the same table
the sources without a reliable fit to their SED for various reasons
(e.g. photometry contaminated by a nearby bright object).

\section{Host galaxy properties}

\subsection{Stellar mass}
\label{stellar_mass}
In this section we estimate the stellar masses of the Type-2 QSOs host
galaxies. Several previous studies of Type-2 AGN host galaxy
properties have assumed that the contamination of the AGN continuum to
their measurements is negligible (e.g. \citealt{silverman08};
\citealt{silverman09}; \citealt{schawinski10};
\citealt{xue10}). However, those studies were focused on moderate
luminosity AGNs, e.g. log(L$_{\rm X}<44$) erg s$^{-1}$. The AGN sample
presented in this work consists of obscured nuclei, but they are all
intrinsically powerful quasars, hence may still contribute in a
non-negligible way to the overall photometry. It is then important to
model this contamination, which would otherwise bias the measurement
of the stellar mass from the SED fitting. To quantify the effect of
the nuclear component, we performed the SED fitting using either the
combination of galaxy and AGN templates described in the previous
section as well as only galaxy templates. In the top panel of
Fig. \ref{mstar_qso2}, we compared the stellar mass estimates obtained
with the two sets of templates. While for the majority of the objects,
the introduction of an AGN component in the SED fit is irrelevant,
there are a $10\%$ of cases where neglecting these components would
cause more than a factor of two overestimation of the stellar masses.

From the stellar masses reported in Table 1, one can immediately see
that the hosts of Type-2 QSOs are mainly massive galaxies. To quantify
this impression, we need to compare the stellar mass distribution of
these hosts with a representative sample of galaxies that matches
their properties. We selected a parent sample of galaxies detected in
the 3.6 $\mu$m IRAC channel from the COSMOS survey using the stellar
masses measured by \cite{ilbert10}. We considered only the redshift
range z=[0.8,1.5], where a high completeness in stellar mass above
$\approx 5 \times 10^9$ M$_{\odot}$ and accurate photometric redshifts
are ensured (see Fig. 8 of \citealt{ilbert10}). The final parent
sample includes 76,061 galaxies. The sensitivity limits of the
XMM-{\it Newton} observations vary spatially across the field (see
Fig. 5 of \citealt{cappelluti09}), hence to derive the fraction of
galaxies hosting a Type-2 QSO as a function of the stellar mass, we
followed the technique discussed in Sec. 3.1 of \citealt{lehmer07}
(see also \citealt{silverman08,silverman09}). The Type-2 QSOs fraction
for each stellar mass bin is determined by summing over the sample of
Type-2 QSOs (N$_{\rm QSO2}$)

\begin{figure}
% \vspace*{-2.0 cm}
\begin{center}
 \includegraphics[width=3.4in]{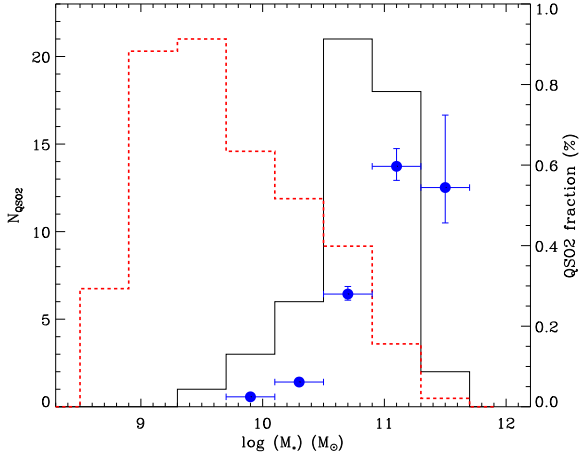} 
% \vspace*{-1.0 cm}
 \caption{Stellar mass distribution of the Type-2 QSOs host galaxies
   (solid histogram) and of the galaxies parent sample (dashed
   histogram) with $0.8<{\rm z}<1.5$. The second histogram has been
   rescaled for graphical reasons by a factor $\sim 1000$. The data
   points represent the fraction of galaxies hosting Type-2 QSOs in
   bin of stellar masses (see Sec. \ref{stellar_mass}).}
   \label{fig_stellar_mass}
\end{center}
\end{figure}

\begin{eqnarray}
f=\sum_{i=1}^{N_{\rm QSO2}}\frac{1}{N_{\rm gal,i}},
\end{eqnarray}

where N$_{\rm gal,i}$ represents the number of galaxies for which we
might have detected an AGN with X-ray luminosity L$_{\rm X,i}$. We
have plotted in Fig. \ref{fig_stellar_mass} the fraction of galaxies
hosting a Type-2 QSOs . This fraction monotonically increases with
stellar masses going from $0.02\%$ at M$_\star\approx 8\times10^{9}$
M$_\odot$ to $\approx 0.6\%$ at M$_\star\approx 10^{11}$ M$_\odot$.

This trend is consistent to what has been found for the overall AGN
population (\citealt{kauffmann03b}; \citealt{best05};
\citealt{silverman09}; \citealt{brusa09}; \citealt{xue10}).

For our sample of Type-2 QSOs, we do not detect broad lines and
therefore we cannot derive ``virial'' black hole masses using the
calibration obtained with reverberation mapping on local AGN
(e.g. \citealt{kaspi00}). To roughly estimate the mass of the black
holes powering these obscured quasars, we used the local M$_{\rm
  BH}$-M$_\star$ scaling relation provided by \citet{haring04}.  Since
this relation was derived locally, we took into account a possible
redshift evolution parametrized by \citet{merloni10}: $\Delta
log(M_{\rm BH}/M_\star)(z)=\delta_2 log(1+z)$, where $\delta_2=0.68
\pm 0.12$. We found that most ($\sim 90\%$) of the Type-2 QSOs have
black hole masses in the range log(M$_{\rm BH})=[7.5,9.5]$
M$_\odot$. Using these estimates of black hole masses and the
bolometric luminosities derived at the end of Sec. \ref{sample}, we
computed the Eddington ratio ($\lambda_{\rm Edd}=L_{\rm bol}/L_{\rm
  Edd}$, where L$_{\rm Edd}=1.38 \times 10^{38} M_{\rm
  BH}/M_\odot$). We found that $\approx 50\%$ of the Type-2 QSOs have
$\lambda_{\rm Edd}>0.1$, the remaining half being in the range
$0.01<\lambda_{\rm Edd}<0.1$.

\subsection{Rest-frame colors}
\label{sec:colors}

Galaxies display a colour bimodality in both the local Universe
(\citealt{strateva01}; \citealt{hogg02}; \citealt{blanton03}) and at
higher redshift (\citealt{bell04}, \citealt{weiner05} up to z$\approx
1$; \citealt{franzetti07}, \citealt{cirasuolo07} up to z$\approx 1.5$;
\citealt{giallongo05}, \citealt{cassata08}, \citealt{williams09} up to
z$\approx 2$). This bimodality in {\it red-sequence} and {\it
  blue-cloud} galaxies has been interpreted as evidence of a dichotomy
in the star-formation and merging histories of galaxies
(e.g. \citealt{menci05}). Color-magnitude diagrams have been used as
tools in galaxy evolution studies, and since many models invoke AGN
feedback as an important player in this evolution, it may be
interesting to locate the hosts of Type-2 quasars in those diagrams.

From the galaxy component of the best fit to the Type-2 QSO SED (see
Sec. \ref{sec:SED}), we derived rest-frame colors for the hosts in the
Johnson-Kron-Cousins system. The inclusion of an AGN component in our
SED fitting procedure prevents contamination from the nuclear emission
to the colors of the host galaxy\footnote{See the discussion in
  Appendix A of \citet{hickox09} about the contribution of the nucleus
  to optical colors of X-ray and radio-selected AGNs if the SED
  fitting is done using only galaxy templates.}.

\begin{figure}
% \vspace*{-2.0 cm}
\begin{center}
 \includegraphics[width=3.4in]{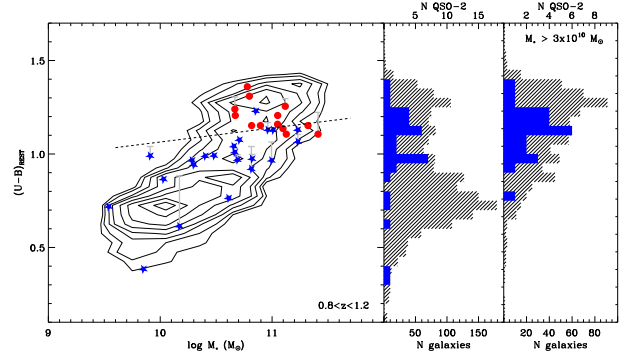} 
% \vspace*{-1.0 cm}
 \caption{{\it Left panel.} Color-mass diagram in the redshift range
   $0.8<z<1.2$ for the comparison sample of normal galaxies (gray
   contours), and the Type-2 QSOs hosts. The rest-frame colors of the
   Type-2 QSOs hosts have been measured after removing the
   contribution from the AGN. Star-forming galaxies are indicated with
   the star symbol, while the circles correspond to passive
   galaxies. The dashed curve is the dividing line between red passive
   and blue star-forming galaxies at z$=1$ according to
   \citet{peng10}. The vertical bars show the variation in rest-frame
   colors if an unobscured AGN component is included in the SED
   fitting procedure (see Sec. 5). {\it Central panel.}  U-B
   rest-frame color distribution for the Type-2 QSOs host galaxies
   (filled histogram) and the comparison sample of normal galaxies
   (hatched histogram) in the redshift range $0.8<z<1.2$. {\it Right
     panel.} As in the central panel but only for sources with
   M$_{\star}>3 \times 10^{10}$ M$_{\odot}$.}
   \label{rest_frame_col}
\end{center}
\end{figure}

From the histogram in the central panel of Fig. \ref{rest_frame_col},
one could conclude that the host galaxies of Type-2 QSOs mainly have
rest-frame colors in-between the blue and red galaxy populations,
similarly to what has already been reported for lower luminosity
X-ray/optically selected AGNs (\citealt{nandra07}; \citealt{martin07};
\citealt{rovilos07}; \citealt{silverman08}; \citealt{georgakakis08};
\citealt{schawinski09}; \citealt{hickox09}). \citet{silverman09}
demonstrated that if one considers stellar-mass rather than luminosity
selected samples of AGN and parent galaxies, the fraction of galaxies
hosting an AGN does not decline towards bluer colors. \citet{xue10}
confirmed that the AGN fraction remains nearly constant as a function
of color for a stellar mass selected sample of AGN from the Chandra
Deep Fields. Both these works focused on AGN with luminosities below
L$_{\rm X}<10^{44}$ erg s$^{-1}$. Our sample of Type-2 QSOs is by
construction brighter and therefore care is required when
extrapolating these previous results. If we select only the objects
with M$_\star>4\times10^{10}$M$_\odot$ for which our COSMOS sample is
mostly complete (see \citealt{ilbert10}), the distribution of
rest-frame colors is modified as shown in the right panel of
Fig. \ref{rest_frame_col}. In a mass-selected sample of galaxies the
distribution of rest-frame colors for active and inactive galaxies is
similar. Since stellar mass is likely playing a major role in
determining the evolution of galaxies, mass-selected samples are more
suitable when comparing the properties of the AGN hosts with those of
inactive galaxies.

If we adopt the dividing line between {\it blue star-forming} and {\it
  red passive} galaxies as suggested by \cite{peng10}

\begin{eqnarray}
(U-B)_{AB}=1.10+0.075 log(\frac{m}{10^{10}M_\odot})-0.18z,\nonumber
\end{eqnarray}

using $z=1.0$, we find that $\approx 28\%$ of the host galaxies of
Type-2 QSOs in the redshift range $0.8<z<1.2$ are {\it red} and
$\approx 72\%$ are {\it blue}. If were to use the (U-B) versus M$_B$
diagram and the \cite{willmer06} color division, the fraction of
Type-2 QSOs hosts that are classified as {\it red} galaxies would
increase to $\approx 58\%$. We stress that the use of rest-frame
colors to study the properties of AGN host galaxies is subject to a
proper treatment of reddening. Objects with intermediate or red colors
could actually be star-forming galaxies obscured by large amounts of
dust on kpc scales (e.g. see \citealt{brammer09};
\citealt{pozzetti10}). As we describe in the next section, we were
able to derive estimates of the SFR of our host galaxies from the SED
fitting. We can therefore use the specific star-formation rate
(SSFR=SFR/M$_\star$) to divide the sample into active and quiescent
galaxies depending on whether log(SSFR/Gyr$^{-1}$) is above or below
-1, i.e., galaxies that would take less or more than 10 Gyr to double
their stellar mass at the present SFR, respectively (see
\citealt{pozzetti10}). We find that $\approx 20\%$ of the {\it red}
galaxies are actively forming stars, and therefore their colors are
due to the presence of dust rather than an old stellar population. Our
result is in agreement with previous studies. \citet{cardamone10},
using rest-frame near-infrared colors of lower luminosity AGN hosts to
separate passive from dusty galaxies, found that $\approx 25\%$ of the
hosts in the red cloud are dusty star-forming galaxies;
\citet{brusa09b} from the SED fitting of a sample of obscured AGN in
the Chandra Deep Field South reached the conclusion that the fraction
of dusty star-forming hosts was up to $\approx 50\%$ of their sample.

\begin{figure}
% \vspace*{-2.0 cm}
\begin{center}
 \includegraphics[width=3.4in]{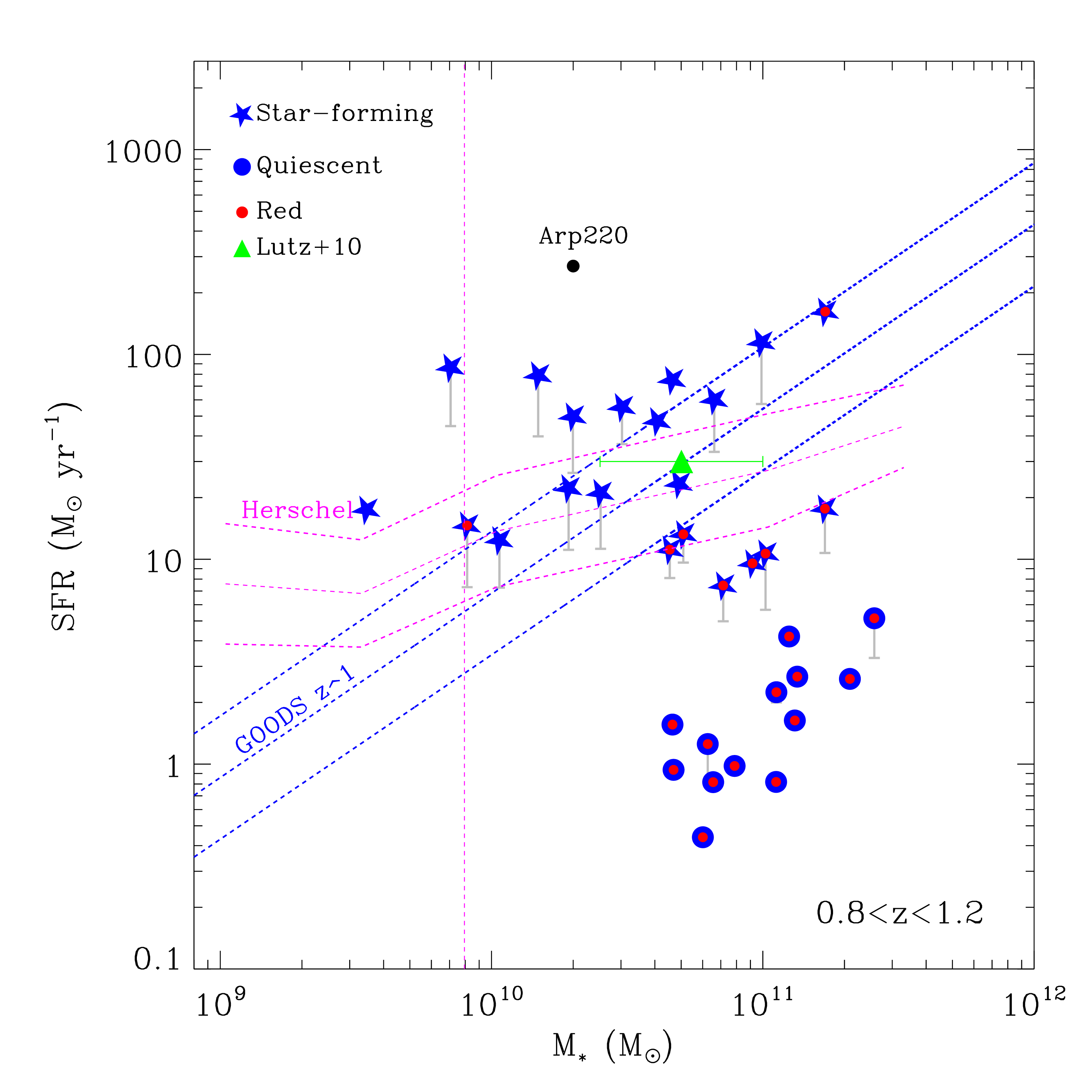} 
% \vspace*{-1.0 cm}
 \caption{SFR versus stellar masses obtained from the SED fitting for
   the host galaxies of the Type-2 QSOs in the redshift range
   $0.8<$z$<1.2$. Galaxies with log(SSFR/Gyr$^{-1}$)$>-1$ are
   indicated with a star symbol, otherwise a circle was used. Type-2
   QSO hosts with red (U-B) rest-frame colors according to the color
   division by \citet{willmer06} have been highlighted with a smaller
   lighter circle. The vertical bars show the variation in SFR if an
   unobscured AGN component is included in the SED fitting procedure
   (see Sec. 5). The dashed lines are the correlation and its
   one-$\sigma$ confidence intervals found for blue star-forming
   galaxies by \citet{elbaz07} and PACS/Herschel sources by
   \citet{rodighiero10} at z$\approx 1$. The curves have been rescaled
   from the original Salpeter IMF to the Chabrier IMF adopted in this
   work. The vertical dashed line indicates the mass completeness
   limit of the PACS sample. The triangle indicates the average SFR
   for the hosts of X-ray selected AGNs estimated by \citet{lutz10} by
   stacking 870 $\mu$m LABOCA/APEX data. }
   \label{sfr_smass1}
\end{center}
\end{figure}

\begin{figure}
% \vspace*{-2.0 cm}
\begin{center}
 \includegraphics[width=3.4in]{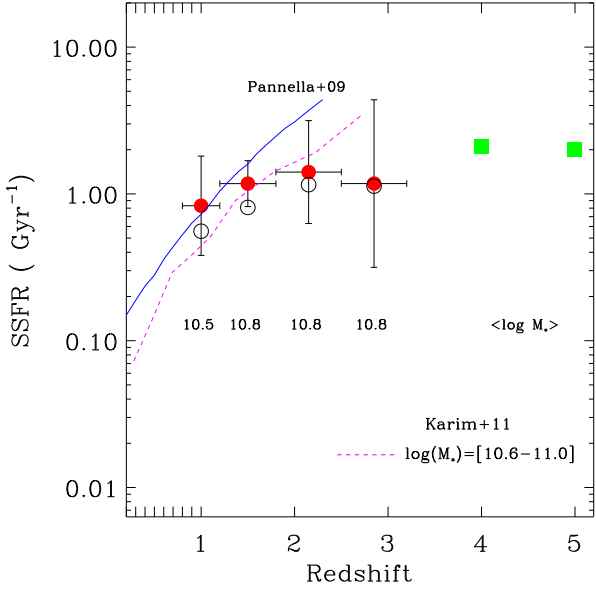} 
% \vspace*{-1.0 cm}
 \caption{Evolution of the average SSFR for Type-2 QSOs hosts as a
   function of redshift. For each redshift bin we report the average
   stellar mass of the host galaxies. Empty circles indicate the
   average SSFR per redshift bin if an unobscured AGN component is
   included in the SED fitting procedure (see Sec. 5). The continuum
   line reproduces the evolution law found by \citet{pannella09} for
   star-forming galaxies up to z$\sim 2$. The dashed line instead is
   the best fit to the SSFR redshift evolution derived by
   \citet{karim11} in the mass range log(M$_{\star}$)=[10.6,11.0]
   M$_{\odot}$. The two squares are the SSFR for normal galaxies at
   z$=[4,5]$ as presented by \citet{gonzalez10} using data from
   \citet{stark09}}
   \label{ssfr_z1}
\end{center}
\end{figure}

 \begin{figure}
% \vspace*{-2.0 cm}
\begin{center}
 \includegraphics[width=3.4in]{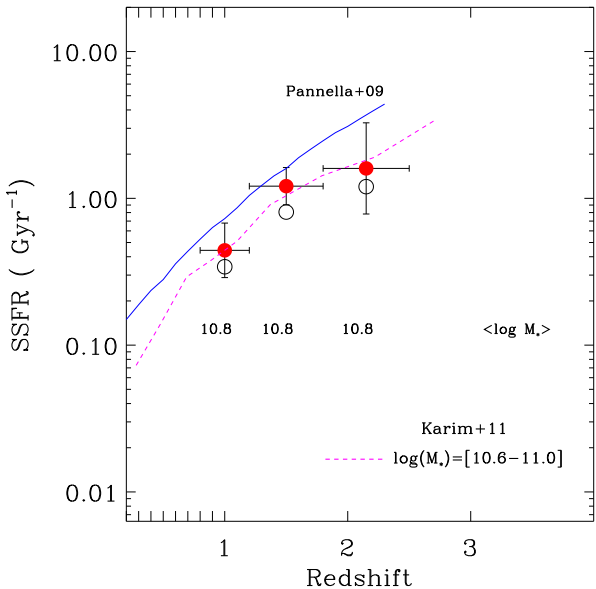} 
% \vspace*{-1.0 cm}
 \caption{Evolution of the average SSFR for Type-2 QSOs hosts with
   log(M$_{\star}$)=[10.6,11.0] M$_{\odot}$ from z$\sim 1$ to z$\sim
   2$. The lines and symbols are the same as in Fig. \ref{ssfr_z1}.}
   \label{ssfr_z2}
\end{center}
\end{figure}

\subsection{SFR}

As described in Sec. \ref{sec:SED}, we estimated the SFR values of the
host galaxies using the SED fitting technique. We stress that to
reliably estimate the SFR of the host galaxy it is crucial to remove
the contribution of the AGN to the observed photometry. As already
discussed, we achieved this by performing the SED fitting with a
combination of AGN and galaxy templates. To quantify the effect of not
modeling the nuclear emission, we compared the SFR estimates obtained
only with galaxy templates and a combination of galaxies and AGN SEDs,
as discussed in Sec. \ref{stellar_mass}. We find that for $\approx
40\%$ of the objects neglecting the AGN component will overestimate
the SFR by more than a factor of two, and up to a factor of fifty for
a few galaxies (see bottom panel of Fig. \ref{mstar_qso2}).\\ For
comparison, we also estimated the SFR using the UV continuum
luminosity. Using the prescriptions of \citet{daddi04} for
star-forming galaxies at $1.4<$z$<2.5$ and assuming a Chabrier IMF, we
have

\begin{eqnarray}
SFR(M_\odot yr^{-1}) & = & L_{1500 \AA}^{corr}/8.85\times10^{27}/1.7,\nonumber\\
L_{1500 \AA}^{corr} & = & L_{1500 \AA}\times 10^{0.4\times
  A_{1500}},\nonumber
\end{eqnarray}

where L$_{1500 \AA}$ is the galaxy luminosity at 1500 $\AA$
(erg/s/Hz), L$_{1500 \AA}^{corr}$ is corrected for extinction by dust,
and A$_{1500}\approx 10 \times E(B-V)$ assuming a Calzetti's law for
the dust extinction, where we use the E(B-V) values derived from the
best SED fit using both galaxies and AGN templates. Considering the
$\approx 60$ QSO-2 hosts that match the \citet{daddi04} selection
criteria, we find that the median ratio (SFR$_{SED}$/SFR$_{UV}$) is
$\approx 1.1$ with a dispersion of $\approx 0.6$.

One of the main goals of this work was to compare the level of star
formation in the hosts of the Type-2 QSOs with that measured for
``non-active'' galaxies. Surveys of the local Universe
(\citealt{brinchmann04}; \citealt{salim05}) found that galaxies can be
divided mainly into two populations: star-forming objects exhibiting a
continuous increase in SFR with M$_{\star}$ along a tight correlation
and {\it quenched} galaxies with little star formation and high
stellar masses. The tight correlation between SFR and M$_{\star}$ for
star-forming galaxies has been confirmed to already be in place at
higher redshifts (\citealt{noeske07}, and \citealt{elbaz07} up to
z$\approx 1$; \citealt{daddi07}, and \citealt{pannella09} up to
z$\approx 2$), also when using PACS/Herschel observations
(\citealt{rodighiero10}). We refer to it as the main sequence (MS) of
star-forming galaxies following \citet{noeske07}. The slopes of the MS
at different redshifts appear to be the same but the normalization has
decreased by a factor $\approx 3.7$ going from z$\approx 2 $ to
z$\approx 1$, and an additional factor $\approx 7$ at z$\approx 0$
(\citealt{daddi07}). This decrease in the activity of star-forming
galaxies of a given stellar mass reflects the decline in the global
SFR density with time, and a likely cause of this is gas exhaustion
(e.g. \citealt{noeske07b}).\\ The galaxy dichotomy found using the
correlation between SFR and M$_{\star}$ is more fundamental than that
based on rest-frame colors: red colors do not necessarily mean that
the galaxy has stopped forming stars (see the discussion in
Sec. \ref{sec:colors}, and the red colors of some MS galaxies at high
M$_{\star}$ in Fig. 17 of \citealt{elbaz07}). Our aim is to compare
the properties of the host galaxies with a parent sample of normal
galaxies in the context of this correlation.\\ In the redshift range
$0.8<$z$<1.2$, we have 35 Type-2 QSOs; and we were able to establish a
good fit to their SED for 34\footnote{Source 5326 has near-IR
  photometry contaminated by a nearby source and will not be
  considered in this analysis.} of which. We find that $\approx 62\%$
of the Type-2 QSO host in this redshift range are actively forming
stars. In the sample of star-forming galaxies used by \citet{elbaz07}
to derive the MS at z$\approx 1$, only ``blue'' galaxies were
considered (see their Fig. 13). For consistency, we derived rest-frame
absolute magnitudes in the standard U and B Johnson's filters from the
galaxy component of the SED fit of our Type-2 QSO hosts, and separated
them into blue and red galaxies using the color division by
\citet{willmer06}. As shown in Fig. \ref{sfr_smass1}, there is a mild
correlation between SFR and M$_{\star}$ for the ``blue'' star-forming
hosts. The Spearman rank correlation coefficient for these sources is
r$=0.38$ (p$=0.2$). For the majority of our ``blue'' star-forming host
galaxies, the level of SSFR at z$\approx 1$ is consistent with the MS
of star-forming galaxies at the same redshift. In
Fig. \ref{sfr_smass1}, we also indicated, the average SFR estimated by
\citet{lutz10} for the hosts of X-ray selected AGN at z$\sim 1$
obtained by stacking 870 $\mu$m LABOCA observations in the Extended
Chandra Deep Field South. Most of these hosts have stellar masses in
the range $10.4<$log(M$_{\star})<11$ M$_{\odot}$
(\citealt{xue10}). From this stacking analysis, the derived value of
SSFR for AGN hosts is also consistent with the MS of star-forming
galaxies at z$\sim 1$.

\begin{figure}
% \vspace*{-2.0 cm}
\begin{center}
 \includegraphics[width=1.in]{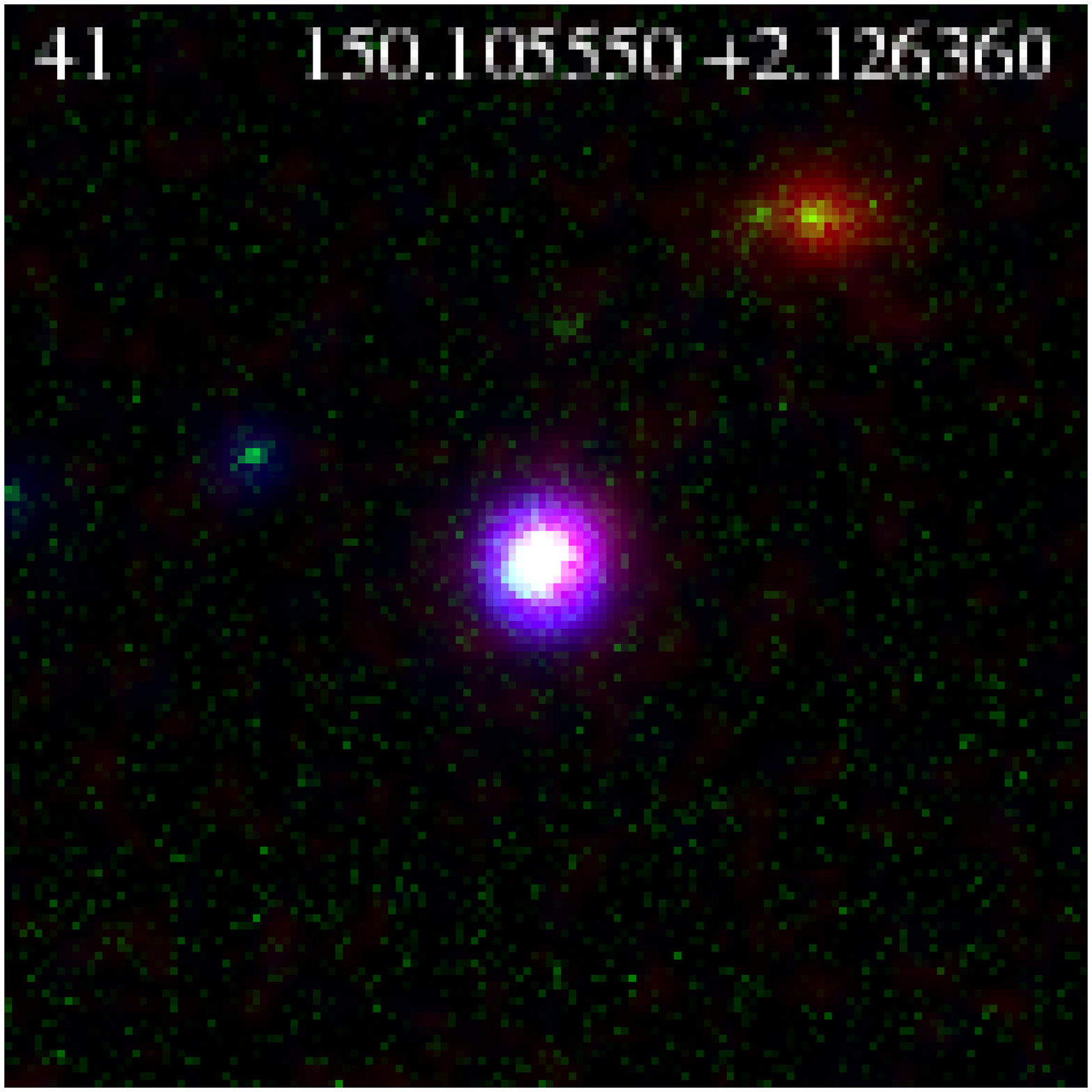}
 \includegraphics[width=1.in]{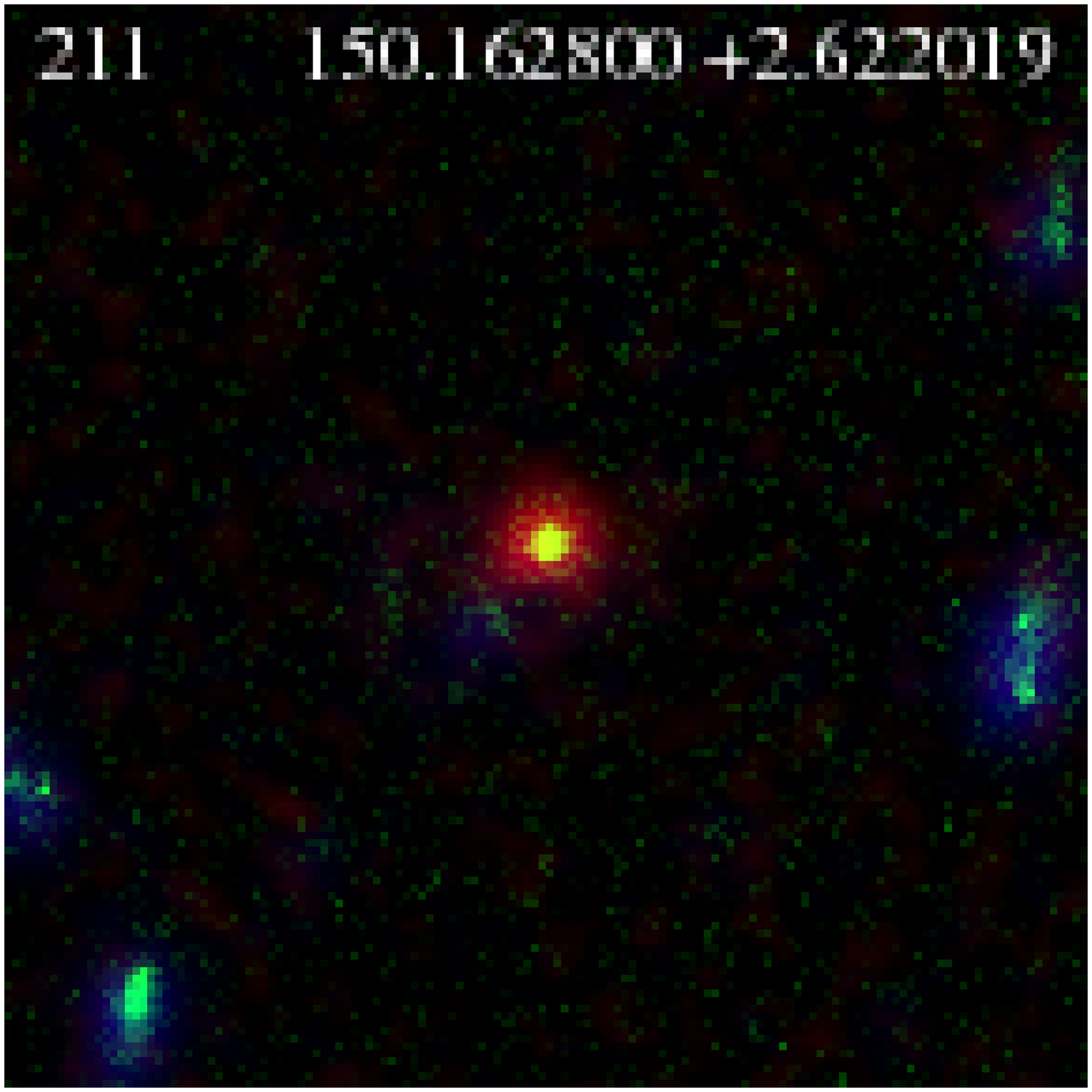} 
 \includegraphics[width=1.in]{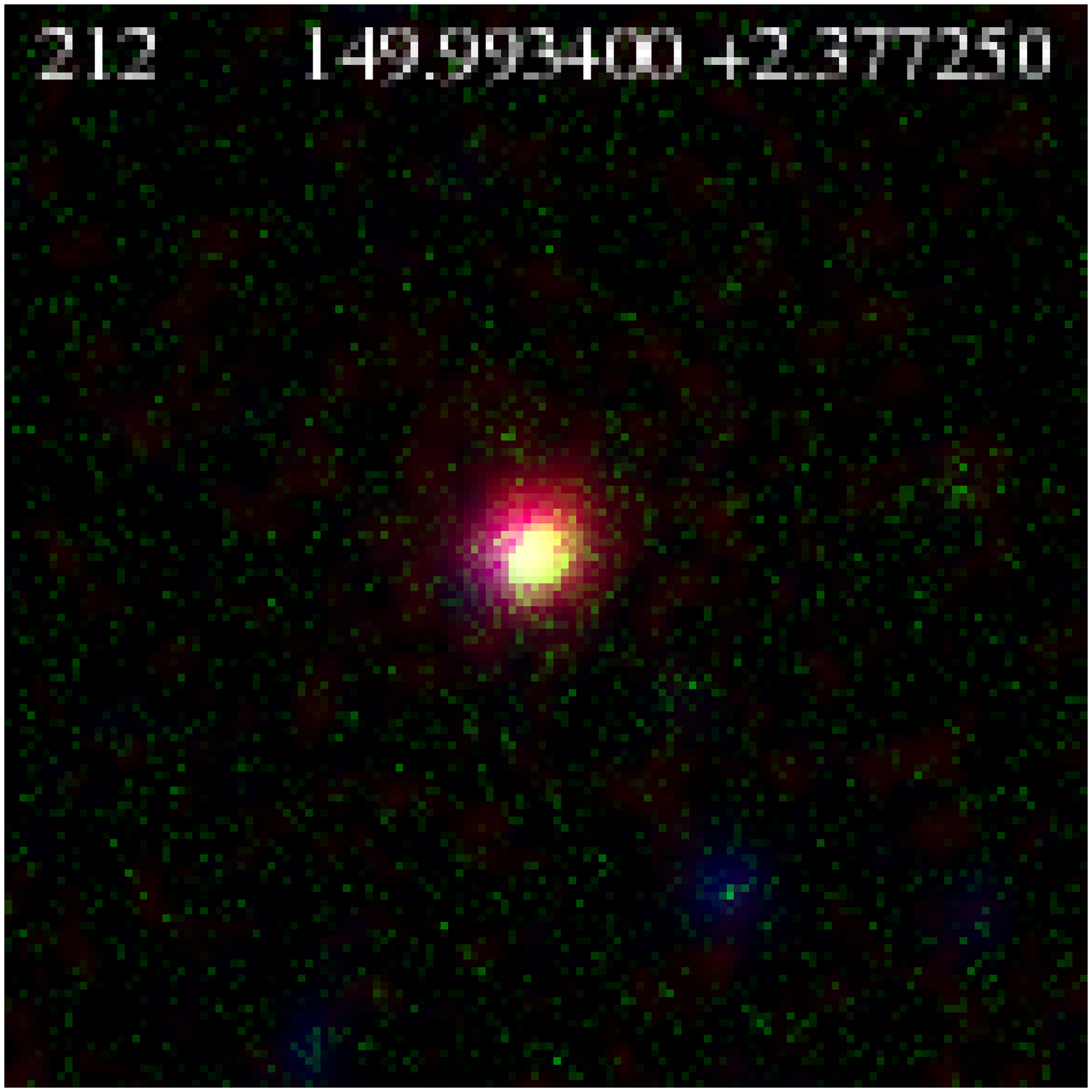} 
 \includegraphics[width=1.in]{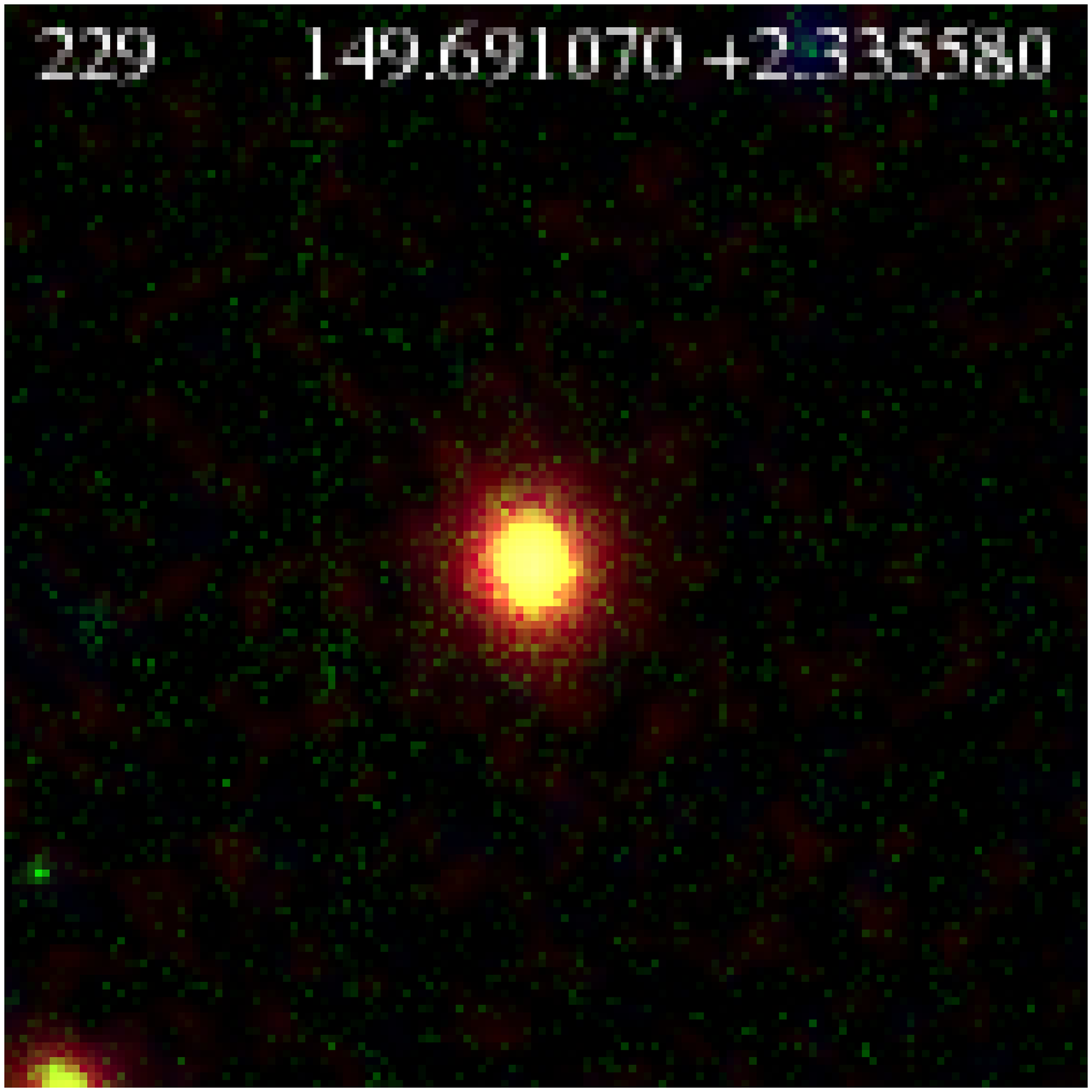} 
 \includegraphics[width=1.in]{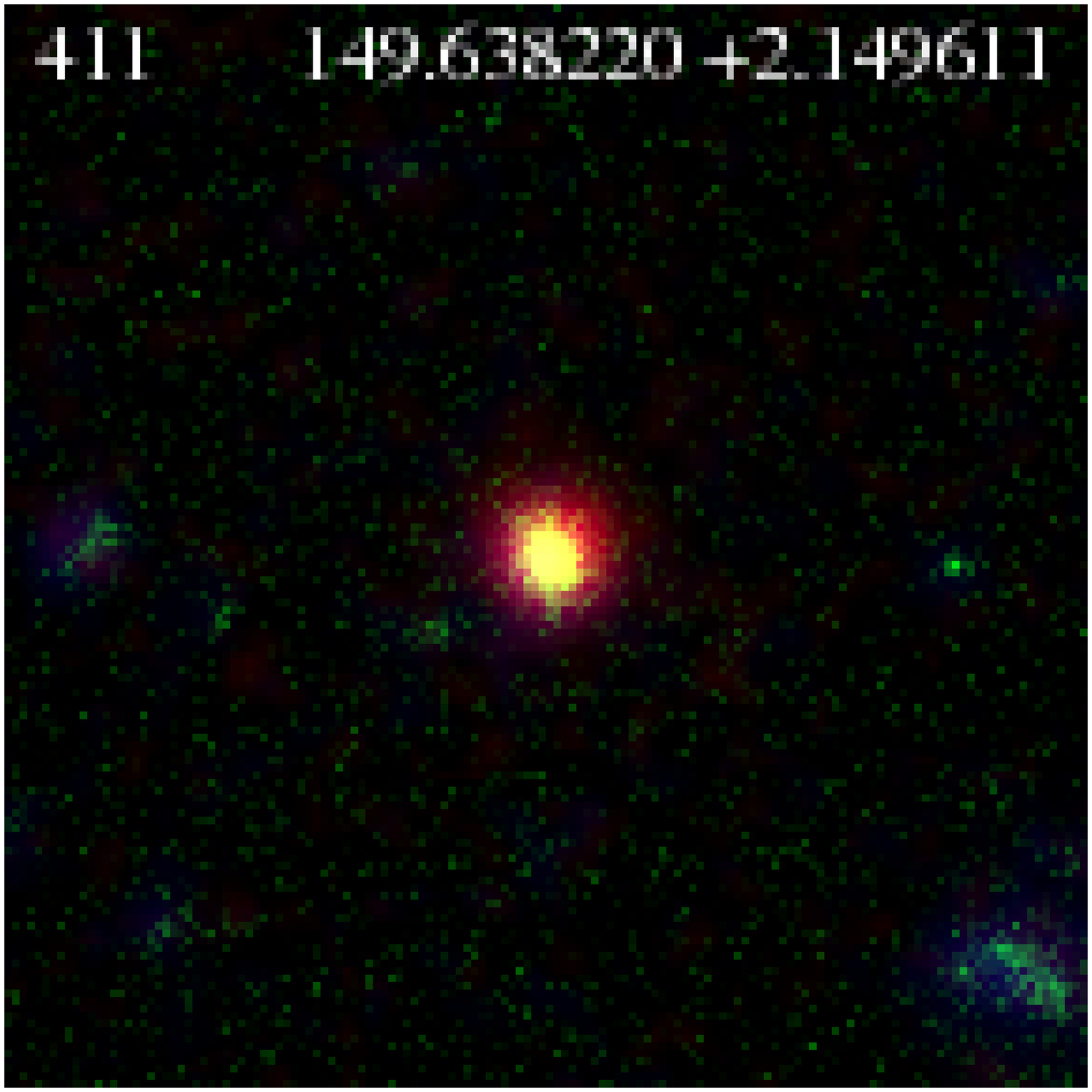} 
\includegraphics[width=1.in]{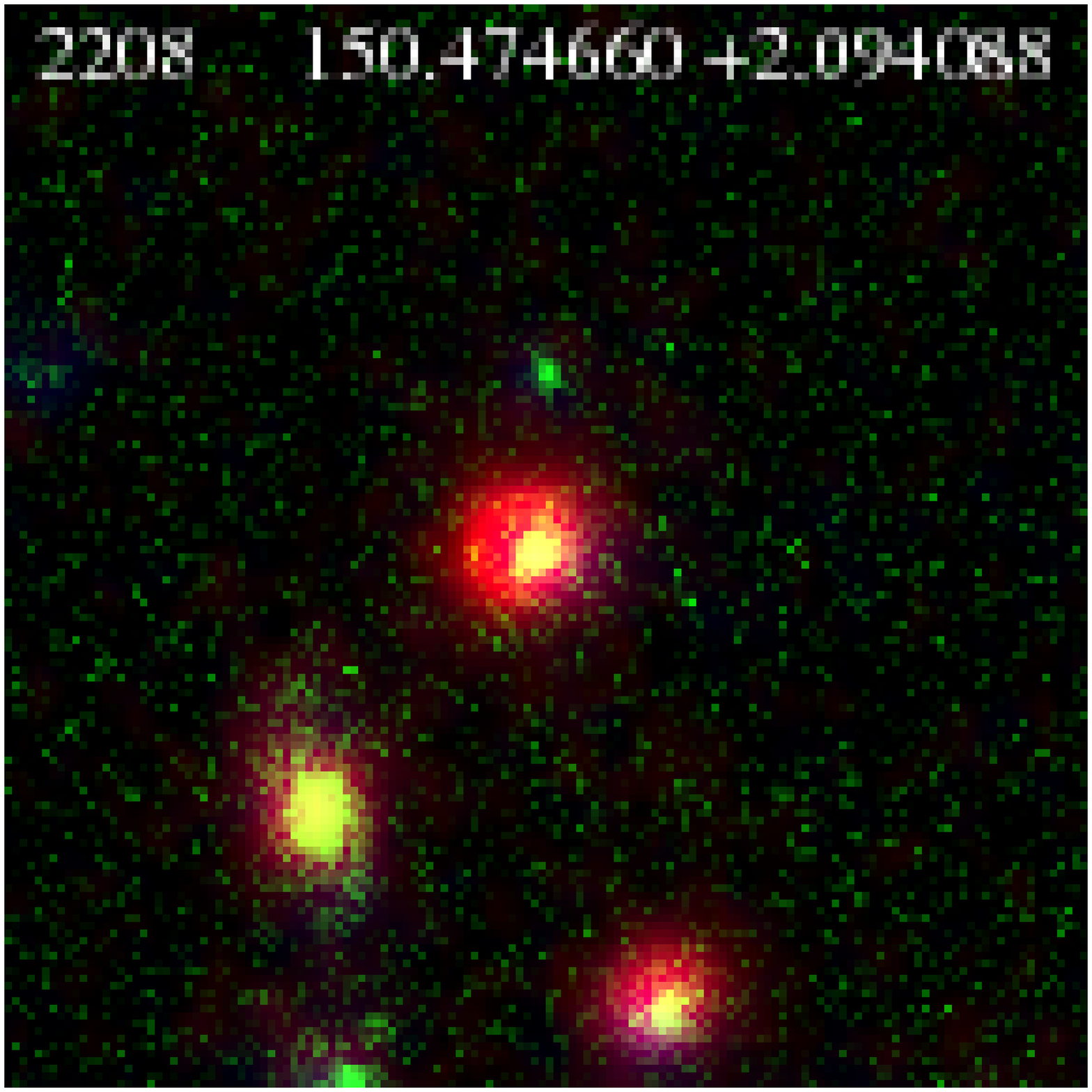}
\includegraphics[width=1.in]{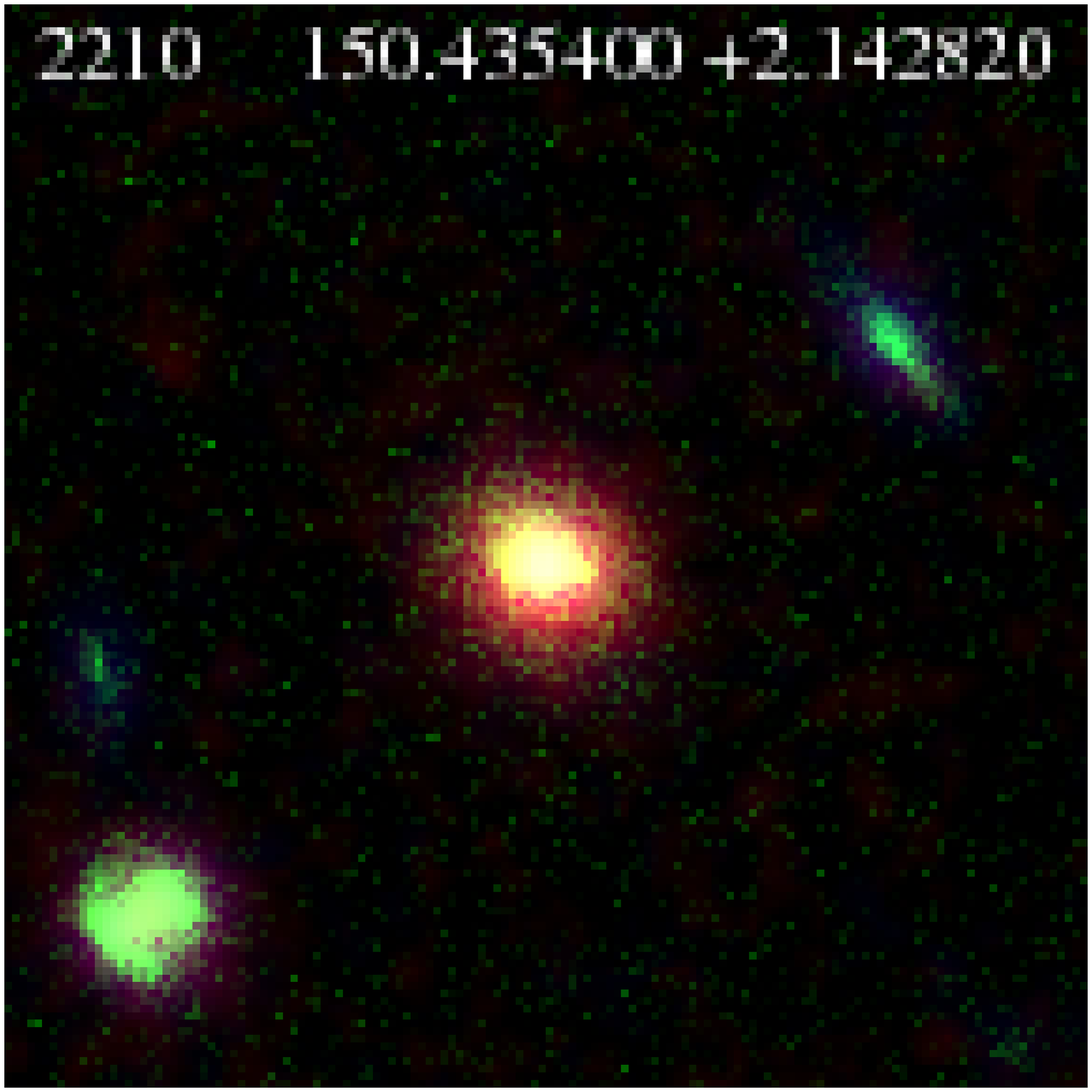}
\includegraphics[width=1.in]{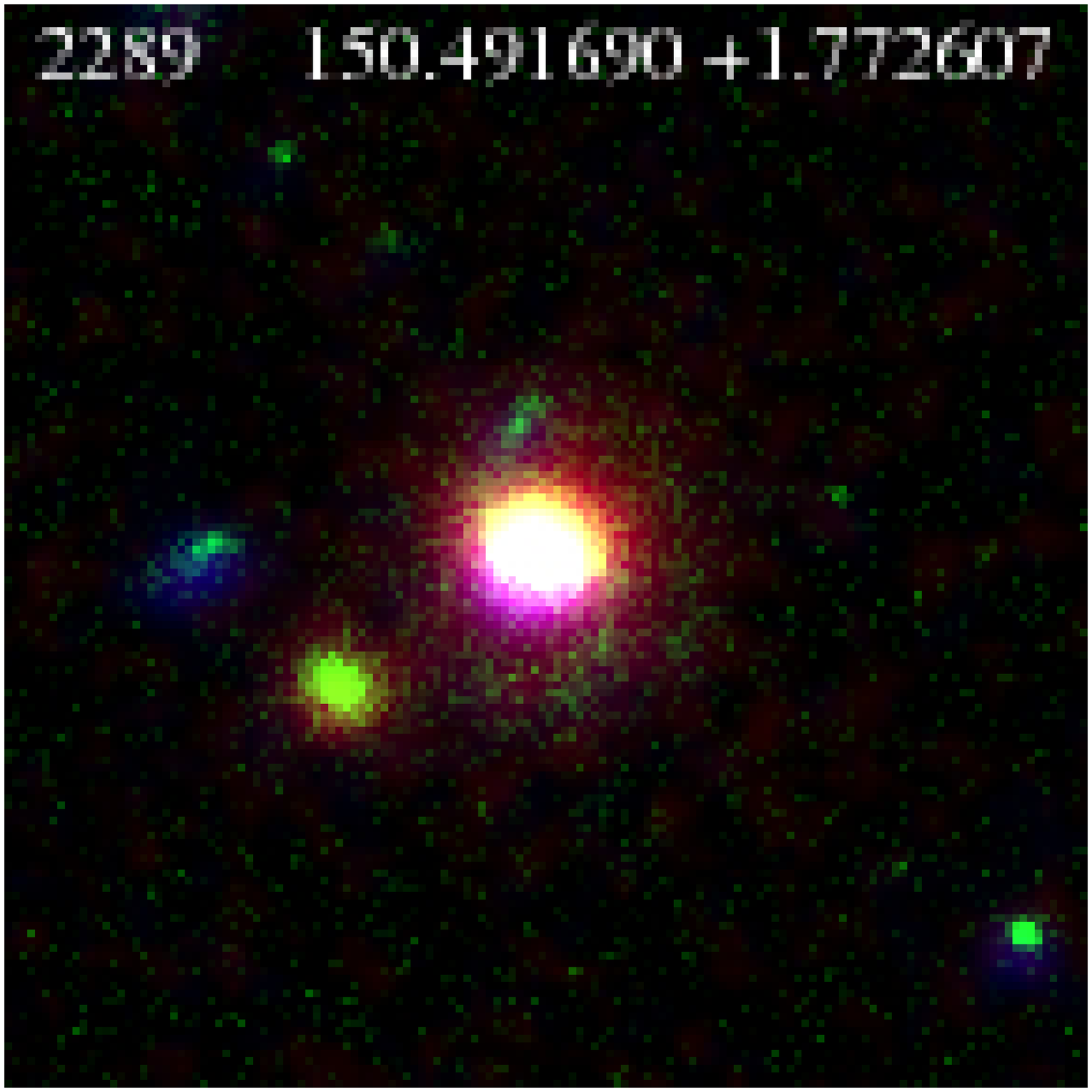}
\includegraphics[width=1.in]{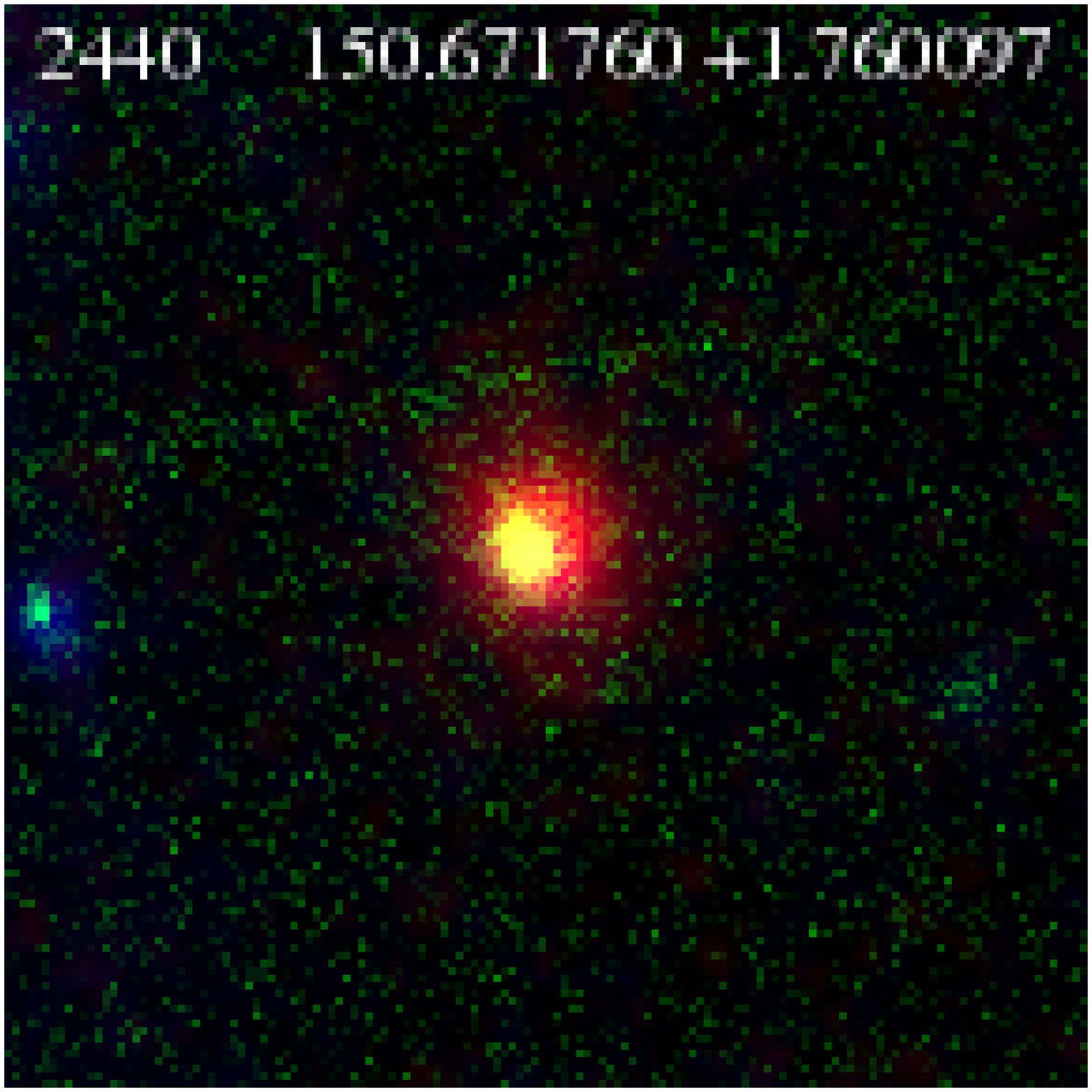}
\includegraphics[width=1.in]{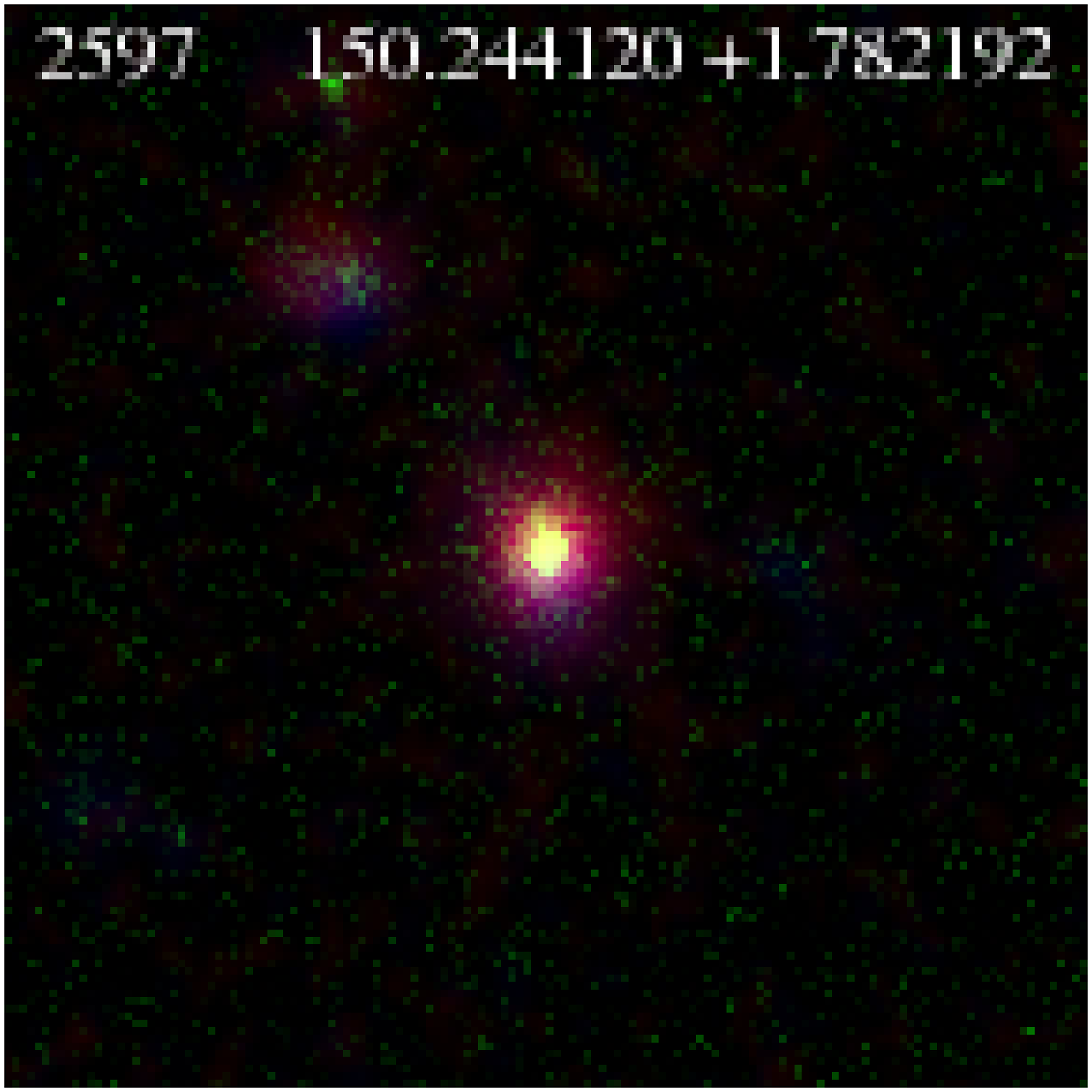}
\includegraphics[width=1.in]{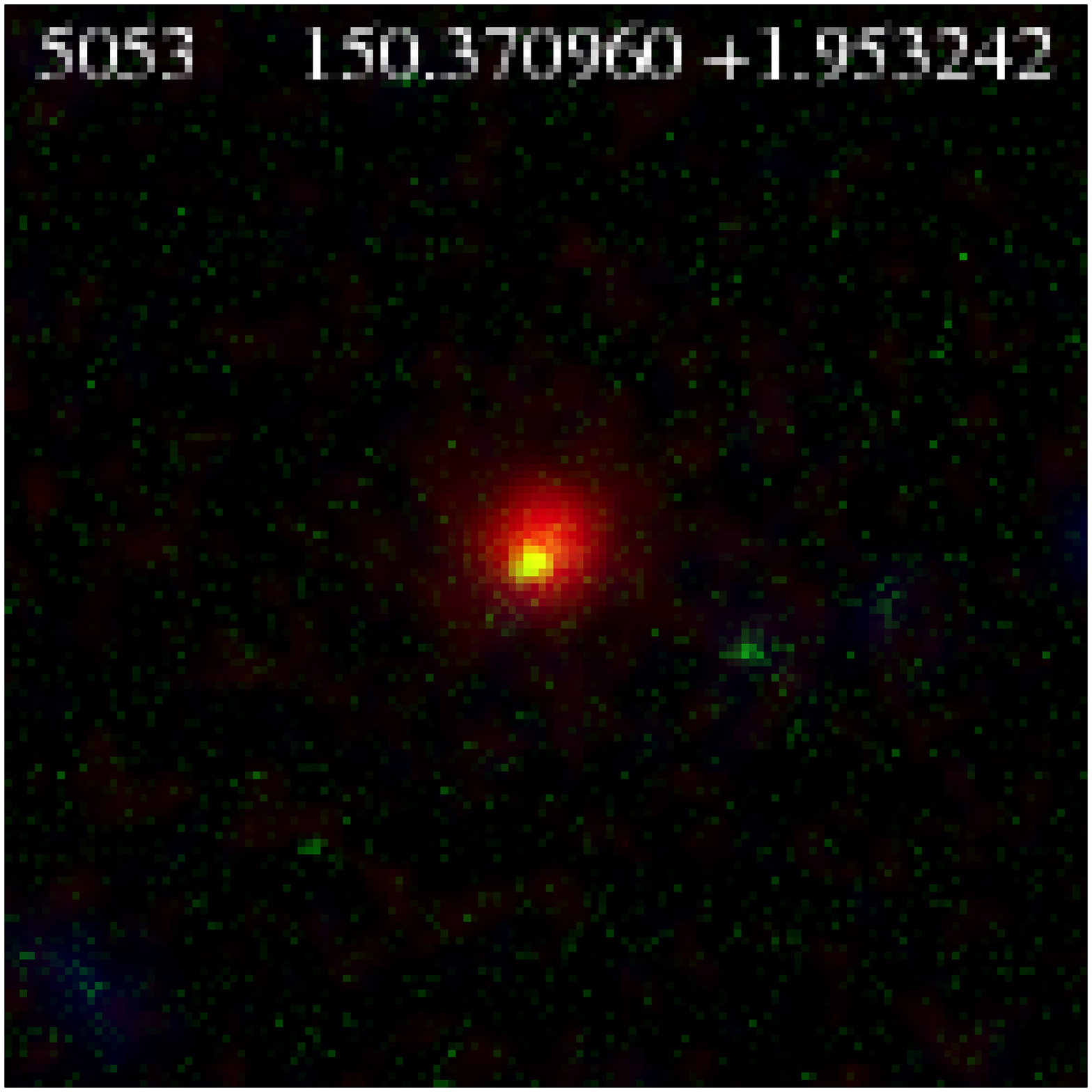}
\includegraphics[width=1.in]{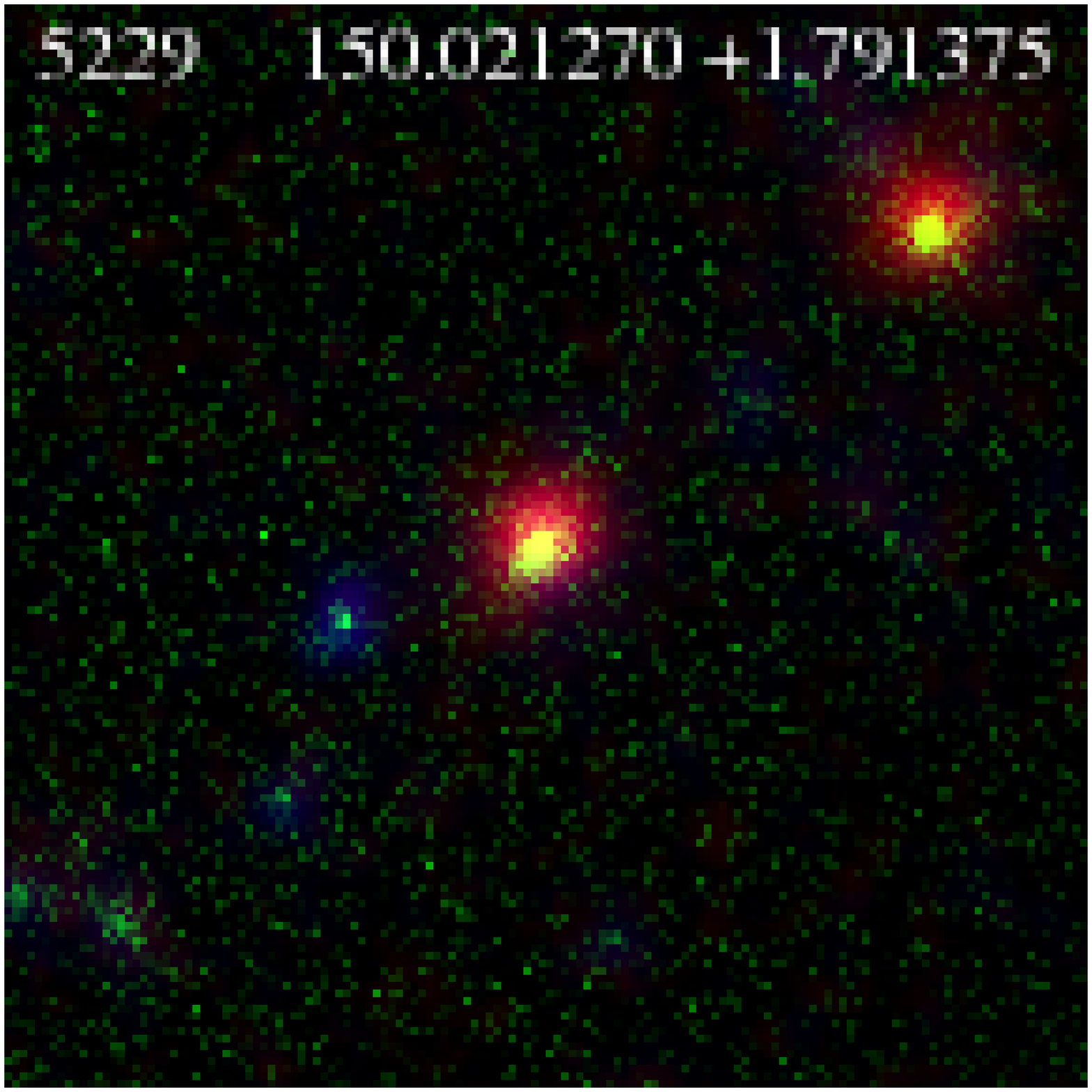}
\includegraphics[width=1.in]{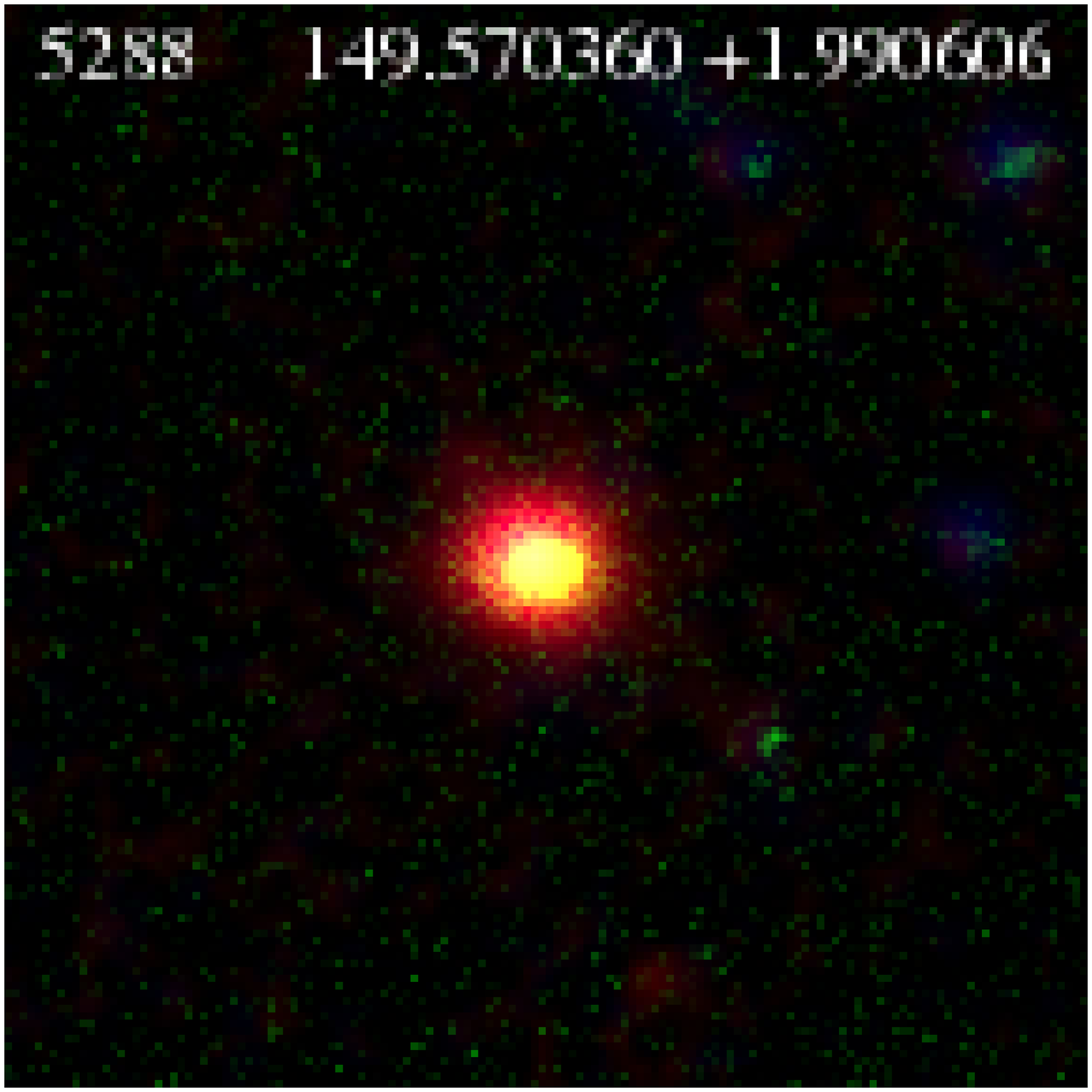}
\includegraphics[width=1.in]{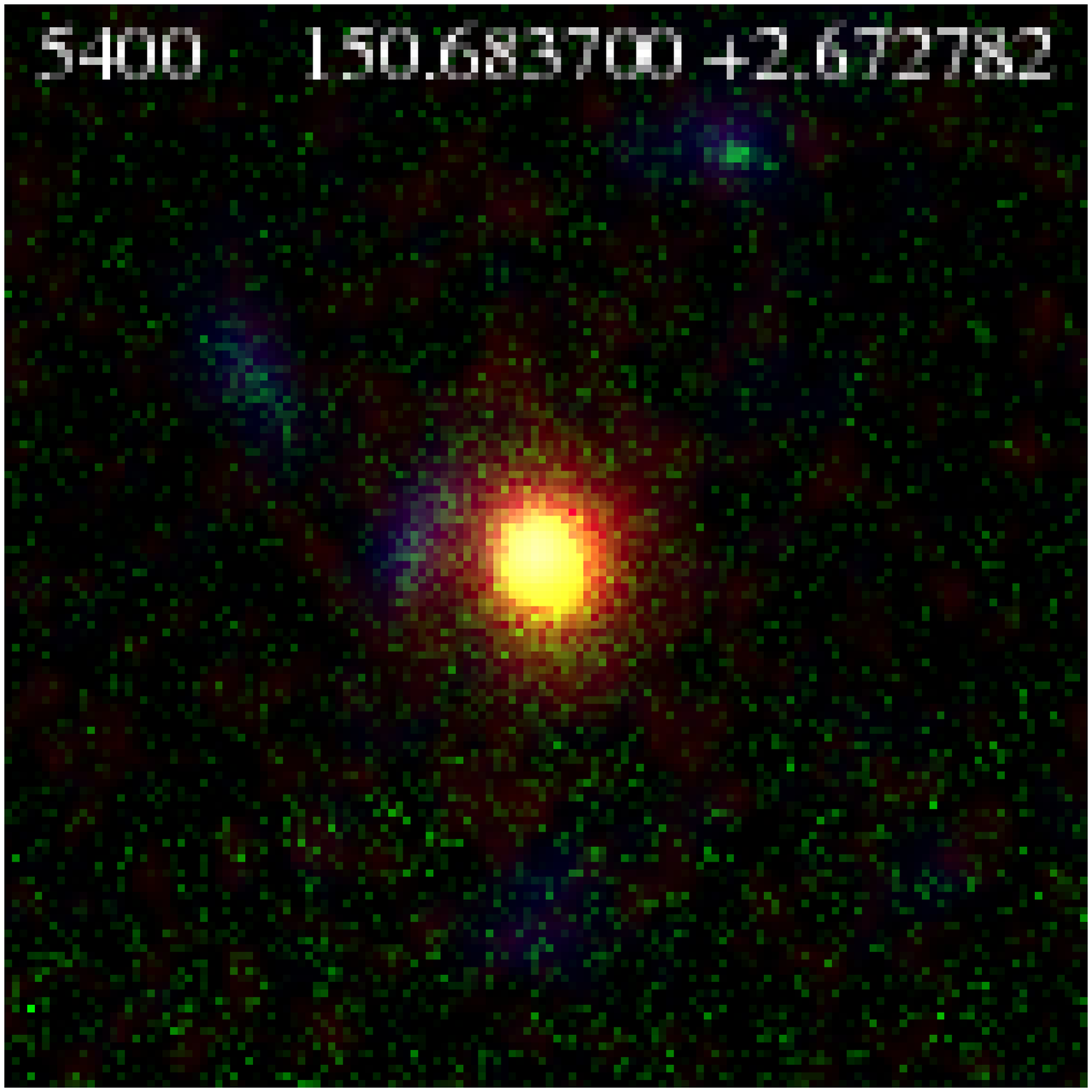}
\includegraphics[width=1.in]{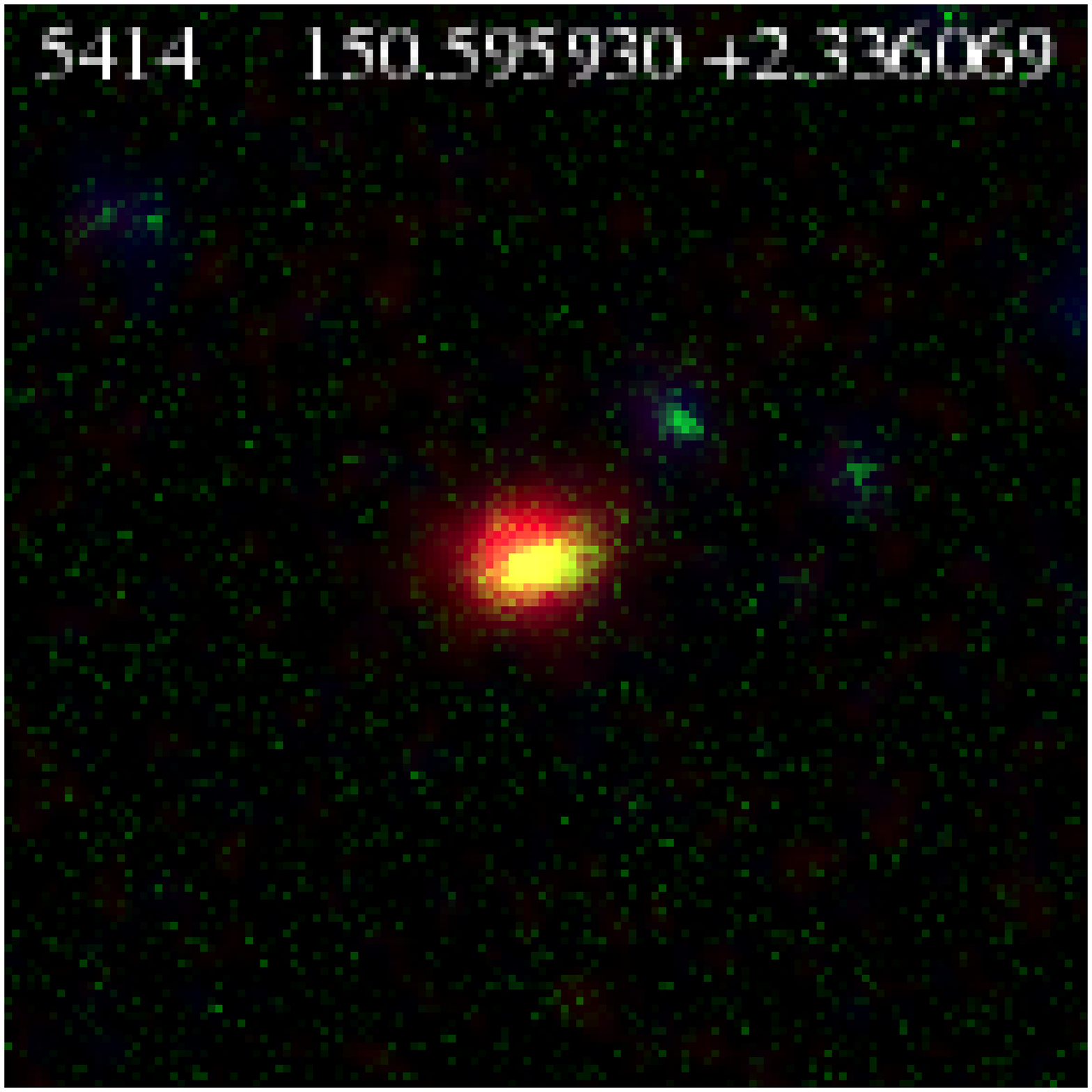}
\includegraphics[width=1.in]{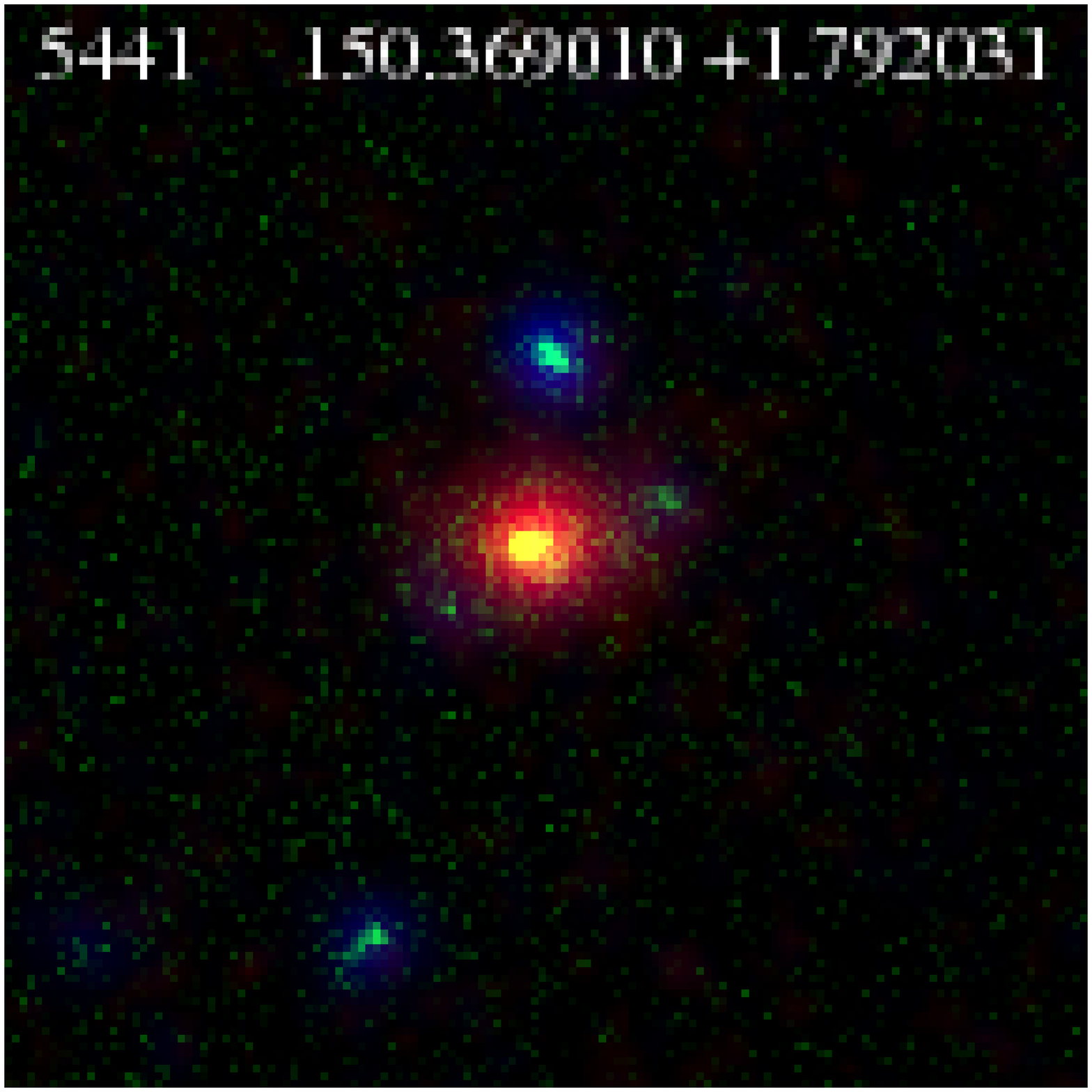}
\includegraphics[width=1.in]{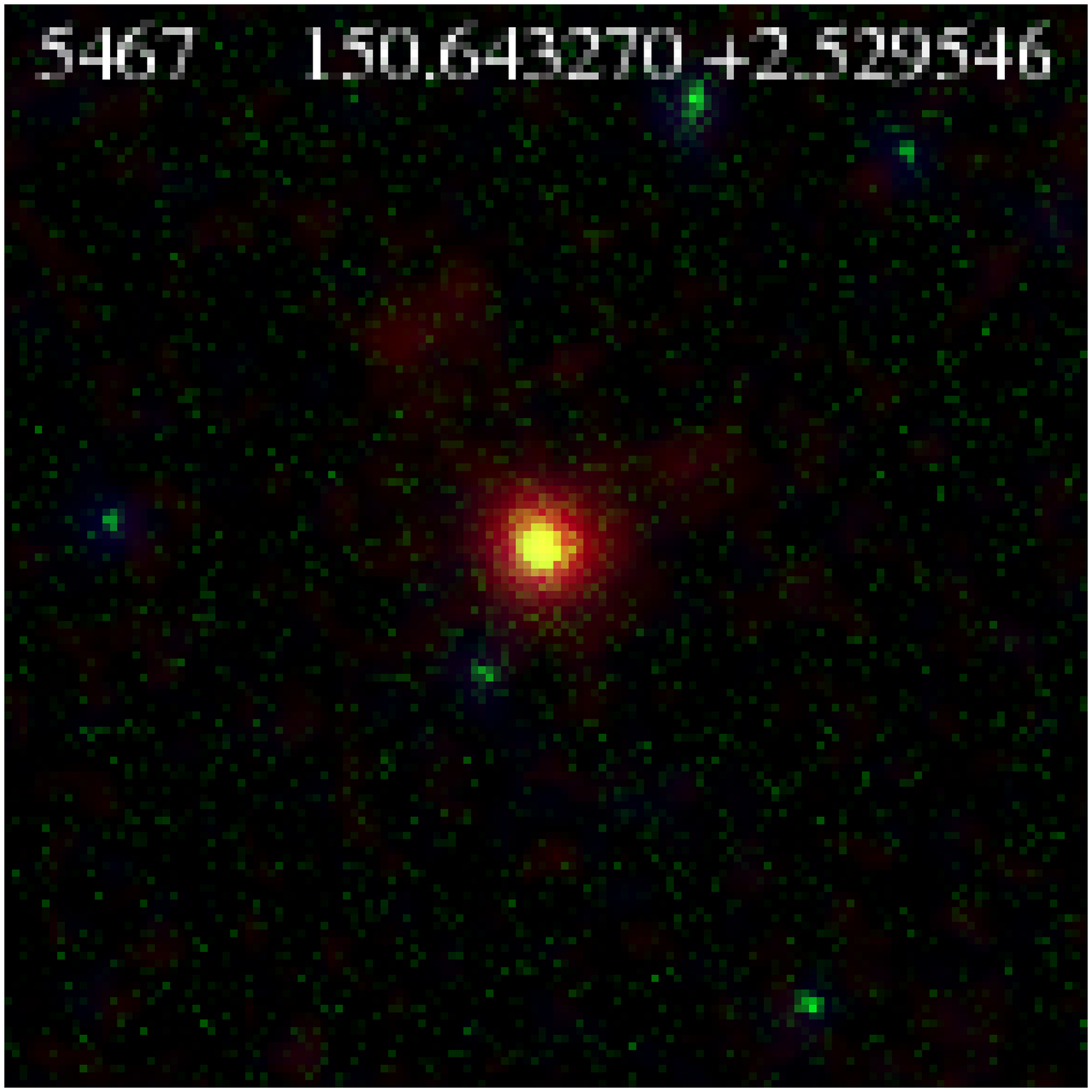}
\includegraphics[width=1.in]{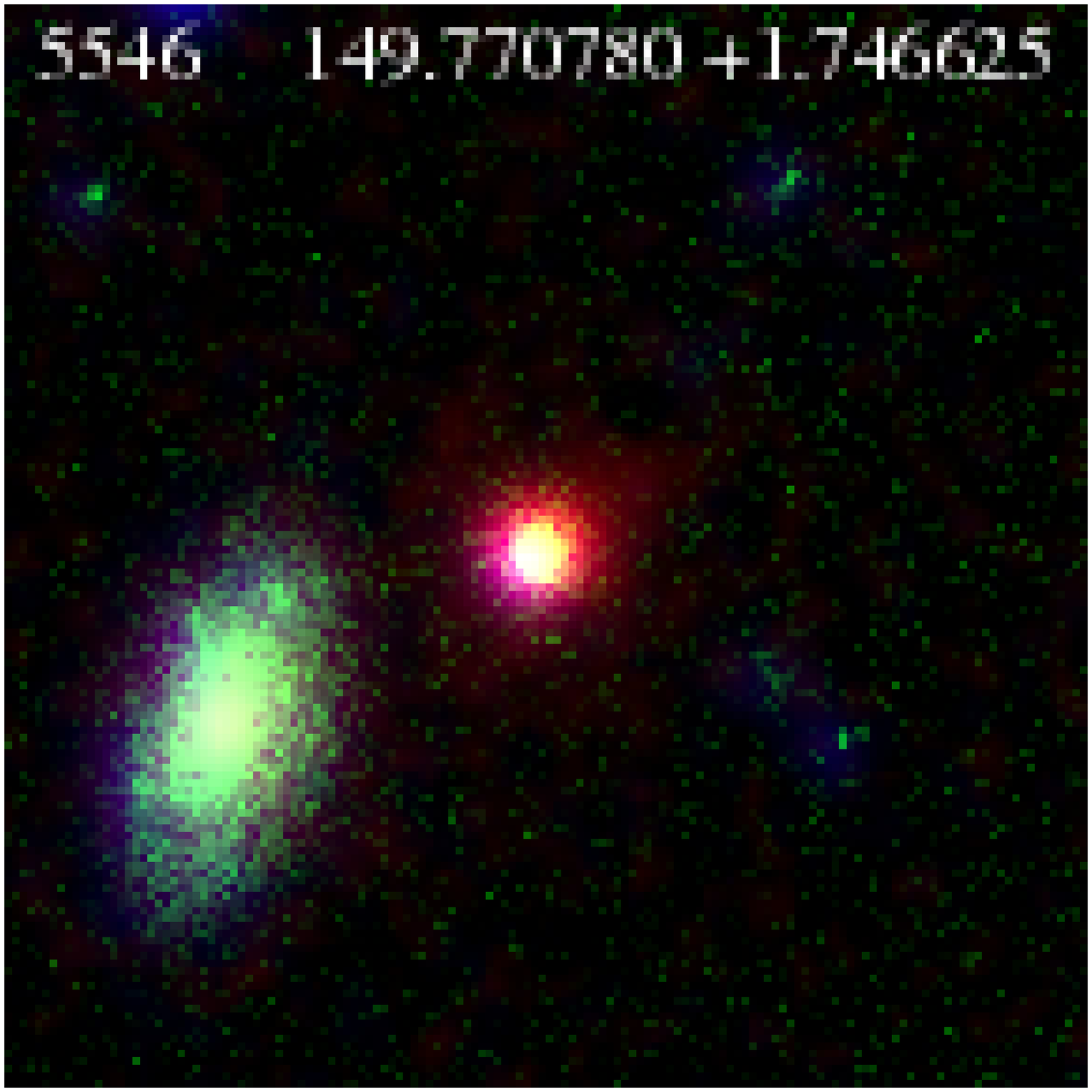}
\includegraphics[width=1.in]{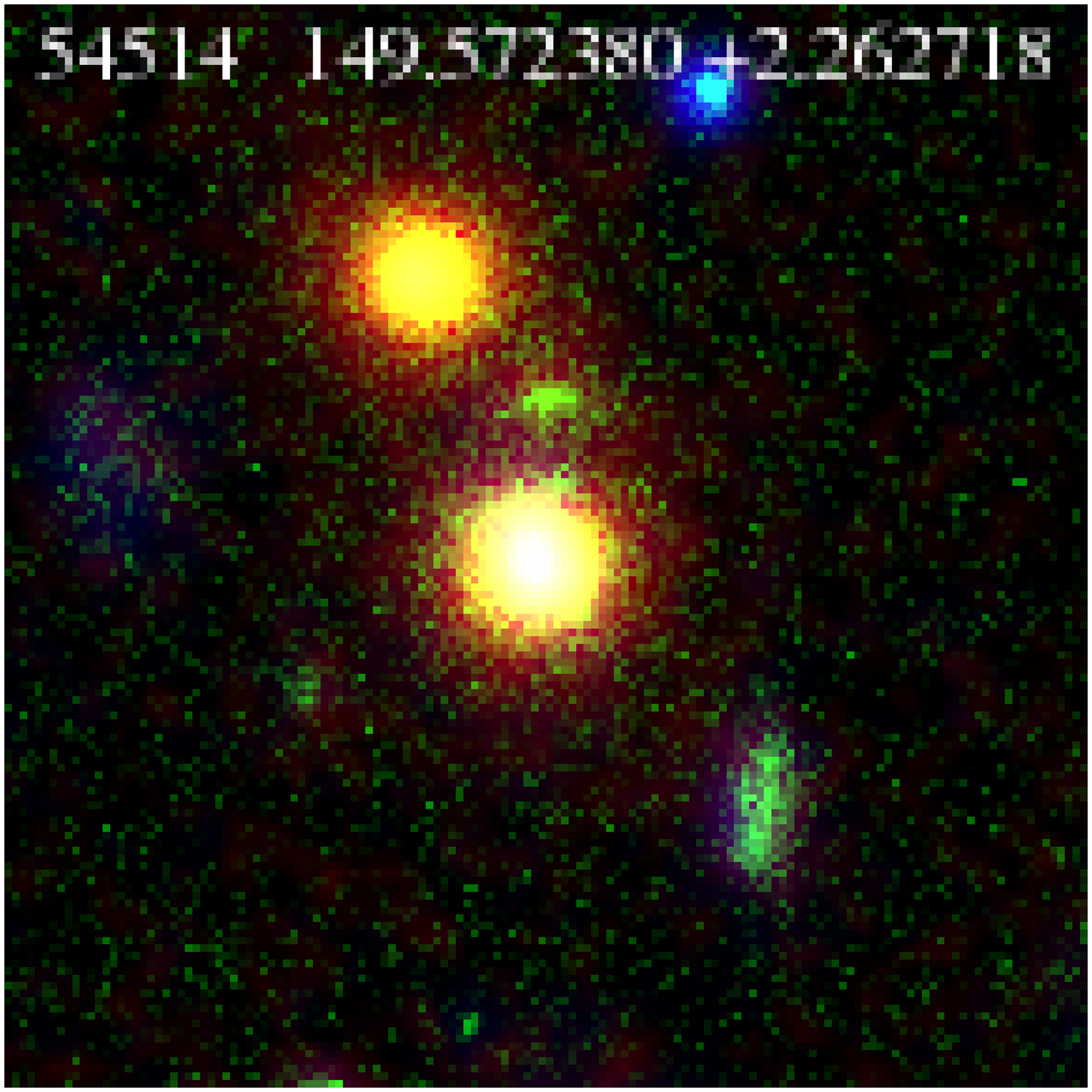}
\includegraphics[width=1.in]{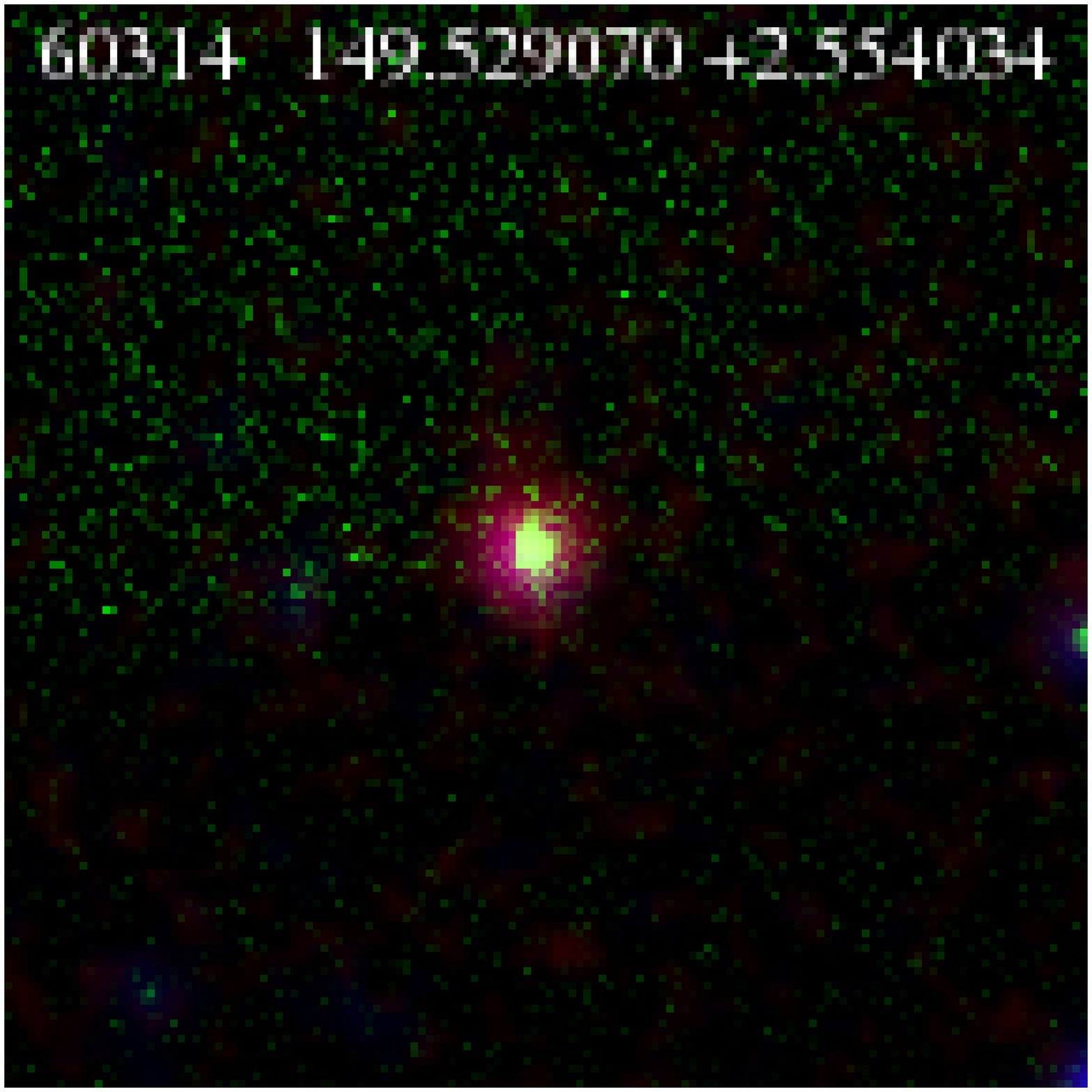}
 \caption{Type-2 QSOs hosts classified as {\it bulge-dominated}
   galaxies. The images are 10\arcsec~ on a side. They were obtained
   by combining Subaru B-band ({\it blue}), ACS F814W ({\it green}),
   and CFHT Ks ({\it red}) images. XID and coordinates of the source
   are reported at the top of each image.}
   \label{morph_qso2_bulge}
\end{center}
\end{figure}

\begin{figure}
% \vspace*{-2.0 cm}
\begin{center}
 \includegraphics[width=1.in]{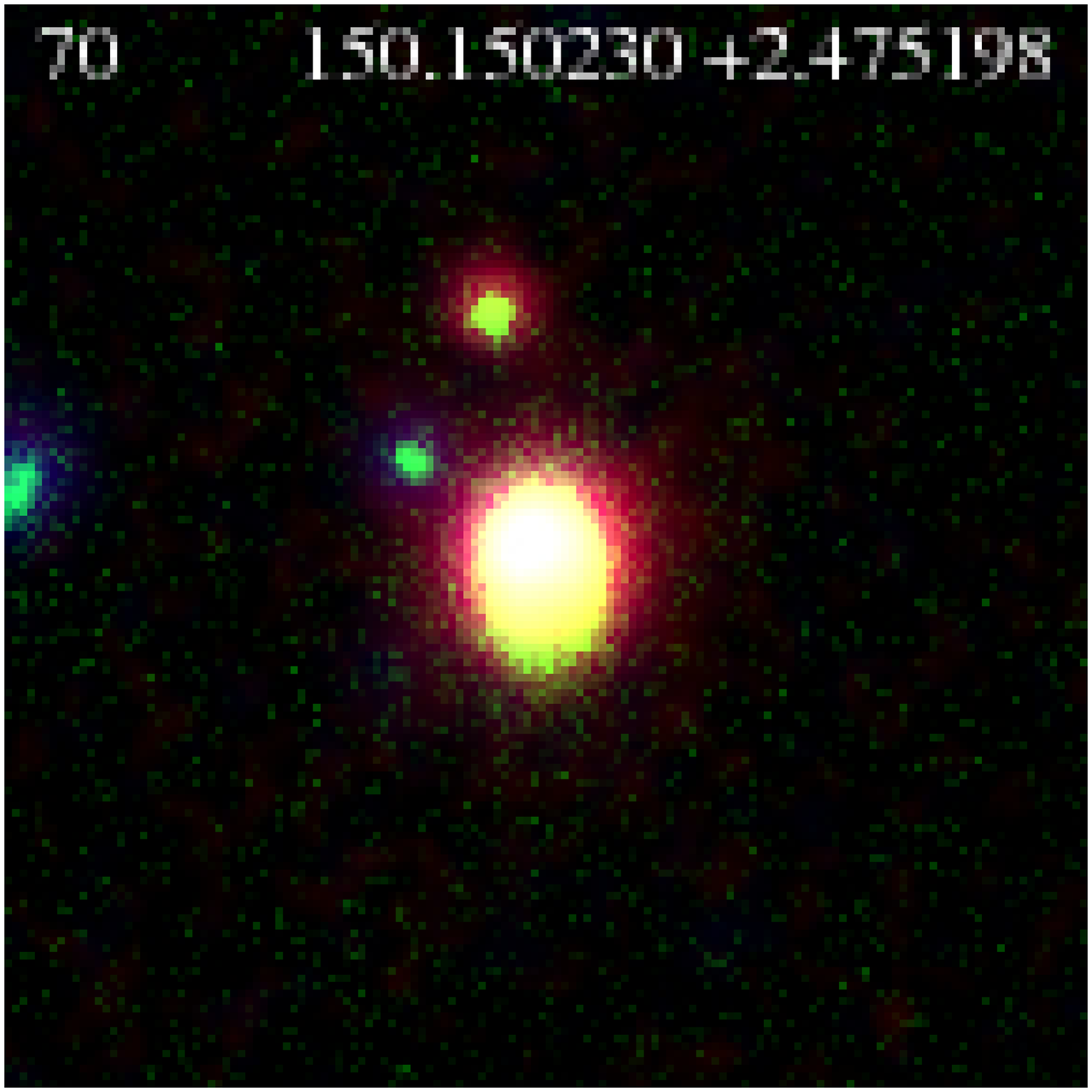} 
 \includegraphics[width=1.in]{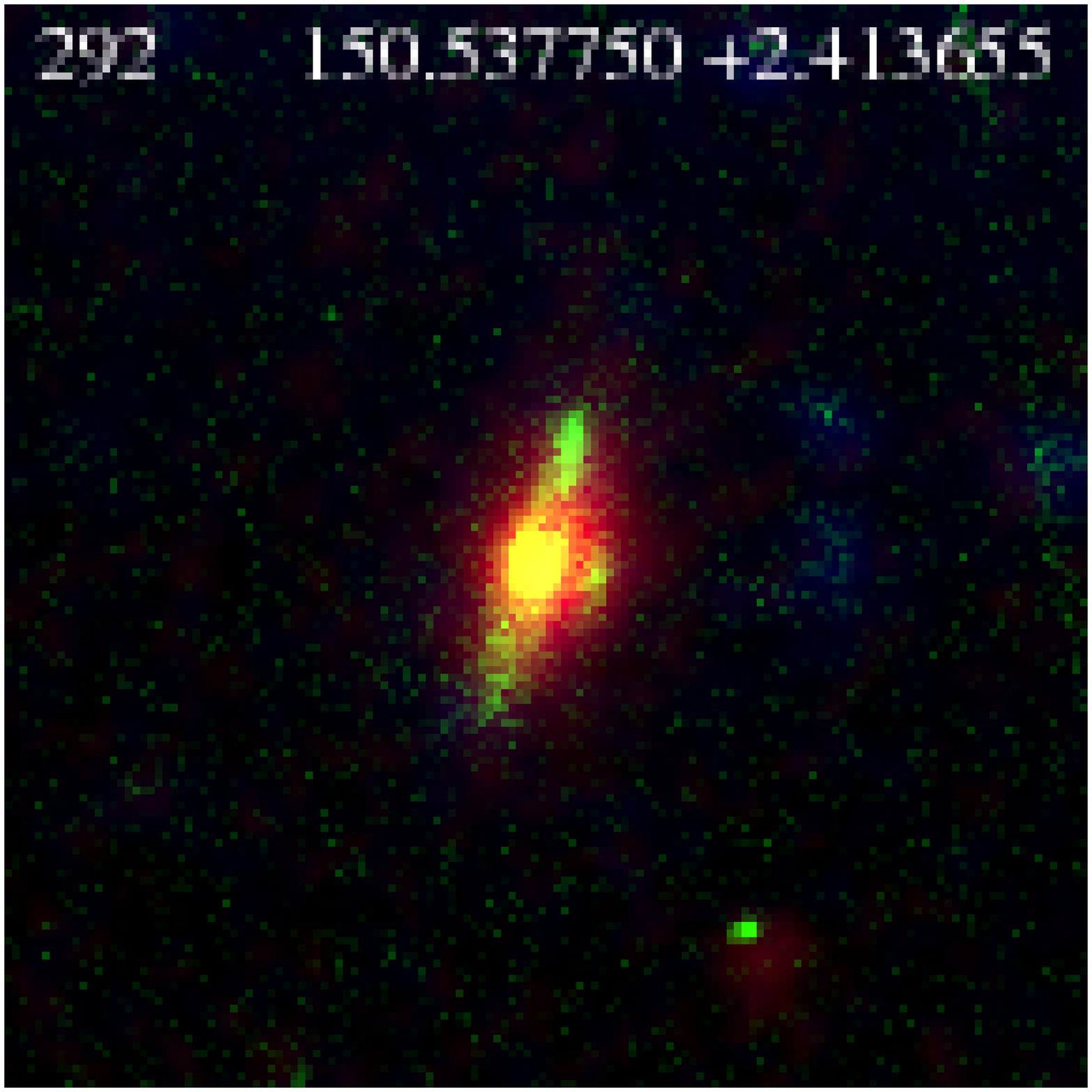} 
 \includegraphics[width=1.in]{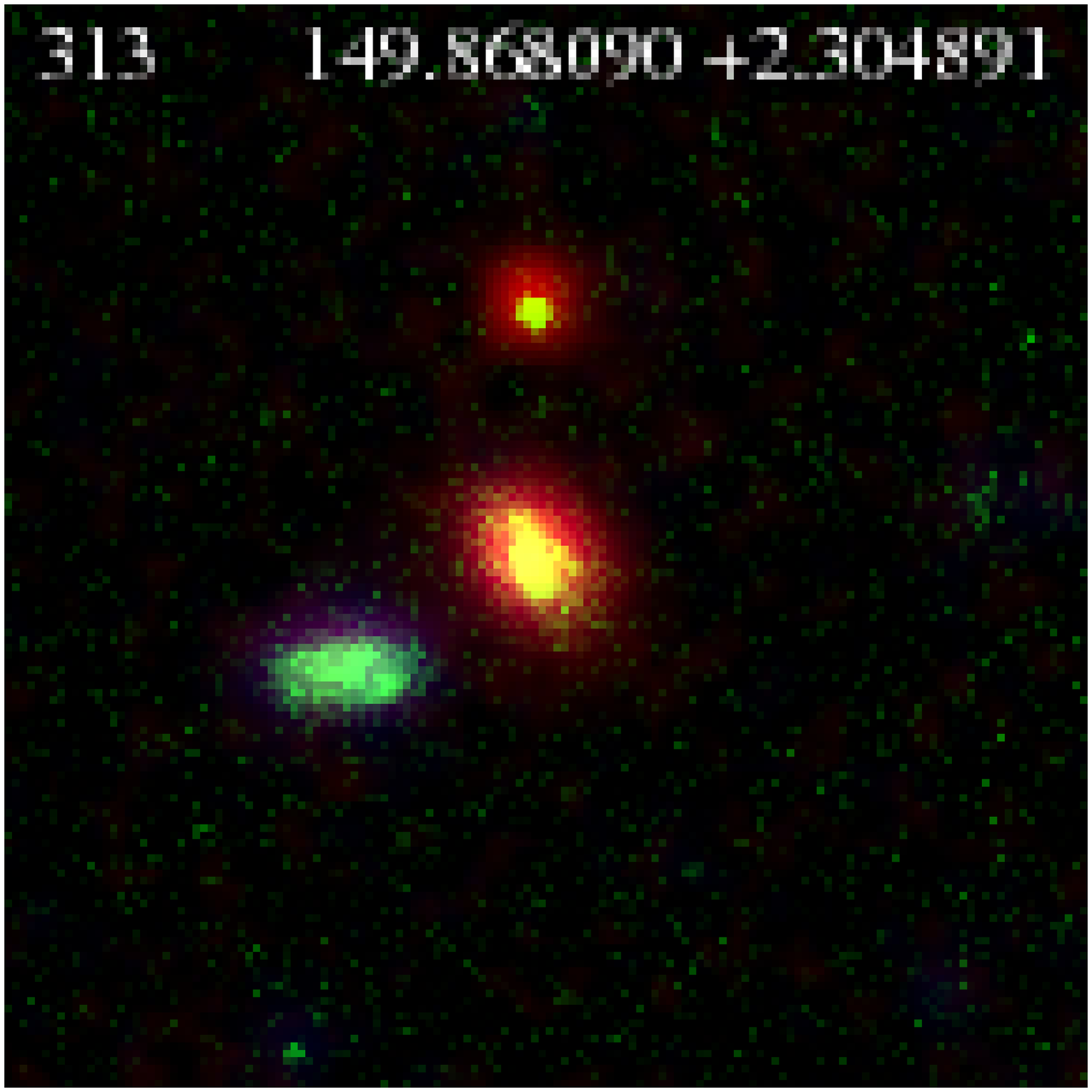} 
 \includegraphics[width=1.in]{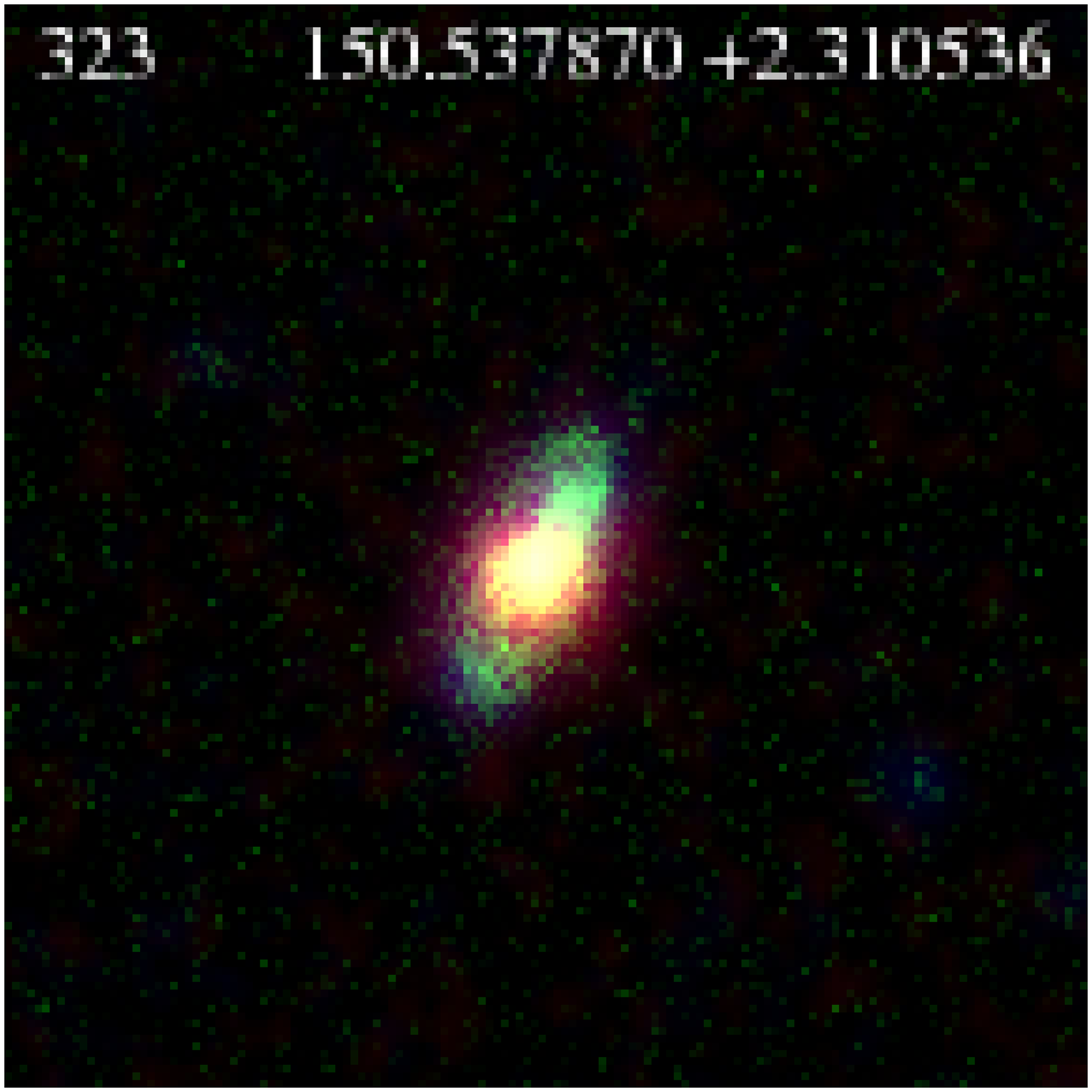} 
\includegraphics[width=1.in]{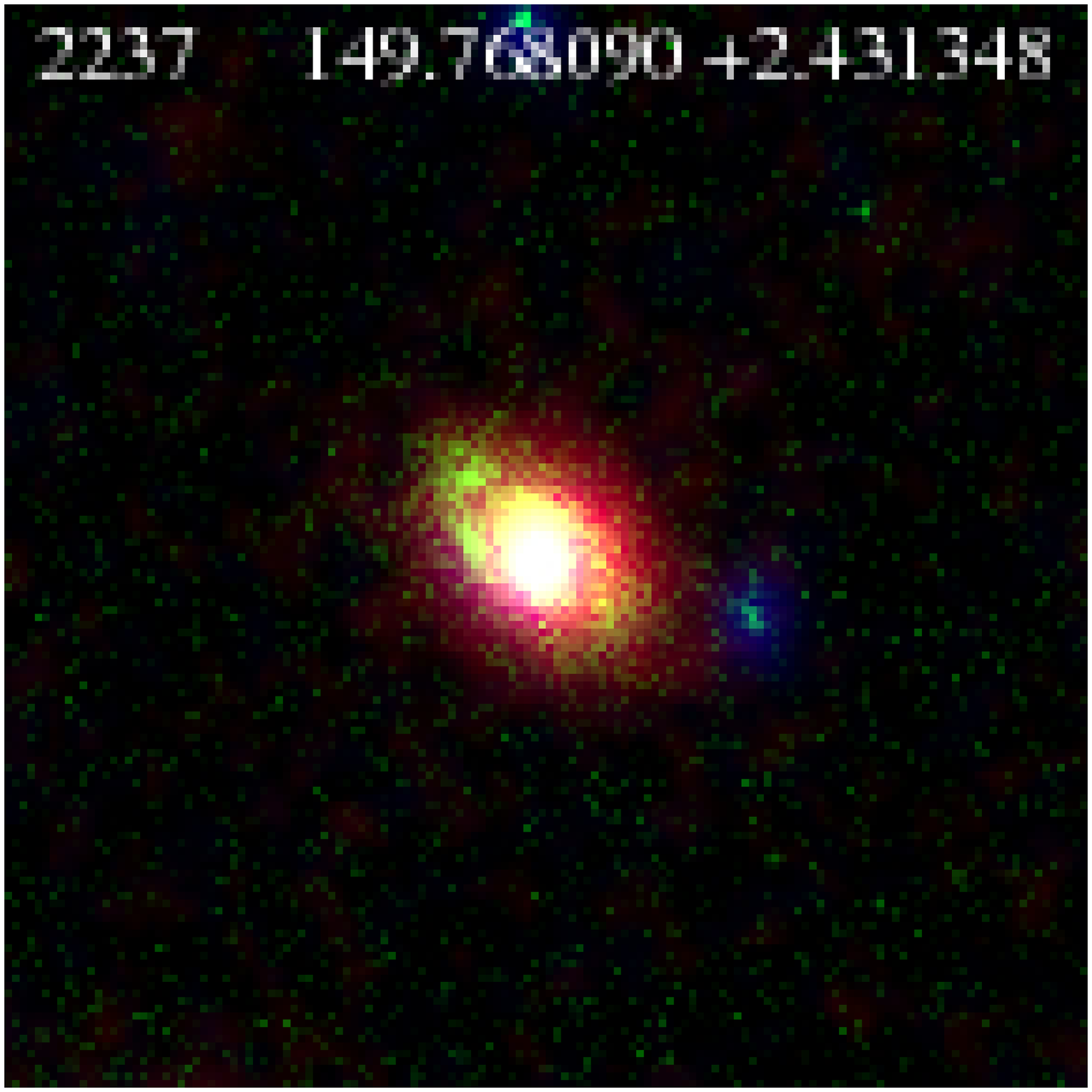}
\includegraphics[width=1.in]{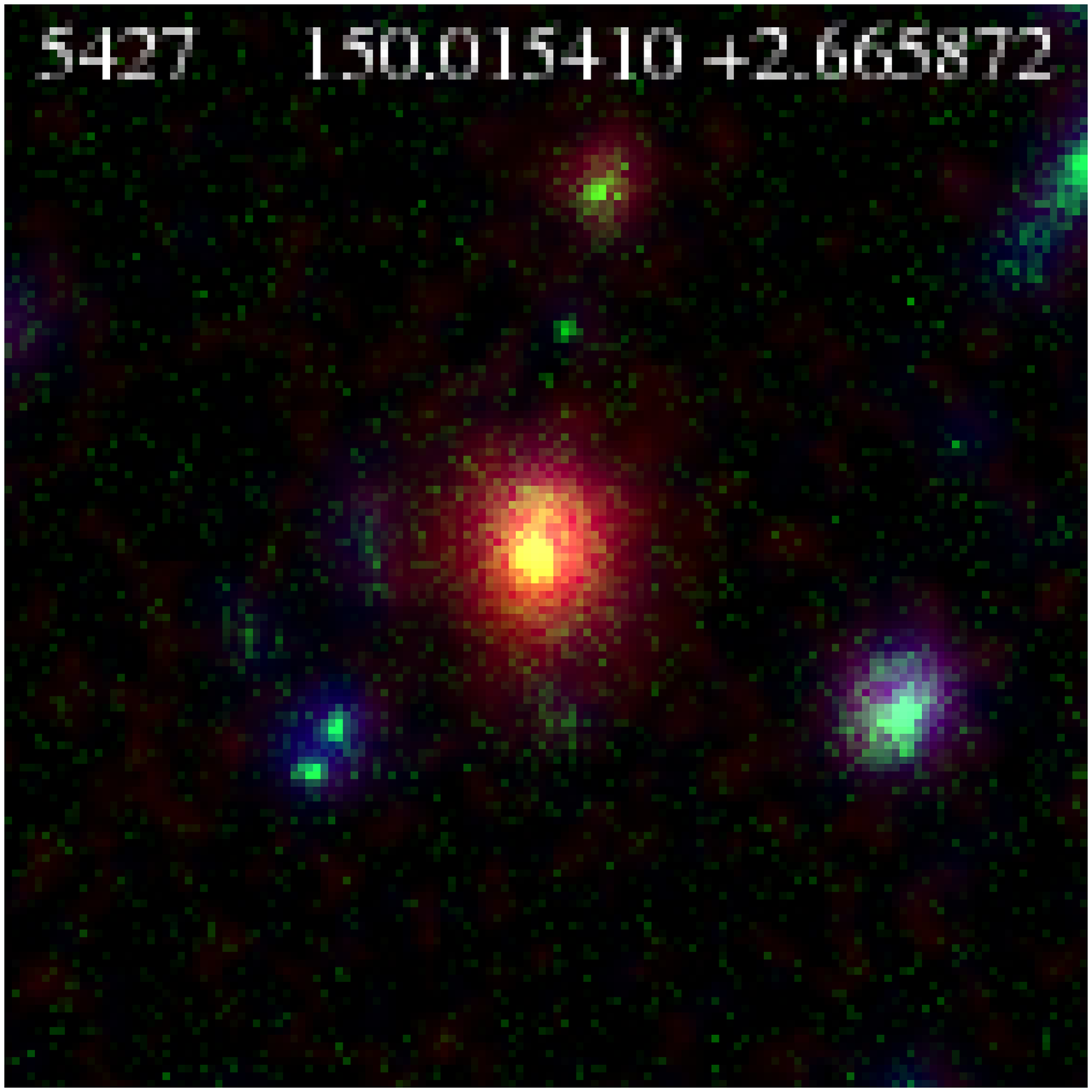}
\includegraphics[width=1.in]{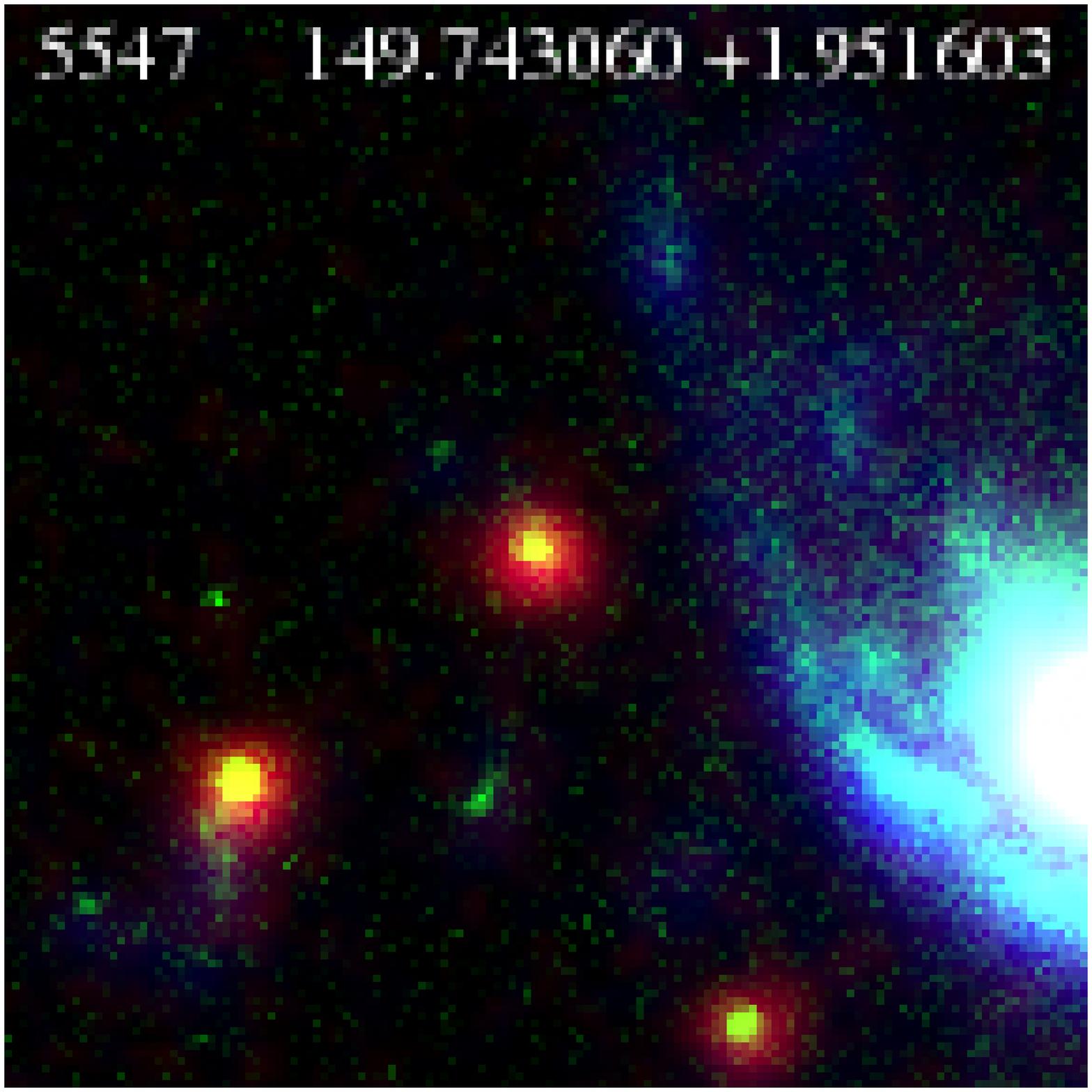}
\includegraphics[width=1.in]{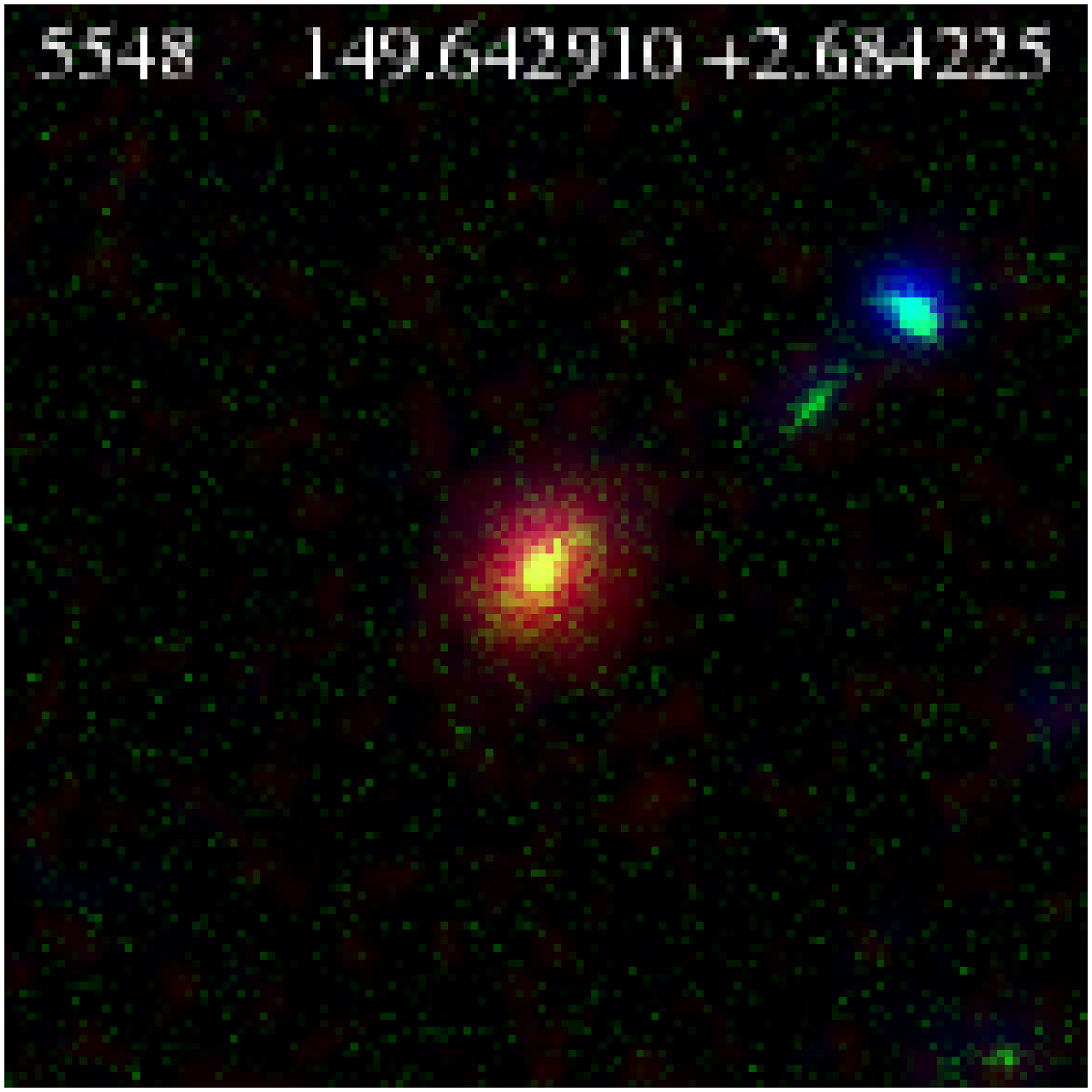}
% \vspace*{-1.0 cm}
 \caption{Type-2 QSOs hosts classified as {\it disk-dominated}
   galaxies. The images are 10\arcsec~ on a side. They were obtained
   by combining Subaru B-band ({\it blue}), ACS F814W ({\it green}),
   and CFHT Ks ({\it red}) images.XID and coordinates of the source
   are reported at the top of each image.}
   \label{morph_qso2_disk}
\end{center}
\end{figure}

\begin{figure}
% \vspace*{-2.0 cm}
\begin{center}
 \includegraphics[width=1.in]{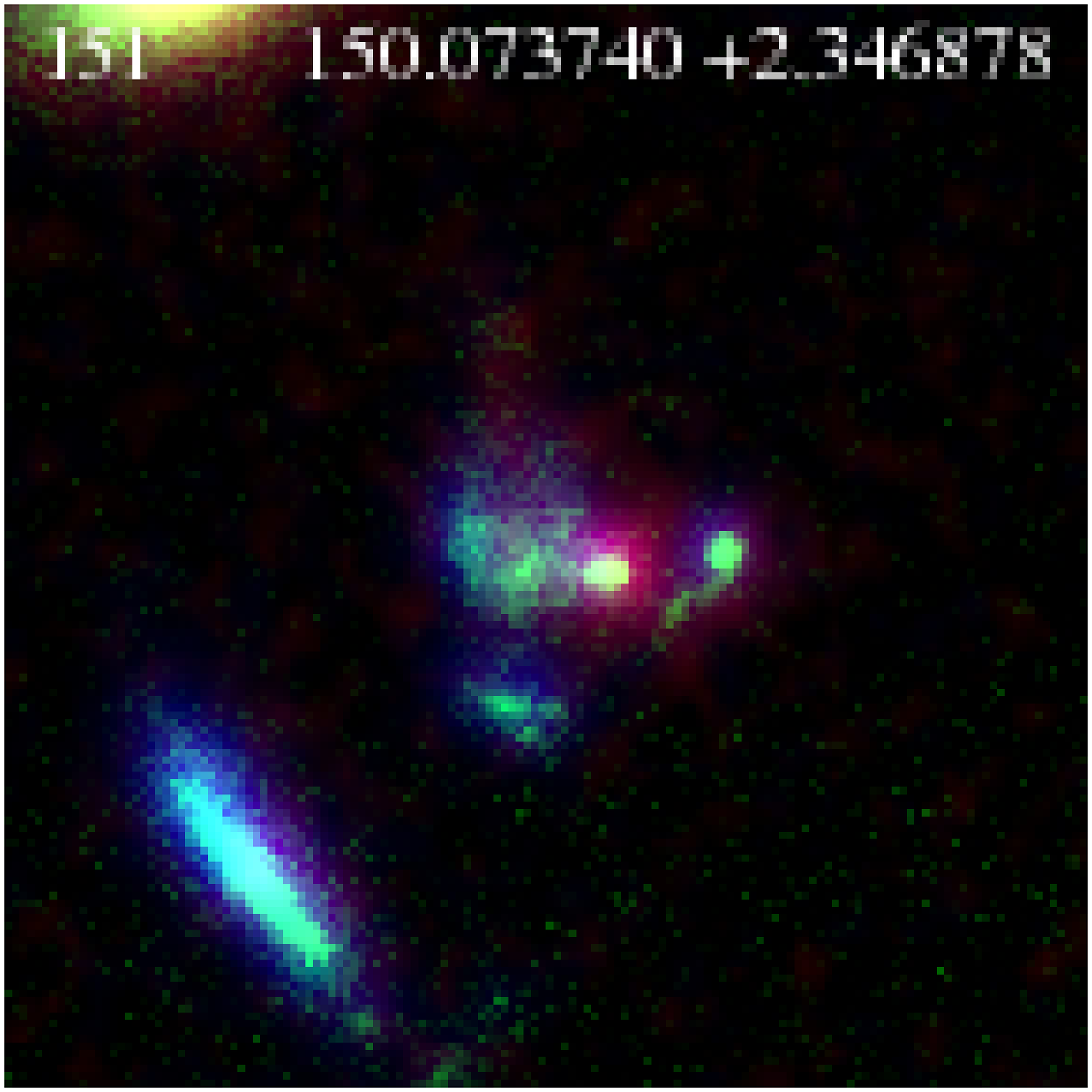} 
 \includegraphics[width=1.in]{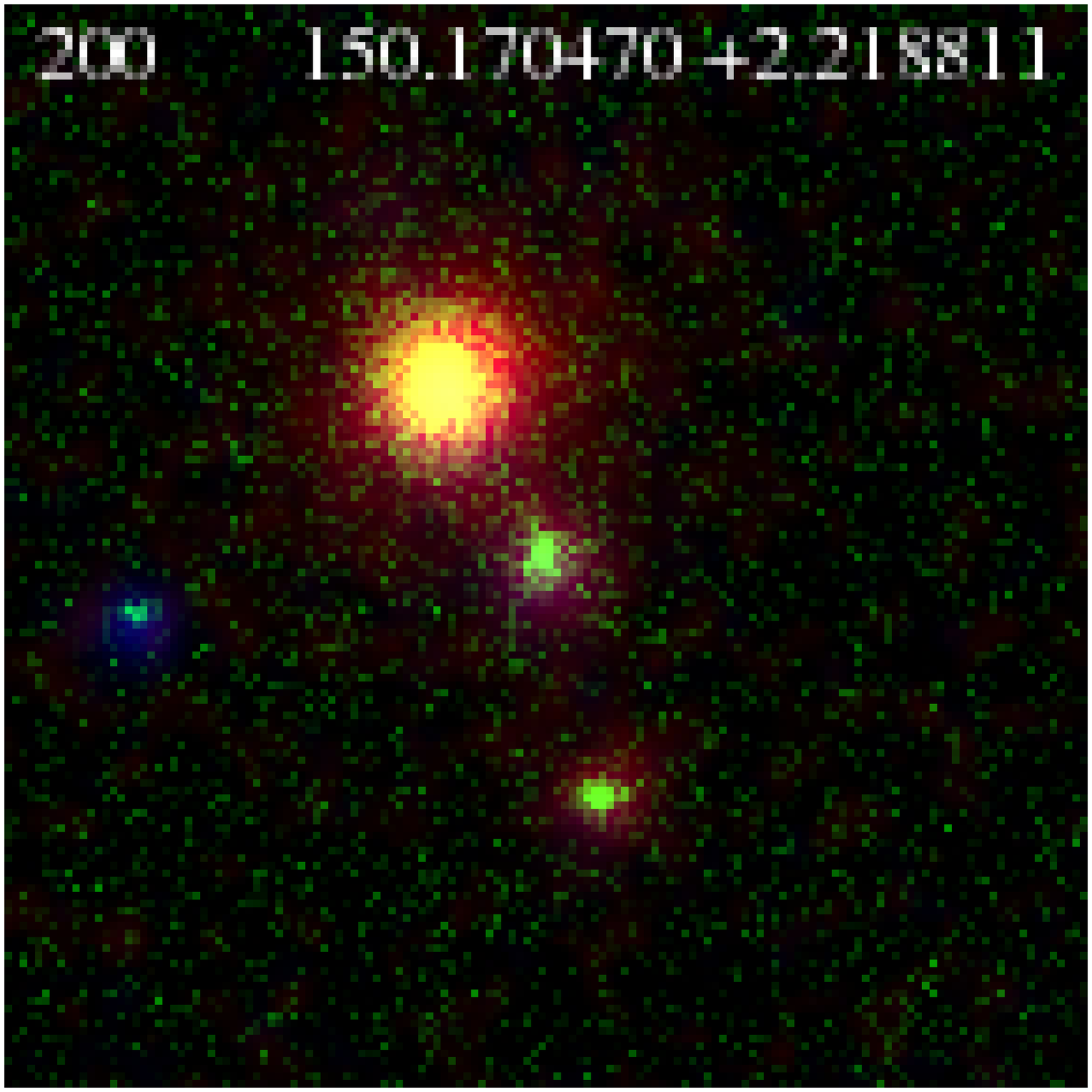} 
\includegraphics[width=1.in]{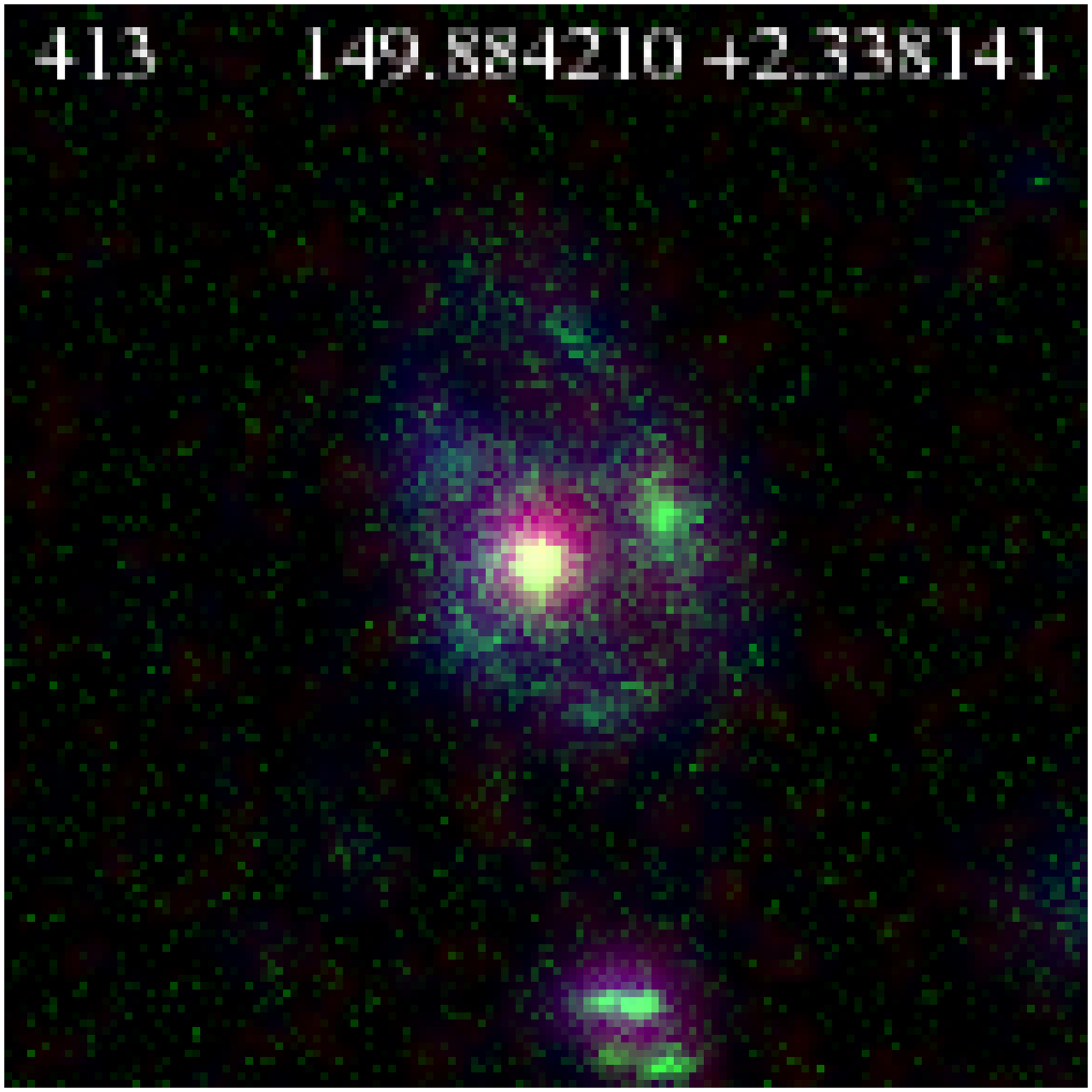}
\includegraphics[width=1.in]{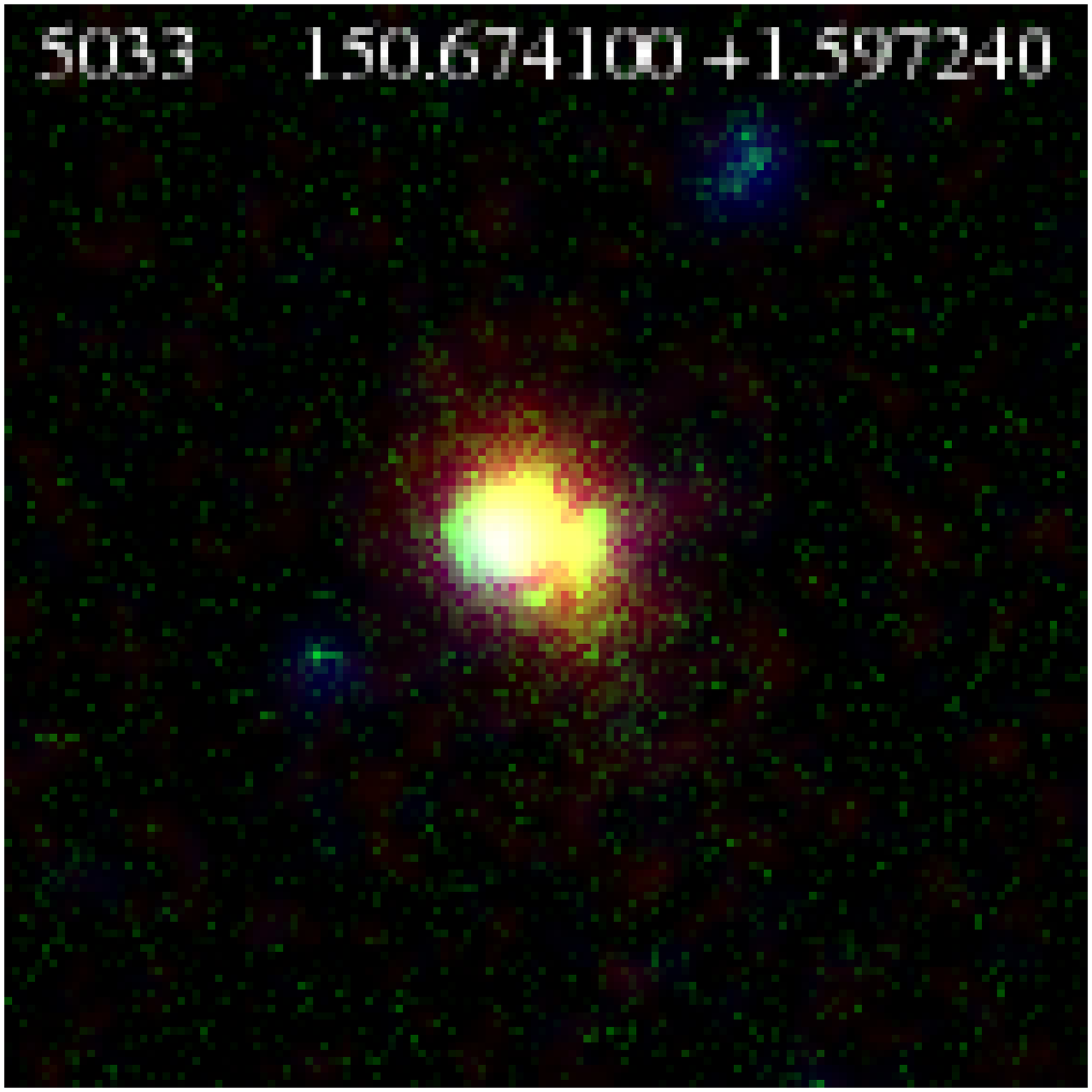}
\includegraphics[width=1.in]{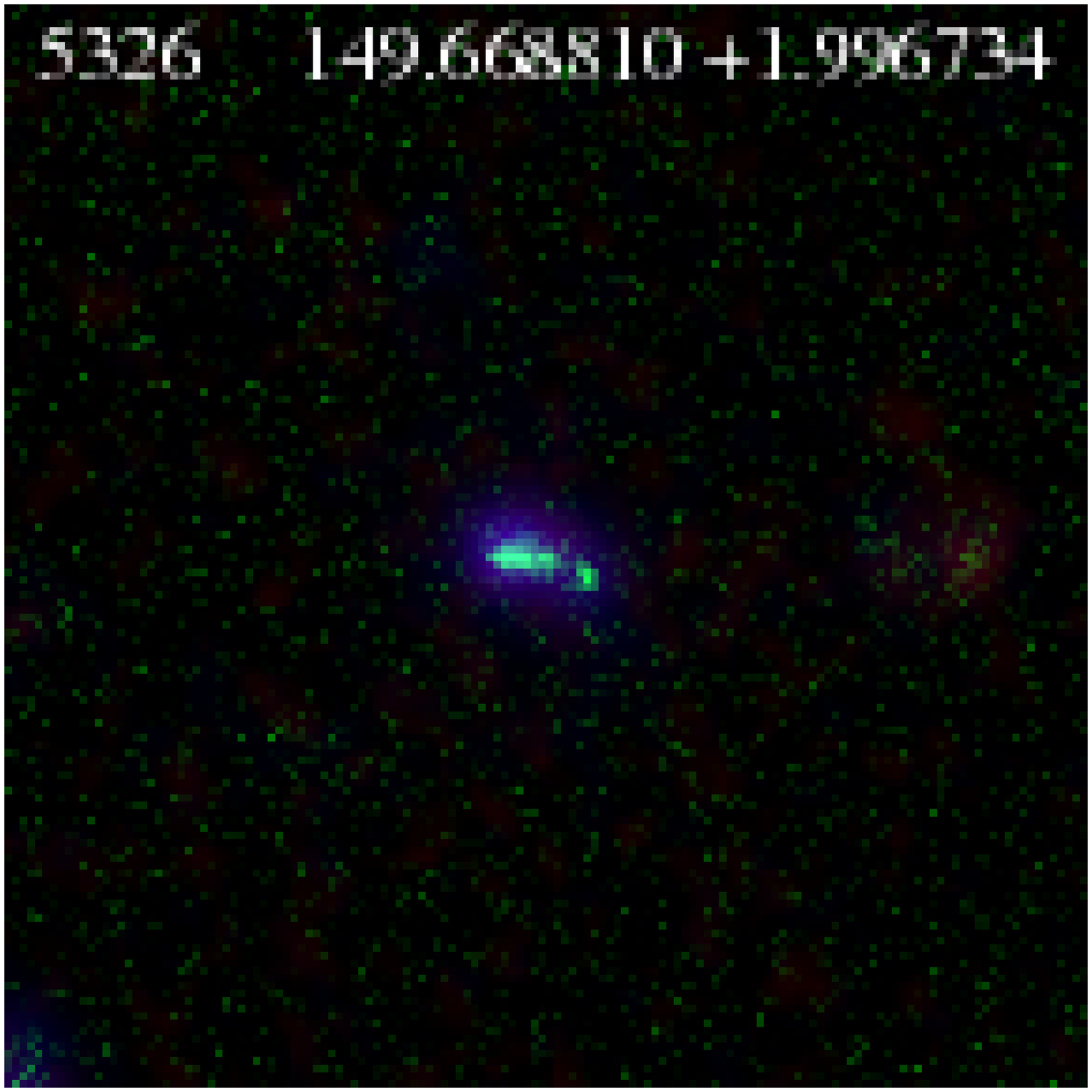}
\includegraphics[width=1.in]{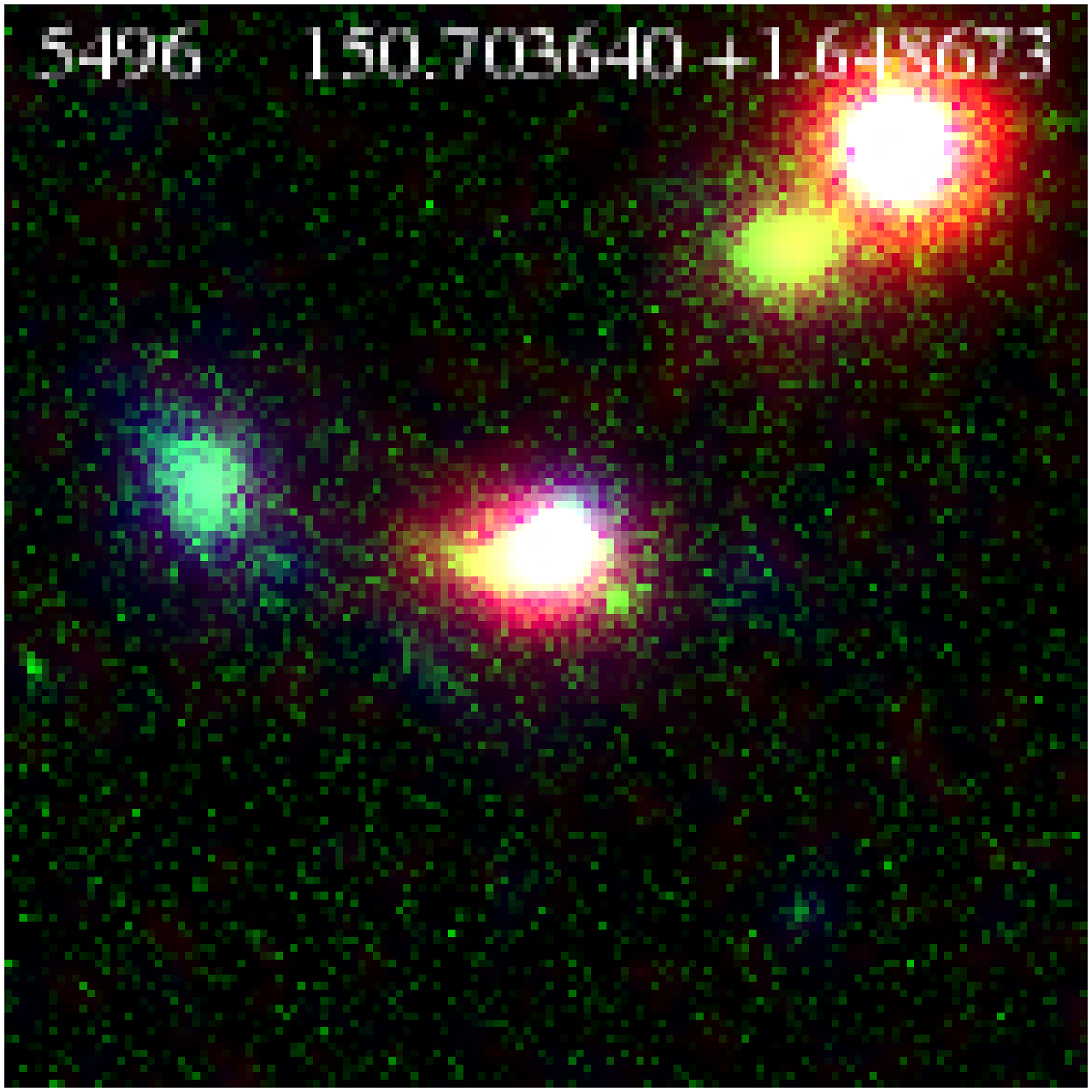}
% \vspace*{-1.0 cm}
 \caption{Type-2 QSOs hosts classified as {\it possible-merger}. The
   images are 10\arcsec~ on a side. They were obtained by combining
   Subaru B-band ({\it blue}), ACS F814W ({\it green}), and CFHT Ks
   ({\it red}) images. XID and coordinates of the source are reported
   at the top of each image.}
   \label{morph_qso2_merger}
\end{center}
\end{figure}

We repeated a similar analysis for the hosts in the three higher
redshifts bins: $1.2<z<1.8$ (36 objects), $1.8<z<2.5$ (28 objects),
and $2.5<z<3.2$ (12 objects). First, we find that the fraction of
star-forming galaxies between our Type-2 QSO hosts increases with
redshift: $66\%$, $71\%$, and $100\%$ respectively. We then looked at
the evolution of the SSFR for Type-2 QSO hosts between redshift
z$=0.8$ and z$=3$ as shown in Fig. \ref{ssfr_z1}. At a first look,
this figure would suggest that there is only a mild evolution, if any,
of the average SSFR for Type-2 QSO hosts going from z$\sim 1$ to
z$\sim 2$, while there is no evolution at all at z$>2$. The latter
agrees with the constant SSFR measured from z$\sim 7$ to z$\sim 2$ for
normal galaxies (e.g. \citealt{gonzalez10}), while the former
disagrees with the substantial drops of the average SSFR below z$\sim
2$ (e.g. \citealt{noeske07}, and \citealt{daddi07}). \citet{shao10}
used Herschel/PACS observations of the GOODS-N field to estimate the
star-formation rate in X-ray selected AGNs. They find that the host
far-infrared luminosity of AGN with L$_{2-10 keV}\approx 10^{44}$ erg
s$^{-1}$ does not significantly increase from z$\sim 1$ to z$\sim 2$
(see their Fig. 4). Since the rest-frame far-infrared emission is
typically dominated by the host rather than the AGN
(e.g. \citealt{netzer07b}) it can be used as a SFR diagnostic and
therefore the \citet{shao10} results are consistent with ours in
Fig. \ref{ssfr_z1}. However, submillimetre surveys of narrow line
radio galaxies, possibily radio loud Type-2 QSOs, found that their FIR
luminosity rises with redshift out to z$\approx 4$, and interpreted it
as an increase in the star formation activity with redshift
(e.g. \citealt{archibald01,reuland04}). We note that the samples of
Type-2 QSOs in the redshift bins plotted in Fig. \ref{ssfr_z1} have
different mean stellar masses: at z$\sim 1$, the average stellar mass
is $<M_{\star}>\sim 3 \times 10^{10}$ M$_{\odot}$, while it is higher,
$<M_{\star}>\sim 6 \times 10^{10}$ M$_{\odot}$, at z$\sim 3$. Also the
curve from \citet{pannella09} plotted in Fig. \ref{ssfr_z1} is most
valid for M$_{\star} \sim 3\times 10^{10}$ M$_{\odot}$. According to
\citet{karim11}, there is a correlation between SSFR and stellar
masses, with higher mass objects having lower SSFRs. For the three
redshifts bins below z$\sim 2$, we now include only the hosts with
log(M$_{\star}$)=[10.6-11.0] M$_{\odot}$ and consider the best fit to
the redshift evolution of the SSFR for normal galaxies in the same
mass range (Table 5 of \citealt{karim11}). As Fig. \ref{ssfr_z2}
shows, the agreement between Type-2 QSO hosts and normal galaxies is
now excellent: the average SSFR for our hosts decreases by a factor
$\approx 3.6$ going from z$\sim 2$ to z$\sim 1$.

\begin{figure}
% \vspace*{-2.0 cm}
\begin{center}
 \includegraphics[width=3.4in]{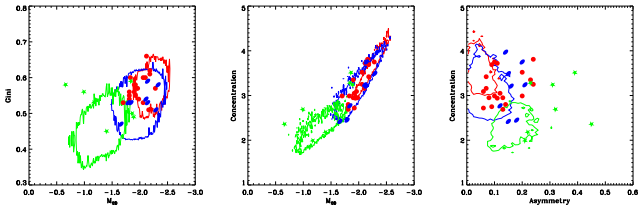} 
% \vspace*{-1.0 cm}
 \caption{Location of the hosts morphologically classified as
   bulge-dominated (filled circles), disk-dominated (filled ellipse),
   possible-mergers (filled stars) in three planes defined using the
   four non-parametric measurements of the light distribution:
   asymmetry, concentration, M20, Gini coefficient. The contours
   include $99\%$ of the \citet{scarlata07} galaxies with
   morphological classification as bulge-dominated (red),
   disk-dominated (blue), and possible-mergers (green).}
   \label{morph_qso2}
\end{center}
\end{figure}

\section{Morphology}

Another important diagnostic of galaxy evolution is their
morphologies, which correlates with their star-formation histories. In
Type-2 QSOs, obscuration on parsec scales prevents the light of the
nuclei from outshining the host galaxy, therefore making it easier,
compared to unobscured Type-1 QSOs, to study the morphological
properties of their hosts. Using the Cycle 12 Hubble Space Telescope
(HST) Advanced Camera for Surveys (ACS) F814W images of the COSMOS
field (\citealt{scoville07}; \citealt{koekemoer07}), we studied the
morphology of 34 Type-2 QSO host galaxies with z $<1.2$ and I$_{\rm
  AB}<24$. We used an upgraded version of the Zurich Estimator of
Structural Types (ZEST; \citealt{scarlata07}) known as ZEST+ (Carollo
et al., in preparation). This new tool, based on a principal component
analysis, includes additional measurements of non-parametric
morphological indices for characterizing both structure and
sub-structures. For each of the 34 host galaxies, we measured the
half-light radius (r$_{1/2}$), the Petrosian radius (R$_{\rm P}$), and
the ellipticity ($\epsilon$). In addition, we derived estimates of
four popular non-parametric measurements of the light distribution in
galaxies (e.g. \citealt{abraham03}; \citealt{lotz04}): concentration
(C), asymmetry (A), second-order moment of the brightest $20\%$ of the
galaxy flux (M$_{20}$), and Gini coefficient (G). The values of these
parameters for each of the 34 Type-2 QSOs hosts are reported in Table
2. Using a combination of this information and after a visual
inspection of the ACS images, we divided our host galaxies into three
classes: {\it bulge-dominated}, {\it disk-dominated}, and {\it
  possible-merger}. The breakdown in morphological classes for the
Type-2 QSO host galaxies at z$\le 1.2$ is the following: $57\%$
bulge-dominated, $23\%$ disk-dominated, and $20\%$ possible
mergers. Color cutouts of the hosts divided by morphological classes
are shown in Fig. \ref{morph_qso2_bulge}, \ref{morph_qso2_disk}, and
\ref{morph_qso2_merger}.  In Fig. \ref{morph_qso2}, we plot the
distribution of our host galaxies in three planes: G vs.  M$_{20}$, C
vs. M$_{20}$, and C vs. A. \citet{scarlata07} obtained reliable
morphological classifications for $\sim 56000$ I$_{\rm AB}\le 24$
galaxies in the COSMOS field using ZEST. For comparison, we added in
Fig. \ref{morph_qso2} the contours enclosing $99\%$ of the
\citet{scarlata07} galaxies belonging to the ZEST morphological
classes T=1 (red), T=2.0 or 2.1 (blue), which correspond to disk
galaxies with a Sersic index $n>1.25$, and T=3 (green). While in the
first two panels our morphological classification agrees quite well
with the large galaxies sample studied by \citet{scarlata07}, in the
third panel a significant fraction of our hosts are outside the
contours mainly due to larger asymmetry values than observed for
galaxies of the same morphological class. This could be interpreted as
an indication of a higher fraction of disturbed morphologies in our
sample of Type-2 QSO host galaxies than normally observed in a
magnitude limited sample of normal galaxies. Furthermore, our hosts
tend to be highly concentrated and $\approx 65\%$ of them have C$>3$,
a range in which elliptical and bulge-dominated galaxies are
located. This is not surprising since they are mostly massive galaxies
that usually have a consistent bulge already in
place. \citet{silverman08} also found that for X-ray selected AGN of
lower luminosity (log L$_{\rm 0.5-10 keV}< 43.7$ erg s$^{-1}$) at
z$<1$ the majority of the host galaxies are bulge dominated. We note
that in the morphological analysis presented in this paper, we did not
attempt to subtract any possible residual point-like source at the
center of the galaxy. If present, these point sources would bias the
concentration measurements towards higher values: \citet{gabor09}
found that this is the case for a sample of $\approx 200$ X-ray
selected AGN at $0.3<$z$<1$ in the COSMOS field (see their
Fig. 12). The AGN sample of \citet{gabor09} was, on average, less
obscured than the N$_{\rm H}>10^{22}$ cm$^{-2}$ objects we present,
hence the likelihood that a residual point-like source is present in
the ACS images is higher. We plan to study this possible bias in
detail in future work.\\ As reported above, we were able to detect
signs of recent mergers in only $\approx 20\%$ of the hosts of Type-2
QSOs at z$<1.2$. The lack of morphological merger signatures for the
majority of the objects should be treated with caution. The ACS images
could simply be not deep enough to detect faint surface brightness
features, as tidal tails or shells, to trace past interactions. For
example \citet{ramos11}, using deep optical images, found that
$\approx 80\%$ of a complete sample of z$<0.7$ narrow line radio
galaxies show signs of interactions based on the detection of faint
morphological features. In addition, some models
(e.g. \citealt{dimatteo05}) predict that the quasar phase occurs
towards the end of the merging process, and at that point the final
galaxy should be mostly relaxed and have low values of asymmetry
(e.g. \citealt{conselice03}). In this scenario, it becomes difficult
to test merging-induced quasar fueling models by studying the
morphological appearance of the host galaxies. At the same time, this
hypothesis also opens another possible approach: a major merger would
most likely form spheroidal galaxies and large disks should usually be
destroyed \footnote{Some simulations (e.g. \citealt{barnes96};
  \citealt{springelhernquist05}) show that mergers of gas-rich
  galaxies could produce a disk-dominated remnant but only for
  particular orientations of the angular momentum vectors. These
  conditions are usually not related to substantial BH growth
  (\citealt{hopkins09}).}. As described above, we find that the
majority of the Type-2 QSO hosts at z$<1.2$ are bulge-dominated
galaxies and that a few of them contain disks. In a study of the
morphology of the hosts of 140 AGN in the XMM-COSMOS survey,
\citet{cisternas11} found that: a) there is no significant difference
in the distortion fractions between active and inactive galaxies; b)
$\sim 65\%$ of the AGN hosts are disk dominated. On the basis of these
findings, they concluded that major mergers are not the trigger of BH
accretion. We note that the vast majority of the \citet{cisternas11}
AGN sample has lower X-ray luminosities than the Type-2 QSOs presented
in this paper: only 18 sources ($13\%$) of their sample have L$_{\rm
  X}>10^{44}$ erg s$^{-1}$ and these are mostly unobscured QSOs. The
different fraction of disk galaxies between the two studies could be
connected to different triggering mechanisms of accretion activity at
low and high AGN luminosities (e.g. \citealt{hopkins09}).\\ In
Fig. \ref{morph_edd}, we compared the distribution of Eddington ratios
for the three morphological classes. The range of values of this
statistic is limited, but bulge-dominated galaxies tend to host AGN
with low Eddington ratios ($\lambda < 0.1$), while disk-dominated or
merging galaxies at their centers contain BHs accreting at high
Eddington ratios ($\lambda > 0.1$). The mean BH mass of the
bulge-dominated sample is $<M_{\rm BH}> \sim 4 \times 10^{8}$
M$_{\odot}$, while for the mergers $<M_{\rm BH}> \sim 1 \times 10^{8}$
M$_{\odot}$. The distributions in Fig. \ref{morph_edd} are therefore
consistent with the general finding that the lowest mass BHs are the
fastest accretors (\citealt{mclure02}; \citealt{netzer07}).

\begin{figure}
% \vspace*{-2.0 cm}
\begin{center}
 \includegraphics[width=3.4in]{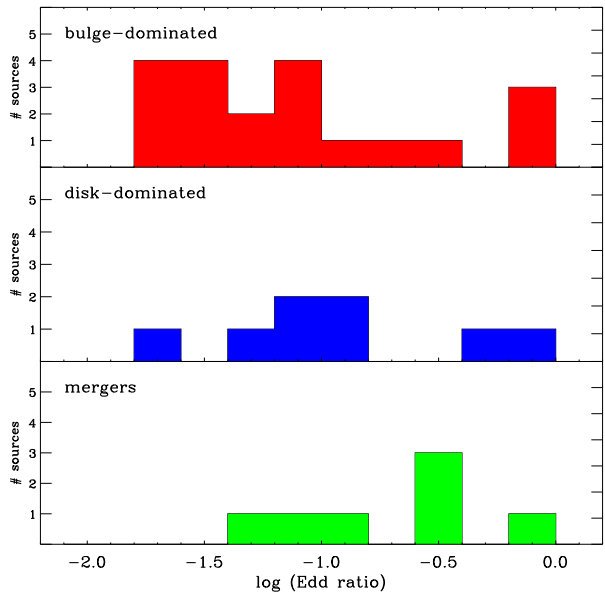} 
% \vspace*{-1.0 cm}
 \caption{Distribution of Eddington ratios for the hosts of Type-2
   QSOs morphologically classified as bulge-dominated (top panel),
   disk-dominated (middle-panel), and possible mergers (bottom
   panel).}
   \label{morph_edd}
\end{center}
\end{figure}

\section{Conclusions}

We have explored the relationships between SMBH accretion in obscured
quasars and the physical properties of their host galaxies. The sample
of 142 Type-2 QSOs was selected from those considered for the X-ray
spectral analysis in the XMM-COSMOS survey. The bulk of this quasar
sample is located in the redshift range $1<z<2.5$, which includes the
peaks of both the global star-formation rate and AGN activity. The
obscured quasars presented in the paper have a mean bolometric
luminosity of L$_{\rm bol}= 8 \times 10^{45}$ erg s$^{-1}$, a regime
where several models invoke major mergers of gas-rich galaxies as the
main fueling mechanism of the central SMBH.\\ A key ingredient of this
work has been the ability to use an AGN+host galaxy SED decomposition
technique to put reliable constraints on the stellar masses and
star-formation rates of the hosts. This has been possible for the
excellent multi-wavelength coverage in the COSMOS field.\\ The main
results of this work are the following:

\begin{itemize}

\item Type-2 QSOs reside almost exclusively in massive galaxies: more
  than $80\%$ of their hosts have M$_\star>10^{10}$ M$_\odot$. The
  fraction of galaxies hosting Type-2 QSOs monotonically increases
  with stellar masses going from $0.02\%$ at M$_\star\approx
  8\times10^{9}$ M$_\odot$ to $\approx 0.6\%$ at M$_\star\approx
  10^{11}$ M$_\odot$.

\item The rest-frame colors of their host galaxies are mostly between
  those of the blue and red galaxy population. We confirm that this is
  the effect of a luminosity-selected rather than a mass-selected
  sample. Furthermore, the classification of the host galaxies based
  on their rest-frame colors is subject to the effect of dust
  reddening: $\approx 20\%$ of the host galaxies with red or
  intermediate rest-frame colors are dusty star-forming objects.

\item At z$\sim 1$, $\approx 62\pm7\%$ of Type-2 QSOs are actively
  forming stars and this fraction increases with redshift, $\approx
  71\pm3\%$ at z$\sim2$, and $100\%$ at z$\sim 3$.

\item We find that the redshift evolution of the specific star
  formation rate for the hosts of Type-2 QSOs is in excellent
  agreement with that observed for star-forming galaxies in the
  redshift range $1<z<3$: the average SSFR increase by a factor
  $\approx 3.6$ going from z$\sim 1$ to z$\sim 2$, and remains
  constant at higher redshifts.

\item A morphological analysis of the ACS images has indicated that
  the majority of the host galaxies are bulge dominated, there being a
  few cases of disk galaxies or mergers. This lack of merger
  signatures cannot be automatically translated as proof that mergers
  are not playing an important role in the accretion history of these
  quasars. Several models indeed assume that the quasar phase occurs
  toward the end of the merging process, and at that point we expect
  the host galaxy to be mostly relaxed with low values of
  asymmetry. Numerical simulations also predict that large disks are
  destroyed during a major merger, therefore detecting large number of
  disks could place some strong constraint on the occurrence of a
  major merger. With the current data, we have detected clear disks in
  only a few cases.

\item We find that bulge-dominated galaxies tend to host Type-2
  QSOs with low Eddington ratios ($\lambda < 0.1$), while
  disk-dominated or merging galaxies have at their centers BHs
  accreting at high Eddington ratios ($\lambda > 0.1$).

\end{itemize}

The morphological analysis presented in this paper is based on F814W
ACS images, which corresponds to the rest-frame UV at the average
redshift of our galaxies. This implies that we are mainly mapping
regions of recent star-formation. It will be extremely interesting to
repeat this analysis using future near-IR observations from space that
will provide high resolution images in the optical rest-frame. The
CANDELS HST Multi-cycle Treasury Program \citep{grogin11,koekemoer11}
will image with WFC3/HST part of the COSMOS field and therefore
provide a less biased view of the morphology of these Type-2 QSOs
hosts. It would be also interesting to estimate the nebular continuum
contribution to the UV SED of Type-2 quasars host galaxies. This would
be possible using NIR spectroscopy to derive H$\beta$ flux and the
line of sight reddening for each quasar (see
e.g. \citealt{tadhunter02}). Finally, the ongoing Herschel survey of
the COSMOS field, as well as other patches of the sky, will improve
the mid and far-infrared coverage, a crucial wavelength range to
measure the SFR of AGN hosts.

\begin{acknowledgements}
  We are grateful to Hagai Netzer for inspiring discussions and to
  Giulia Rodighiero for providing the Herschel curves in
  Fig. \ref{sfr_smass1} in electronic format. We thank the referee,
  Montse Villar-Martin, for a careful reading of the manuscript and
  useful comments. This work is based on observations made with ESO
  Telescopes at the La Silla/Paranal Observatories under the zCOSMOS
  Large Programme 175.A-0839; the XMM--{\it Newton} satellite, an ESA
  science mission with instruments and contributions directly funded
  by ESA Member States and the US (NASA); the Magellan Telescope,
  operated by the Carnegie Observatorie and the MMT Observatory, a
  joint facility of the University of Arizona and the Smithsonian
  Institution; the Subaru Telescope, operated by the National
  Astronomical Observatory of Japan; and the NASA/ESA Hubble Space
  Telescope, operated at the Space Telescope Science Institute, which
  is operated by AURA Inc, under NASA contract NAS 5-26555. Support
  from the Italian Space Agency (ASI) under the contracts ASI-INAF
  I/088/06/0 and I/009/10/0 is acknowledged. G.H. and M.S. acknowledge
  support by the German Deutsche Forschungsgemeinschaft, DFG Leibniz
  Prize (FKZ HA 1850/28-1).
\end{acknowledgements}

\twocolumn

\longtab{1}{
\begin{longtable}{r c c c c r r r r}
\caption{SED of Type-2 QSOs. The legend for the column named {\it Survey} is the following:  s=spectroscopic redshift; p=photometric redshift.}\\
\hline\hline
 XID & log L$_{\rm X}$[0.5-10 keV] & log N$_{\rm H}$ & z & Survey & log M$\star$ & SFR & M$_{\rm U}$ & M$_{\rm B}$\\
 & (erg s$^{-1}$) & (cm$^{-2}$) & & & (M$_\odot$) &  (M$_\odot$ yr$^{-1}$) & (AB) & (AB)\\
\hline
\endfirsthead
\caption{continued.}\\
\hline\hline
 XID & log L$_{\rm X}$[0.5-10 keV] & log N$_{\rm H}$ & z & Survey & log M$\star$ & SFR & M$_{\rm U}$ & M$_{\rm B}$ \\
 & (erg s$^{-1}$) & (cm$^{-2}$) & & & (M$_\odot$) &  (M$_\odot$ yr$^{-1}$) & (AB) & (AB) \\
\hline
\endhead
\hline
\endfoot
41 & 45.20 & 22.1$^{22.5}_{22.1}$ & 0.962 & s & 10.2$^{10.2}_{10.2}$ & 79.1$^{79.4}_{78.3}$ & -21.4 & -22.0 \\\smallskip
50 & 45.44 & 22.5$^{22.6}_{22.3}$ & 2.941 & s & 9.8$^{9.8}_{9.8}$ & 157.8$^{158.9}_{157.8}$ & -21.8 & -22.2 \\\smallskip
70 & 44.51 & 23.2$^{23.4}_{23.1}$ & 0.688 & s & 10.8$^{10.8}_{10.8}$ & 94.8$^{94.8}_{94.8}$ & -20.6 & -21.7 \\\smallskip
97 & 44.53 & 22.4$^{22.5}_{22.2}$ & 1.300 & p & 10.7$^{10.7}_{10.7}$ & 64.3$^{64.4}_{64.3}$ & -20.9 & -21.8 \\\smallskip
105 & 44.68 & 22.7$^{22.9}_{22.3}$ & 1.600 & p & 10.7$^{10.7}_{10.7}$ & 4.3$^{4.3}_{4.3}$ & -19.8 & -20.9 \\\smallskip
106 & 44.16 & 22.6$^{22.8}_{22.5}$ & 1.260 & p & 10.8$^{10.8}_{10.8}$ & 0.5$^{0.5}_{0.5}$ & -19.1 & -20.5 \\\smallskip
107 & 44.77 & 22.0$^{22.5}_{21.3}$ & 1.530 & p & 10.4$^{10.4}_{10.3}$ & 33.0$^{33.0}_{32.2}$ & -21.9 & -22.5 \\\smallskip
108 & 44.44 & 22.3$^{22.6}_{21.4}$ & 1.570 & p & 10.5$^{10.5}_{10.5}$ & 84.7$^{86.5}_{83.8}$ & -20.4 & -21.4 \\\smallskip
122 & 45.12 & 23.0$^{23.3}_{22.8}$ & 2.418 & s & 10.6$^{10.6}_{10.6}$ & 0.7$^{0.8}_{0.7}$ & -20.1 & -21.2 \\\smallskip
138 & 44.85 & 22.7$^{22.9}_{22.6}$ & 2.610 & p & 11.0$^{11.1}_{10.6}$ & 305.5$^{946.2}_{296.5}$ & -22.4 & -23.2 \\\smallskip
144 & 44.96 & 22.6$^{22.9}_{22.3}$ & 2.930 & p & 11.5$^{11.5}_{11.5}$ & 32.7$^{32.7}_{32.5}$ & -22.2 & -23.3 \\\smallskip
151 & 44.10 & 22.7$^{22.9}_{22.4}$ & 0.797 & s & 10.6$^{10.6}_{10.6}$ & 113.2$^{114.0}_{110.9}$ & -21.9 & -22.7 \\\smallskip
159 & 44.43 & 22.1$^{22.5}_{21.2}$ & 2.230 & p & 11.5$^{11.5}_{11.5}$ & 116.9$^{118.3}_{116.1}$ & -21.0 & -22.2 \\\smallskip
181 & 44.58 & 22.2$^{22.4}_{22.0}$ & 1.720 & p & 11.1$^{11.2}_{11.1}$ & 63.1$^{63.5}_{62.4}$ & -20.4 & -21.6 \\\smallskip
182 & 44.91 & 22.9$^{23.0}_{22.7}$ & 2.040 & p & 10.9$^{11.1}_{10.9}$ & 0.6$^{1.0}_{0.6}$ & -21.1 & -22.3 \\\smallskip
194 & 44.40 & 22.7$^{23.0}_{22.4}$ & 1.456 & s & 10.6$^{10.6}_{10.4}$ & 65.3$^{67.5}_{64.7}$ & -20.8 & -21.7 \\\smallskip
200 & 44.07 & 22.7$^{23.0}_{22.3}$ & 1.156 & s & 10.3$^{10.3}_{10.3}$ & 22.2$^{22.5}_{22.2}$ & -19.7 & -20.7 \\\smallskip
209 & 44.55 & 22.9$^{23.3}_{22.5}$ & 2.000 & p & 11.0$^{11.0}_{10.8}$ & 33.0$^{33.7}_{14.2}$ & -20.3 & -21.4 \\\smallskip
211 & 44.02 & 22.2$^{22.4}_{22.0}$ & 1.188 & s & 10.5$^{10.5}_{10.5}$ & 55.3$^{55.3}_{54.7}$ & -20.0 & -21.0 \\\smallskip
212 & 44.55 & 23.4$^{24.3}_{23.2}$ & 0.931 & s & 10.7$^{10.7}_{10.4}$ & 13.2$^{13.3}_{2.7}$ & -20.1 & -21.2 \\\smallskip
229 & 44.03 & 22.2$^{22.5}_{21.8}$ & 0.864 & s & 11.1$^{11.1}_{11.1}$ & 1.6$^{1.6}_{1.6}$ & -20.2 & -21.5 \\\smallskip
233 & 44.59 & 22.7$^{22.9}_{22.5}$ & 1.930 & p & 10.6$^{10.8}_{10.5}$ & 76.6$^{84.9}_{33.5}$ & -20.4 & -21.4 \\\smallskip
235 & 44.79 & 22.6$^{22.8}_{22.4}$ & 2.220 & p & 11.5$^{11.5}_{11.5}$ & 370.7$^{376.7}_{370.7}$ & -22.2 & -23.3 \\\smallskip
247 & 44.49 & 23.0$^{23.3}_{22.6}$ & 1.950 & p & 10.8$^{10.8}_{10.5}$ & 19.2$^{19.5}_{7.9}$ & -20.3 & -21.3 \\\smallskip
276\footnote{The photometry is contaminated by the halo of a bright nearby star.} & 45.28 & 22.8$^{23.0}_{22.7}$ & 2.620 & p & 10.8$^{10.8}_{10.8}$ & 790.7$^{794.3}_{787.0}$ & -24.9 & -25.2 \\\smallskip
292 & 44.06 & 22.8$^{23.0}_{22.5}$ & 0.618 & s & 10.8$^{10.8}_{10.8}$ & 13.6$^{13.6}_{13.6}$ & -19.4 & -20.6 \\\smallskip
313 & 44.03 & 22.7$^{23.0}_{21.9}$ & 0.970 & p & 10.9$^{10.9}_{10.9}$ & 7.4$^{7.5}_{7.4}$ & -20.0 & -21.2 \\\smallskip
323 & 44.27 & 22.3$^{22.4}_{22.1}$ & 0.839 & s & 10.8$^{10.8}_{10.8}$ & 60.0$^{60.5}_{58.2}$ & -20.8 & -21.8 \\\smallskip
327 & 44.31 & 22.9$^{23.4}_{22.2}$ & 1.910 & p & 11.2$^{11.2}_{11.2}$ & 220.8$^{220.8}_{220.3}$ & -22.2 & -23.1 \\\smallskip
331 & 44.50 & 22.3$^{22.7}_{21.9}$ & 2.390 & p & 11.0$^{11.1}_{10.9}$ & 89.5$^{89.7}_{36.5}$ & -21.2 & -22.2 \\\smallskip
333 & 44.74 & 23.2$^{-1.0}_{20.4}$ & 3.070 & p & 11.0$^{11.0}_{11.0}$ & 150.0$^{150.0}_{149.6}$ & -23.0 & -23.7 \\\smallskip
337 & 44.61 & 22.7$^{22.9}_{22.4}$ & 1.720 & p & 10.9$^{10.9}_{10.9}$ & 179.9$^{180.7}_{177.8}$ & -21.3 & -22.2 \\\smallskip
338 & 44.35 & 23.0$^{23.4}_{22.4}$ & 1.550 & p & 11.0$^{11.0}_{11.0}$ & 3.2$^{3.2}_{3.2}$ & -19.9 & -21.1 \\\smallskip
351 & 44.72 & 22.9$^{-1.0}_{20.4}$ & 2.200 & p & 11.2$^{11.2}_{11.2}$ & 17.7$^{18.2}_{17.1}$ & -20.9 & -22.1 \\\smallskip
362 & 44.22 & 22.8$^{23.3}_{22.3}$ & 1.490 & p & 10.3$^{10.3}_{10.2}$ & 4.6$^{4.7}_{4.5}$ & -20.1 & -20.9 \\\smallskip
390 & 44.67 & 22.8$^{23.2}_{22.4}$ & 1.740 & p & 11.4$^{11.4}_{11.4}$ & 98.4$^{102.8}_{98.4}$ & -20.9 & -22.0 \\\smallskip
403 & 44.59 & 22.2$^{22.5}_{21.6}$ & 2.510 & p & 9.9$^{9.9}_{9.9}$ & 12.5$^{12.9}_{11.6}$ & -20.2 & -20.9 \\\smallskip
406 & 44.68 & 23.0$^{23.3}_{22.7}$ & 2.921 & s & 10.9$^{10.9}_{10.9}$ & 37.2$^{37.4}_{36.9}$ & -21.7 & -22.5 \\\smallskip
411 & 44.09 & 22.4$^{22.6}_{21.9}$ & 0.952 & s & 10.8$^{10.8}_{10.8}$ & 0.8$^{0.8}_{0.8}$ & -20.1 & -21.3 \\\smallskip
412 & 44.42 & 22.8$^{23.0}_{22.7}$ & 1.300 & p & 10.6$^{10.6}_{10.6}$ & 108.1$^{109.1}_{107.6}$ & -20.7 & -21.6 \\\smallskip
413 & 44.80 & 23.6$^{23.9}_{23.3}$ & 1.023 & s & 10.6$^{10.6}_{10.6}$ & 47.2$^{47.3}_{47.2}$ & -21.7 & -22.5 \\\smallskip
428 & 44.52 & 22.2$^{22.6}_{21.7}$ & 2.010 & p & 11.1$^{11.1}_{11.1}$ & 1.0$^{1.0}_{1.0}$ & -21.7 & -22.8 \\\smallskip
435 & 44.49 & 22.5$^{22.9}_{22.2}$ & 2.080 & p & 10.3$^{10.3}_{10.3}$ & 32.9$^{33.6}_{32.8}$ & -21.9 & -22.6 \\\smallskip
440 & 44.25 & 22.3$^{22.7}_{20.7}$ & 1.800 & p & 11.0$^{11.0}_{11.0}$ & 98.2$^{99.5}_{97.3}$ & -20.7 & -21.8 \\\smallskip
463 & 44.32 & 22.6$^{23.0}_{22.3}$ & 1.870 & p & 11.2$^{11.2}_{11.2}$ & 3.4$^{3.4}_{3.4}$ & -20.5 & -21.8 \\\smallskip
470 & 44.31 & 22.7$^{23.4}_{21.8}$ & 2.070 & p & 10.9$^{10.9}_{10.9}$ & 1.7$^{1.7}_{1.7}$ & -19.8 & -21.1 \\\smallskip
2206 & 45.42 & 23.0$^{23.1}_{22.9}$ & 2.410 & p & 11.0$^{11.0}_{11.0}$ & 160.7$^{161.4}_{160.0}$ & -21.8 & -22.8 \\\smallskip
2208 & 45.08 & 23.0$^{23.0}_{22.9}$ & 1.130 & p & 10.7$^{10.7}_{10.6}$ & 75.0$^{77.4}_{73.5}$ & -20.4 & -21.4 \\\smallskip
2210 & 44.08 & 22.8$^{23.0}_{22.6}$ & 0.968 & s & 11.0$^{11.0}_{11.0}$ & 10.7$^{10.7}_{10.6}$ & -20.9 & -22.1 \\\smallskip
2236 & 44.70 & 22.0$^{22.3}_{21.7}$ & 1.800 & p & 10.4$^{10.4}_{10.4}$ & 289.1$^{291.7}_{289.1}$ & -21.4 & -22.1 \\\smallskip
2237 & 44.48 & 22.4$^{22.6}_{22.3}$ & 0.944 & s & 11.2$^{11.2}_{11.2}$ & 17.7$^{17.7}_{17.6}$ & -21.5 & -22.6 \\\smallskip
2289 & 44.17 & 22.6$^{22.8}_{22.3}$ & 0.833 & s & 11.0$^{11.0}_{11.0}$ & 114.6$^{114.8}_{114.0}$ & -21.5 & -22.5 \\\smallskip
2336 & 44.67 & 22.8$^{23.2}_{22.5}$ & 1.760 & p & 10.9$^{10.9}_{10.8}$ & 6.6$^{6.6}_{6.4}$ & -20.9 & -21.9 \\\smallskip
2390\footnote{The ACS image shows a source at 0.5" from the chosen counterpart, therefore the IR photometry is the blending of the emission of these two sources.} & 44.94 & 23.3$^{23.6}_{23.1}$ & 2.830 & p & 10.6$^{10.6}_{10.6}$ & 223.4$^{224.9}_{220.8}$ & -23.8 & -24.2 \\\smallskip
2408 & 44.24 & 23.0$^{23.2}_{22.7}$ & 1.270 & s & 11.1$^{11.1}_{11.1}$ & 317.0$^{317.0}_{317.0}$ & -21.9 & -22.8 \\\smallskip
2414 & 44.55 & 22.2$^{22.4}_{22.0}$ & 1.640 & p & 11.2$^{11.2}_{11.2}$ & 68.4$^{71.9}_{68.2}$ & -21.7 & -22.7 \\\smallskip
2440 & 44.02 & 22.3$^{22.5}_{22.0}$ & 1.175 & s & 11.0$^{11.1}_{11.0}$ & 0.8$^{0.8}_{0.8}$ & -21.5 & -22.7 \\\smallskip
2464 & 44.11 & 22.4$^{22.6}_{22.2}$ & 1.260 & p & 10.9$^{10.9}_{10.9}$ & 0.6$^{0.6}_{0.6}$ & -21.2 & -22.4 \\\smallskip
2483 & 44.62 & 22.4$^{22.7}_{22.0}$ & 1.650 & p & 10.8$^{10.8}_{10.7}$ & 98.6$^{101.2}_{96.4}$ & -21.3 & -22.2 \\\smallskip
2507 & 44.40 & 22.6$^{22.9}_{22.2}$ & 0.937 & s & 9.8$^{9.8}_{9.8}$ & 86.3$^{87.1}_{86.3}$ & -21.9 & -22.3 \\\smallskip
2518 & 44.90 & 23.1$^{23.4}_{22.7}$ & 3.176 & s & 11.5$^{11.5}_{11.5}$ & 77.4$^{77.4}_{77.4}$ & -22.7 & -23.6 \\\smallskip
2528 & 44.78 & 22.3$^{22.6}_{21.8}$ & 2.050 & p & 9.9$^{9.9}_{9.9}$ & 195.9$^{197.2}_{194.5}$ & -22.0 & -22.4 \\\smallskip
2530 & 45.02 & 22.4$^{-1.0}_{20.4}$ & 2.640 & p & 11.8$^{11.8}_{11.8}$ & 3026.9$^{3026.9}_{2992.3}$ & -24.8 & -25.5 \\\smallskip
2544 & 44.65 & 23.3$^{23.6}_{23.0}$ & 0.826 & s & 11.3$^{11.3}_{11.3}$ & 2.6$^{2.6}_{2.6}$ & -21.4 & -22.5 \\\smallskip
2582\footnote{For this objects we have photometry only in five filters.} & 44.25 & 22.5$^{22.8}_{22.0}$ & 2.090 & p & 11.3$^{11.3}_{11.2}$ & 0.0$^{3.4}_{0.0}$ & -18.3 & -20.0 \\\smallskip
2597 & 44.01 & 22.3$^{22.6}_{21.8}$ & 1.119 & s & 10.7$^{10.7}_{10.7}$ & 23.4$^{23.4}_{23.3}$ & -20.5 & -21.5 \\\smallskip
2668 & 45.32 & 23.2$^{23.6}_{20.4}$ & 0.886 & s & 10.8$^{10.8}_{10.8}$ & 0.4$^{0.4}_{0.4}$ & -19.6 & -21.0 \\\smallskip
2684 & 44.47 & 22.5$^{22.7}_{22.2}$ & 1.810 & p & 11.1$^{11.1}_{11.1}$ & 12.1$^{12.1}_{12.1}$ & -20.9 & -22.0 \\\smallskip
2772 & 44.65 & 22.4$^{22.7}_{21.8}$ & 1.600 & p & 11.2$^{11.2}_{11.2}$ & 1.1$^{1.1}_{1.1}$ & -21.8 & -23.0 \\\smallskip
5006 & 45.52 & 24.1$^{24.4}_{23.8}$ & 2.417 & s & 11.3$^{11.3}_{11.3}$ & 540.8$^{543.3}_{538.3}$ & -22.5 & -23.4 \\\smallskip
5007 & 44.99 & 23.8$^{24.5}_{20.4}$ & 2.110 & p & 10.5$^{10.6}_{10.5}$ & 21.9$^{24.9}_{10.3}$ & -19.2 & -20.3 \\\smallskip
5033 & 44.62 & 23.6$^{24.6}_{23.6}$ & 0.670 & p & 10.7$^{10.7}_{10.7}$ & 1.0$^{1.0}_{1.0}$ & -20.3 & -21.4 \\\smallskip
5039 & 44.40 & 22.5$^{22.9}_{21.8}$ & 1.750 & p & 11.2$^{11.2}_{11.2}$ & 81.7$^{83.8}_{79.6}$ & -20.6 & -21.7 \\\smallskip
5043 & 44.14 & 22.9$^{24.2}_{20.4}$ & 1.380 & p & 10.8$^{10.8}_{10.8}$ & 65.5$^{65.5}_{65.5}$ & -20.3 & -21.4 \\\smallskip
5053 & 44.30 & 23.0$^{23.4}_{22.5}$ & 1.200 & p & 11.0$^{11.0}_{11.0}$ & 0.8$^{0.8}_{0.8}$ & -20.9 & -22.1 \\\smallskip
5054 & 44.54 & 22.0$^{22.3}_{21.7}$ & 1.420 & p & 10.8$^{10.8}_{10.8}$ & 38.9$^{39.5}_{38.5}$ & -19.8 & -21.0 \\\smallskip
5056 & 44.19 & 22.5$^{22.8}_{22.3}$ & 1.220 & p & 11.0$^{11.0}_{11.0}$ & 0.7$^{0.7}_{0.7}$ & -20.7 & -21.9 \\\smallskip
5057 & 44.19 & 22.2$^{22.7}_{20.8}$ & 1.690 & p & 11.1$^{11.1}_{11.0}$ & 115.1$^{116.7}_{115.1}$ & -21.5 & -22.5 \\\smallskip
5066 & 44.42 & 22.4$^{22.7}_{21.9}$ & 0.970 & p & 11.0$^{11.0}_{11.0}$ & 9.5$^{9.6}_{9.5}$ & -20.8 & -22.0 \\\smallskip
5070\footnote{The photometry is contaminated by a nearby bright star.} & 44.77 & 23.2$^{23.4}_{23.1}$ & 1.450 & p & 10.3$^{10.3}_{10.3}$ & 462.4$^{465.6}_{461.3}$ & -24.7 & -24.9 \\\smallskip
5111 & 44.63 & 22.9$^{23.1}_{22.6}$ & 1.080 & p & 10.0$^{10.1}_{10.0}$ & 12.4$^{12.4}_{12.4}$ & -19.7 & -20.6 \\\smallskip
5120 & 44.98 & 22.4$^{22.6}_{21.5}$ & 3.650 & p & ... & ... & 0.0 & 0.0\\\smallskip
5149 & 44.07 & 22.2$^{22.8}_{20.4}$ & 1.560 & p & 10.6$^{10.7}_{10.6}$ & 93.8$^{94.6}_{87.5}$ & -20.6 & -21.5 \\\smallskip
5153 & 44.13 & 22.9$^{23.1}_{22.7}$ & 0.787 & s & 10.0$^{10.0}_{10.0}$ & 47.1$^{47.9}_{47.1}$ & -19.6 & -20.5 \\\smallskip
5158 & 44.22 & 22.4$^{22.6}_{22.0}$ & 1.440 & p & 10.9$^{10.9}_{10.9}$ & 22.9$^{22.9}_{22.8}$ & -21.7 & -22.5 \\\smallskip
5161 & 45.54 & 22.9$^{23.7}_{20.4}$ & 3.160 & p & 11.1$^{11.1}_{11.1}$ & 177.4$^{177.4}_{175.8}$ & -22.6 & -23.4 \\\smallskip
5162 & 44.94 & 22.5$^{22.9}_{20.4}$ & 3.524 & s & 10.5$^{10.5}_{10.5}$ & 699.8$^{699.8}_{695.0}$ & -23.4 & -23.8 \\\smallskip
5178 & 44.43 & 23.1$^{23.4}_{22.8}$ & 1.720 & p & 11.3$^{11.3}_{11.3}$ & 1.3$^{1.3}_{1.3}$ & -21.4 & -22.7 \\\smallskip
5186 & 45.16 & 22.6$^{22.7}_{22.4}$ & 2.540 & p & 9.9$^{9.9}_{9.9}$ & 195.4$^{195.4}_{188.8}$ & -20.8 & -21.4 \\\smallskip
5229 & 44.37 & 22.6$^{23.1}_{22.0}$ & 1.030 & p & 10.7$^{10.8}_{10.7}$ & 11.1$^{11.7}_{11.1}$ & -19.8 & -20.9 \\\smallskip
5232 & 44.73 & 22.9$^{23.2}_{21.8}$ & 2.830 & p & 11.2$^{11.2}_{11.1}$ & 72.8$^{89.7}_{70.8}$ & -20.6 & -21.7 \\\smallskip
5266 & 44.72 & 23.1$^{23.3}_{22.8}$ & 1.950 & p & 10.8$^{10.8}_{10.8}$ & 54.6$^{56.1}_{54.6}$ & -20.7 & -21.7 \\\smallskip
5278 & 44.94 & 23.2$^{23.4}_{22.6}$ & 1.620 & p & 10.6$^{10.6}_{10.6}$ & 32.6$^{32.8}_{32.1}$ & -20.7 & -21.6 \\\smallskip
5288 & 44.15 & 22.2$^{22.4}_{21.9}$ & 1.000 & s & 11.1$^{11.1}_{11.1}$ & 2.2$^{2.2}_{2.2}$ & -20.7 & -21.9 \\\smallskip
5292 & 44.57 & 22.4$^{23.0}_{20.4}$ & 2.730 & p & 11.0$^{11.0}_{10.9}$ & 27.0$^{37.3}_{27.0}$ & -19.7 & -21.0 \\\smallskip
5305\footnote{The quality of the SED fit is poor.} & 44.17 & 22.4$^{23.0}_{21.4}$ & 2.500 & p & 10.1$^{10.1}_{10.1}$ & 161.8$^{164.1}_{158.9}$ & -21.4 & -21.9 \\\smallskip
5326\footnote{The IR photometry is contaminated by a nearby source.} & 44.35 & 22.7$^{23.2}_{22.0}$ & 0.950 & p & 8.6$^{8.6}_{8.6}$ & 9.2$^{9.3}_{9.2}$ & -19.3 & -19.6 \\\smallskip
5347 & 44.65 & 23.1$^{24.3}_{20.4}$ & 3.120 & p & 11.4$^{11.4}_{11.4}$ & 212.8$^{224.9}_{208.4}$ & -22.2 & -23.2 \\\smallskip
5357 & 44.83 & 23.6$^{23.9}_{23.4}$ & 1.670 & p & 10.7$^{10.7}_{10.7}$ & 1.6$^{1.7}_{1.6}$ & -20.4 & -21.4 \\\smallskip
5368 & 44.17 & 22.2$^{22.4}_{21.9}$ & 1.230 & p & 11.2$^{11.2}_{11.2}$ & 105.9$^{105.9}_{105.9}$ & -20.9 & -22.1 \\\smallskip
5369 & 44.16 & 22.1$^{22.4}_{21.8}$ & 0.950 & p & 10.4$^{10.4}_{10.4}$ & 21.1$^{21.7}_{21.0}$ & -19.7 & -20.7 \\\smallskip
5400 & 44.19 & 22.7$^{22.9}_{22.6}$ & 0.680 & p & 10.8$^{10.8}_{10.8}$ & 1.2$^{1.2}_{1.2}$ & -20.0 & -21.2 \\\smallskip
5414 & 44.10 & 22.4$^{22.5}_{22.2}$ & 0.910 & p & 10.7$^{10.7}_{10.7}$ & 0.9$^{1.0}_{0.9}$ & -19.7 & -20.9 \\\smallskip
5417 & 44.62 & 22.4$^{22.8}_{21.9}$ & 2.050 & p & 11.0$^{11.0}_{11.0}$ & 1153.5$^{1153.5}_{1153.5}$ & -24.1 & -24.6 \\\smallskip
5427 & 44.86 & 23.6$^{23.8}_{23.4}$ & 1.177 & s & 11.1$^{11.1}_{11.1}$ & 2.7$^{2.7}_{2.7}$ & -21.5 & -22.6 \\\smallskip
5441 & 44.11 & 22.1$^{22.5}_{21.7}$ & 1.170 & p & 11.1$^{11.1}_{11.1}$ & 4.2$^{4.2}_{4.2}$ & -20.8 & -21.9 \\\smallskip
5448 & 45.30 & 22.8$^{23.0}_{22.5}$ & 2.060 & p & 11.0$^{11.0}_{11.0}$ & 43.0$^{43.1}_{42.8}$ & -21.8 & -22.7 \\\smallskip
5452 & 44.40 & 22.0$^{22.4}_{21.6}$ & 1.920 & p & 10.0$^{10.0}_{10.0}$ & 131.5$^{133.0}_{131.5}$ & -21.2 & -21.7 \\\smallskip
5467 & 44.14 & 23.2$^{24.6}_{20.4}$ & 0.870 & p & 10.7$^{10.7}_{10.7}$ & 1.6$^{1.6}_{1.6}$ & -19.1 & -20.3 \\\smallskip
5488 & 44.36 & 23.1$^{23.6}_{22.2}$ & 1.280 & p & 11.0$^{11.0}_{11.0}$ & 163.7$^{164.4}_{163.3}$ & -21.8 & -22.8 \\\smallskip
5496 & 44.60 & 23.6$^{25.0}_{23.6}$ & 0.694 & s & 10.6$^{10.6}_{10.6}$ & 54.5$^{54.5}_{54.5}$ & -20.6 & -21.6 \\\smallskip
5512 & 44.31 & 22.1$^{22.6}_{20.6}$ & 1.130 & p & 11.4$^{11.4}_{11.4}$ & 5.2$^{5.2}_{5.1}$ & -22.2 & -23.3 \\\smallskip
5546 & 44.62 & 22.8$^{-1.0}_{20.4}$ & 1.200 & p & 11.1$^{11.1}_{11.1}$ & 2.4$^{2.4}_{2.4}$ & -21.3 & -22.5 \\\smallskip
5547 & 44.07 & 22.4$^{22.6}_{22.0}$ & 0.960 & p & 9.5$^{9.5}_{9.5}$ & 17.5$^{17.8}_{17.5}$ & -19.2 & -19.9 \\\smallskip
5548 & 45.01 & 23.1$^{23.5}_{22.8}$ & 1.060 & p & 10.9$^{10.9}_{10.9}$ & 1.0$^{1.0}_{1.0}$ & -20.3 & -21.5 \\\smallskip
5553 & 44.51 & 23.1$^{-1.0}_{20.4}$ & 2.390 & p & 11.2$^{11.2}_{11.1}$ & 1.1$^{1.1}_{0.9}$ & -20.6 & -22.0 \\\smallskip
5563 & 44.88 & 22.8$^{23.0}_{22.6}$ & 3.410 & p & 10.3$^{10.3}_{10.3}$ & 473.2$^{493.2}_{443.6}$ & -21.7 & -22.3 \\\smallskip
5569 & 45.04 & 22.7$^{23.2}_{22.0}$ & 1.810 & p & 11.2$^{11.2}_{11.2}$ & 3.3$^{3.3}_{3.3}$ & -21.1 & -22.3 \\\smallskip
5573 & 44.88 & 22.2$^{22.4}_{22.0}$ & 1.660 & p & 11.2$^{11.2}_{11.2}$ & 182.0$^{188.8}_{181.1}$ & -21.4 & -22.5 \\\smallskip
5582 & 44.60 & 22.9$^{23.1}_{22.7}$ & 1.490 & p & 10.7$^{10.7}_{10.7}$ & 0.9$^{0.9}_{0.9}$ & -20.3 & -21.4 \\\smallskip
5606 & 44.73 & 22.6$^{23.2}_{20.8}$ & 4.166 & s & 10.2$^{10.2}_{10.2}$ & 401.8$^{401.8}_{399.0}$ & -23.4 & -23.7 \\\smallskip
10094 & 44.69 & 22.9$^{23.2}_{20.4}$ & 2.670 & p & 10.7$^{10.7}_{10.7}$ & 5.0$^{5.0}_{5.0}$ & -20.7 & -21.8 \\\smallskip
10499 & 44.11 & 22.5$^{23.0}_{21.7}$ & 1.990 & p & 11.2$^{11.2}_{11.2}$ & 98.9$^{98.9}_{98.9}$ & -20.9 & -22.0 \\\smallskip
10690 & 44.52 & 22.7$^{23.3}_{21.9}$ & 3.100 & p & 10.2$^{10.2}_{10.2}$ & 177.8$^{179.1}_{176.2}$ & -22.7 & -23.1 \\\smallskip
10719 & 44.68 & 22.5$^{22.9}_{22.1}$ & 1.700 & p & 10.5$^{10.5}_{10.5}$ & 360.6$^{364.8}_{360.6}$ & -22.9 & -23.3 \\\smallskip
30292 & 44.37 & 23.0$^{-1.0}_{20.9}$ & 2.550 & p & 10.0$^{10.0}_{10.0}$ & 1.1$^{1.1}_{1.1}$ & -19.7 & -20.6 \\\smallskip
30361 & 44.19 & 22.8$^{23.4}_{20.4}$ & 1.910 & p & 11.1$^{11.1}_{11.1}$ & 156.0$^{156.0}_{151.4}$ & -21.2 & -22.2 \\\smallskip
30789 & 44.99 & 22.8$^{23.3}_{20.4}$ & 1.810 & p & 11.1$^{11.1}_{11.1}$ & 2.7$^{2.7}_{2.7}$ & -20.9 & -22.1 \\\smallskip
31357 & 44.63 & 23.2$^{23.5}_{22.9}$ & 1.700 & p & 11.4$^{11.4}_{11.4}$ & 1.7$^{1.7}_{1.7}$ & -21.7 & -23.0 \\\smallskip
31419 & 44.35 & 23.6$^{-1.0}_{20.4}$ & 1.130 & p & 9.9$^{10.0}_{9.9}$ & 14.6$^{14.7}_{7.0}$ & -18.6 & -19.6 \\\smallskip
53351 & 45.80 & 23.2$^{-1.0}_{20.4}$ & 3.000 & p & 10.3$^{10.3}_{10.1}$ & 53.7$^{66.5}_{53.2}$ & -21.1 & -21.9 \\\smallskip
54316 & 44.26 & 22.5$^{22.9}_{22.0}$ & 1.980 & p & 11.2$^{11.2}_{11.0}$ & 68.7$^{161.4}_{68.7}$ & -21.1 & -22.2 \\\smallskip
54514 & 44.44 & 23.6$^{-1.0}_{20.4}$ & 0.707 & s & 11.1$^{11.1}_{11.1}$ & 2.5$^{2.5}_{2.5}$ & -20.8 & -22.0 \\\smallskip
60075 & 44.15 & 22.9$^{23.4}_{22.4}$ & 1.080 & p & 10.8$^{10.8}_{10.8}$ & 1.3$^{1.3}_{1.3}$ & -19.4 & -20.7 \\\smallskip
60144 & 44.07 & 23.0$^{-1.0}_{20.4}$ & 1.500 & p & 10.9$^{10.9}_{10.9}$ & 0.5$^{0.5}_{0.5}$ & -21.1 & -22.2 \\\smallskip
60168 & 44.84 & 22.6$^{22.9}_{22.3}$ & 2.140 & p & 10.9$^{10.9}_{10.9}$ & 380.2$^{383.7}_{372.4}$ & -21.9 & -22.7 \\\smallskip
60214 & 44.17 & 22.8$^{-1.0}_{20.4}$ & 1.640 & p & 10.9$^{10.9}_{10.8}$ & 0.3$^{0.5}_{0.3}$ & -19.8 & -21.1 \\\smallskip
60255 & 44.15 & 22.8$^{-1.0}_{20.4}$ & 0.890 & p & 11.2$^{11.2}_{11.2}$ & 161.8$^{161.8}_{161.8}$ & -21.3 & -22.3 \\\smallskip
60314 & 44.59 & 23.5$^{24.0}_{20.4}$ & 1.080 & p & 10.3$^{10.3}_{10.3}$ & 49.8$^{50.8}_{49.5}$ & -19.9 & -20.8 \\\smallskip
60421 & 44.24 & 22.6$^{23.1}_{20.4}$ & 1.600 & p & 10.0$^{10.0}_{9.8}$ & 8.3$^{18.5}_{8.0}$ & -19.9 & -20.7 \\\smallskip
60422 & 45.17 & 23.0$^{23.3}_{22.6}$ & 1.890 & p & 9.8$^{9.8}_{9.8}$ & 71.3$^{71.8}_{71.0}$ & -21.1 & -21.6 \\\smallskip
60423 & 44.34 & 22.2$^{22.8}_{20.4}$ & 1.560 & p & 10.9$^{10.9}_{10.9}$ & 387.3$^{393.6}_{387.3}$ & -22.6 & -23.3 \\\smallskip
\end{longtable}
}

%\longtab{3}{
\begin{table*}
\caption{Morphological parameters of Type-2 QSO hosts at z$\le 1.2$. The morphological classes in column 4 are : 1 = bulge dominated; 2 = disk dominated; 3 = possible merger.}
\label{table:3}
\centering       
\begin{tabular}{r c c c c c c c c c c}
\noalign{\smallskip}
\hline\hline   
\noalign{\smallskip}
 XID & z & magI & Class & r$_{1/2}$\footnote{Half-light radius (r$_{1/2}$) and Petrosian radius (R$_{\rm P}$) are expressed in pixels, where 1 pxl$=0.03\arcsec$}    & R$_{\rm P}$ & $\epsilon$ & C & A & G & M$_{20}$\\
 & & AB & & pixels & pixels & & & &  &  \\
\noalign{\smallskip}
\hline 
\noalign{\smallskip}
41 & 0.962 & 21.73  & 1 & 3.01  & 6.01  & 0.10 & 2.99 & 0.20 & 0.53 & -1.70 \\
70 & 0.688 & 20.51  & 2 & 15.14 & 30.35 & 0.28 & 2.42 & 0.15 & 0.47 &  -1.66 \\
151 & 0.797 & 21.99 & 3 & 21.17 & 51.81 & 0.42 & 2.36 & 0.45 & 0.58 & -0.66 \\
200 & 1.156 & 23.34 & 3 & 11.18 & 31.52 & 0.25 & 3.24 & 0.31 & 0.49 & -1.89 \\
211 & 1.188 & 23.94 & 1 & 4.91 & 13.06 & 0.01 & 2.99 & 0.10 & 0.60 & -1.83 \\
212 & 0.931 & 22.47 & 1 & 7.73 & 19.87 & 0.17 & 3.16 & 0.07 & 0.60 & -2.10 \\
229 & 0.864 & 21.67 & 1 & 8.23 & 18.33 & 0.20 & 2.72 & 0.06 & 0.55 & -1.92 \\
292 & 0.618 & 21.68 & 2 & 21.47 & 41.24 & 0.67 & 2.45 & 0.18 & 0.53 & -1.79 \\
313 & 0.970 & 22.51 & 2 & 10.03 & 22.59 & 0.40 & 2.77 & 0.12 & 0.53 & -1.84 \\
323 & 0.839 & 21.29 & 2 & 16.77 & 36.11 & 0.56 & 3.49 & 0.15 & 0.51 & -2.29 \\
411 & 0.952 & 22.56 & 1 & 6.52 & 15.84 & 0.23 & 2.95 & 0.08 & 0.54 & -1.98 \\
413 & 1.023 & 21.46 & 3 & 31.00 & 83.23 & 0.31 & 3.51 & 0.39 & 0.50 & -1.86 \\
2208 & 1.130 & 23.04 & 1 & 7.99 & 21.86 & 0.25 & 3.35 & 0.23 & 0.58 & -1.90 \\
2210 & 0.968 & 21.48 & 1 & 13.38 & 35.99 & 0.18 & 3.75 & 0.10 & 0.57 & -2.28 \\
2237 & 0.944 & 20.94 & 2 & 12.57 & 34.64 & 0.44 & 3.97 & 0.14 & 0.58 & -2.33 \\
2289 & 0.833 & 20.79 & 1 & 6.68 & 15.90 & 0.29 & 3.69 & 0.09 & 0.61 & -2.18 \\
2440 & 1.175 & 22.27 & 1 & 9.62 & 22.94 & 0.19 & 3.05 & 0.12 & 0.58 & -1.92 \\
2597 & 1.119 & 22.89 & 1 & 9.60 & 29.59 & 0.20 & 3.52 & 0.20 & 0.60 & -2.06 \\
5033 & 0.670 & 21.02 & 3 & 18.95 & 39.91 & 0.20 & 1.99 & 0.21 & 0.53 & -1.17 \\
5053 & 1.200 & 23.83 & 1 & 8.78 & 29.95 & 0.29 & 3.25 & 0.24 & 0.53 & -2.06 \\
5229 & 1.030 & 23.32 & 1 & 6.46 & 14.65 & 0.23 & 2.95 & 0.13 & 0.56 & -1.90 \\
5288 & 1.000 & 21.98 & 1 & 7.52 & 18.34 & 0.26 & 3.09 & 0.06 & 0.57 & -1.96 \\
5326 & 0.950 & 23.97 & 3 & 7.76 & 17.84 & 0.52 & 2.69 & 0.32 & 0.56 & -0.99 \\
5400 & 0.680 & 21.18 & 1 & 10.04 & 23.91 & 0.20 & 3.42 & 0.07 & 0.62 & -2.18 \\
5414 & 0.910 & 22.59 & 1 &  8.60 & 19.75 & 0.46 & 2.75 & 0.09 & 0.54 &  -1.81 \\
5427 & 1.177 & 22.13 & 2 & 18.08 & 45.91 & 0.26 & 3.24 & 0.21 & 0.54 & -2.11 \\
5441 & 1.170 & 22.77 & 1 & 14.84 & 38.39 & 0.30 & 3.81 & 0.24 & 0.51 & -2.16 \\
5467 & 0.870 & 23.13 & 1 & 6.63 & 15.94 & 0.11 & 2.94 & 0.11 & 0.57 & -1.94 \\
5496 & 0.694 & 21.06 & 3 & 12.10 & 27.63 & 0.41 & 3.28 & 0.23 & 0.59 & -1.84 \\
5546 & 1.200 & 22.25 & 1 & 4.16 & 8.85 & 0.10 & 2.80 & 0.14 & 0.56 & -1.81 \\
5547 & 0.960 & 23.77 & 2 & 6.37 & 20.50 & 0.15 & 3.75 & 0.20 & 0.59 & -2.14 \\
5548 & 1.060 & 22.77 & 2 & 13.83 & 30.79 & 0.48 & 3.21 & 0.16 & 0.53 & -2.14 \\
54514 & 0.707 & 20.41 & 1 & 13.76 & 42.68 & 0.32 & 3.72 & 0.11 & 0.66 & -2.12 \\
60314 &	1.080 & 23.26 & 1 &  6.51 & 15.56 & 0.24 & 2.69 & 0.13 & 0.55 &  -1.81 \\     
\hline               
\end{tabular}
\end{table*}

\onecolumn
\newpage
\begin{appendix}
\section{SED fits}

SED decomposition for the sample of Type-2 QSOs. Black circles are the
observed photometry in the rest-frame. Purple and blue lines
correspond respectively to the galaxy and AGN template found as the
best fit solution, while the black line shows their sum. The triangle
is the predicted AGN emission at 12.3 $\mu m$ based on the
\citet{gandhi09} correlation. We tried to anchor the AGN template
to this value allowing an uncertainty egual to the dispersion observed
by \citet{gandhi09} in their correlation.

\end{appendix}

\end{document}